\documentclass[twoside,12pt,graphicx,psfig]{article}
\usepackage{epsfig}
\usepackage{graphicx}
\usepackage{amssymb}

\newcommand{\br}{{\bf r}}

\newcommand{\be}{\begin{equation}}
\newcommand{\ee}{\end{equation}}
\newcommand{\bea}{\begin{eqnarray}}
\newcommand{\eea}{\end{eqnarray}}

\begin{document}

\title{ \vspace{1cm} 
Tensor interaction
in mean-field and density functional theory approaches to nuclear
structure
}

\author{Hiroyuki Sagawa,$^{1,2}$ Gianluca Col\`o,$^{3,4}$
\\
$^1$Center for Mathematics and Physics, University of Aizu\\
Aizu-Wakamatsu, Fukushima 965-8560, Japan\\
$^2$Nishina Center, RIKEN, Wako, Japan\\
$^3$Dipartimento di Fisica, Universit\`a
degli Studi di Milano,\\ via Celoria 16, 20133 Milano, Italy\\
$^4$INFN, Sezione di Milano, via Celoria 16, 20133 Milano, Italy}

\maketitle

\begin{abstract} 
The importance of the tensor force for nuclear structure has
been recognized long ago. Recently, the interest for this topic
has been revived by the study of the evolution of nuclear
properties far from the stability line. However, in the
context of the effective theories that describe medium-heavy
nuclei, the role of the tensor force is still debated. This
review focuses on ground-state properties like masses and
deformation, on single-particle states, and on excited
vibrational and rotational modes. The goal is to assess
which properties, if any, can bring clear signatures 
of the tensor force 
within the mean-field or density functional theory framework.
It will be concluded that, while evidences for a clear
neutron-proton tensor force exist despite quantitative
uncertainties, the role of the tensor force among equal
particles is less well established.
\end{abstract}

\section{Introduction}

As discussed in all nuclear physics textbooks, the tensor force
is one of the important components of the nucleon-nucleon
(NN) interaction. The long-range part of this interaction
is associated with the exchange of the lightest meson, namely
the pion, and it has a tensor character; 
the one pion exchange potential (OPEP)  
is the most well known part of the NN interaction yet at the same 
time it is 
not strong enough 
to bind two nucleons, because its expectation
value is of the same order of magnitude of the kinetic energy
associated with the relative motion. However, the tensor force is 
 of paramount importance for the nuclear binding since
the exchange of two pions, i.e., the second order effect 
of 
the tensor force, is providing a strong central  attraction 
in the isospin zero ($T=0$) channel which is responsible for
the deuteron binding. At the same time, the electric quadrupole
moment of the deuteron is a signature of the pure tensor
component. The introduction of the tensor force dates back to
the early 1940s \cite{Bethe,Rarita1,Rarita2}, not so long after the birth 
of nuclear physics (see also \cite{BW}). 

All this belongs to conventional nuclear physics wisdom.
More qualitative and especially quantitative understanding 
has been obtained later concerning the role of tensor
terms in the bare NN interaction, since sophisticated 
interactions have been built that can explain at the same
time the deuteron and many high-precision scattering data
with a $\chi^2/$datum of the order of $\approx$ 1. This can
be done within the traditional picture of the NN force
and, to some extent, also within effective field theories
(EFTs) based on the chiral symmetry and its breaking
(see, e.g., \cite{Epelbaum} for recent reviews). 

However, our focus is different and concerns the role
of tensor terms when the interaction in the nuclear medium
is considered, in particular within the framework of
those models that can be applied throughout the whole
periodic table like self-consistent mean-field or
density functional theory (DFT) based methods 
\cite{Bender}. There is a new blooming of studies of
the possible tensor terms, in general because of the
tremendous progress achieved by these theoretical methods 
in the last decades but also for specific reasons that
we shall briefly illustrate.

A very important motivation to revive the study of the
tensor force in nuclear physics is related to the new
domain of exotic, unstable nuclei. Generally speaking, 
nuclei far from the stability valley open a new test 
ground for nuclear models. Recently, many experimental 
and theoretical efforts have been devoted to the study of the structure 
and the reaction mechanisms in nuclei near the drip lines. 
Modern radioactive ion beam facilities (RIBFs) and experimental 
detectors reveal several unexpected phenomena in unstable nuclei 
such as the existence of haloes and skins~\cite{Tani},
the modifications of shell closures~\cite{Ozawa}
and the so-called pygmy resonances in electric dipole 
transitions~\cite{Pigmy}. The tensor force plays a role, 
in particular, in the shell evolution of nuclei far from 
the stability line \cite{Otsuka}. This fact has motivated 
thoroughly studies of the effect of the tensor force 
on the shell structure of both stable and unstable nuclei,
with emphasis also on masses, single-particle states
and sometimes on the onset of deformation.

Another context in which the tensor force is expected
to be crucial are the properties of spin and spin-isospin
states. Such modes of nuclear excitation, like the 
Gamow-Teller or spin-dipole states, are not only of interest
for nuclear structure but also for nuclear astrophysics
and particle physics (in connection with $\beta$ and 
$\beta\beta$ decays, neutrino mass and its possible Majorana
nature). Review papers have been devoted to this topic 
\cite{Osterfeld,Ichimura}. Many of the currently employed 
mean-field or DFT-based methods are not well calibrated 
in the spin-isospin channel and/or suffer from 
spin and spin-isospin instabilities (i.e., spontaneous
magnetization) above some critical density (see, e.g.,
\cite{Navarro} and references therein). There is
obviously a strong need to improve the predictive power of the
models as far as the spin-isospin states are concerned,
and it must be established to which extent the tensor
terms are important in this channel.

Our review is somehow timely because most of this discussion
concerning the tensor force in stable and unstable nuclei,
and its role for collective excitations (mainly spin and
spin-isospin modes but also density modes and rotations
of deformed nuclei), has already produced a considerable 
number of results, and yet some of the key questions are not 
completely solved.

One of the fundamental issues is related to whether the effective
tensor force that governs shell evolution and excited states
in complex nuclei keeps a close resemblance with the original
bare tensor force or not. In some of the works that we shall
discuss, the adopted point of view is that at least the 
proton-neutron tensor force is only slightly renormalized
in the nuclear medium because of its long-range
character; in other words, some authors believe that 
finite nuclei still bear signatures of the tail of the 
OPEP. In other cases it is 
assumed that the potential is actually renormalized but the
bare OPEP (together with the tensor components associated with
exchange of other mesons) still must be taken as guideline.
Another, complementary point of view is that in the 
mean-field or DFT description the effective interaction
does not need to bear any connection with the interaction
in the vacuum. When its parameters are fitted, the effect
of the bare tensor force is likely to be reabsorbed by
the central force parameters because, as we recalled
at the start of this Introduction, the 
second order effect 
of the  tensor interaction gives rise to central terms. In this
context, the search for an appropriate ``remnant'' of
the tensor force in nuclei (or nuclear matter, or neutron
stars) is mainly driven by a careful attempt to reproduce
the data.

Using these arguments as a red thread throughout our paper,
we will review the recent attempts to implement tensor
terms and compare with experimental data. In particular,
the outline of our paper is as follows. In Sec. 2 we
will discuss the formal aspects related to the tensor
force, starting from a brief summary of the origin of
the bare tensor force. We first discuss relativistic
theories like Relativistic Mean Field (RMF) or Relativistic
Hartree-Fock (RHF) based on the meson-exchange picture,
and move then to nonrelativistic models by distinguishing
the zero-range Skyrme interactions and the finite-range
ones. In Sec. 3 we analyze the results obtained by the
various groups, starting from ground-state properties 
and moving then to excited states. A specific section
(Sec. 4) will be devoted to the role of the tensor
force to drive instabilities (mainly the ferromagnetic
instability we have mentioned above, but not only) in nuclear matter.
Since instabilities are not present in realistic
calculations one should remove them or push them to
high densities, but at present the main effort is devoted
to identifying these instabilities and the role of the
tensor force in producing them. We will draw some
conclusions in Sec. 5.

\section{Bare and effective tensor forces}
\subsection{Discussion of the bare meson exchange potentials}

\begin{figure}[htp*]
\centering
\includegraphics[width=16cm,clip,bb=0.0 0.0 720.0 540.0]{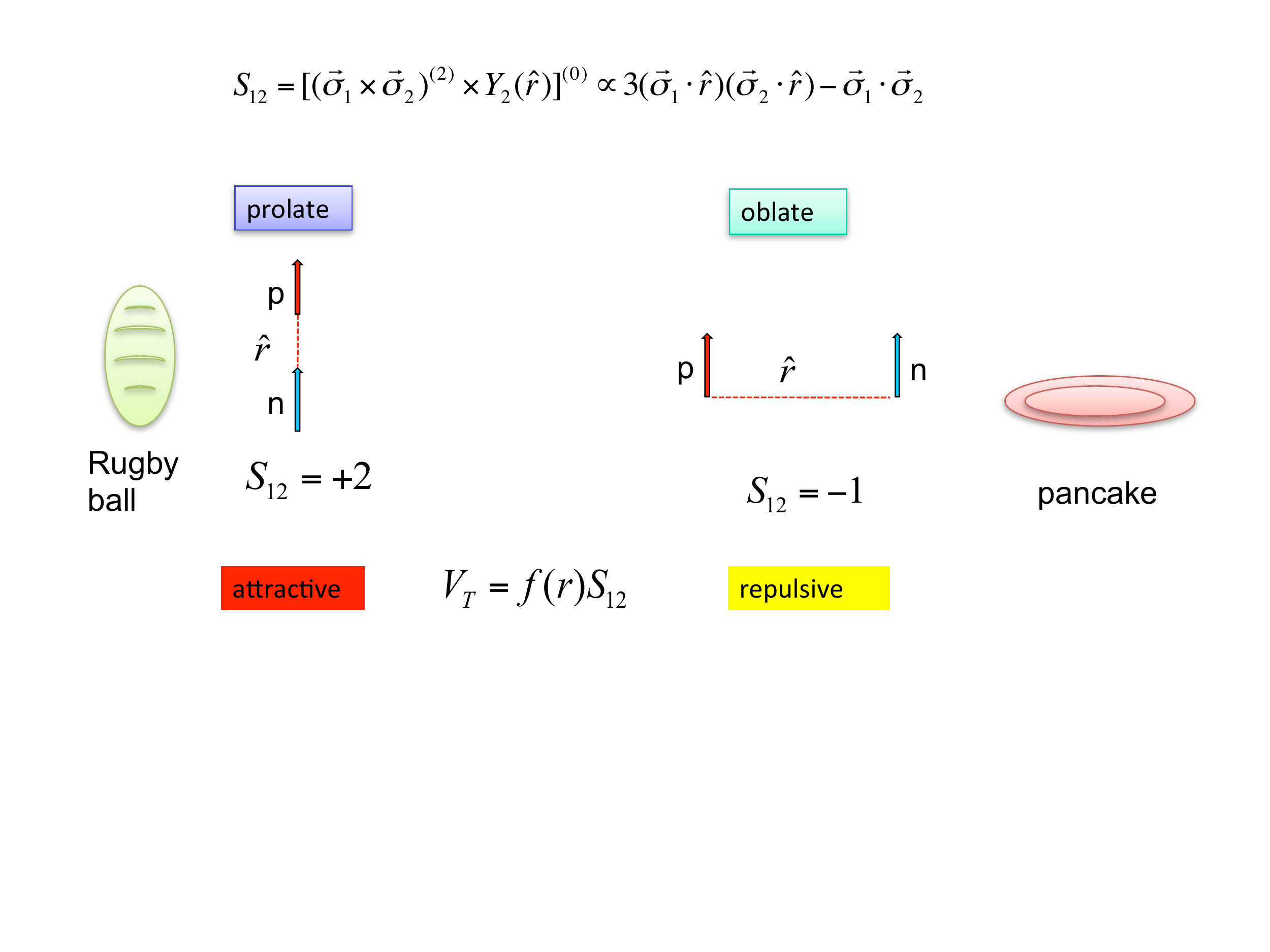}
\vspace*{-1cm}
\caption{\label{fig:a1}
Schematic picture of the expectation values of the tensor
operator $S_{12}$ when the spins are either aligned with (prolate
configuration) or perpendicular to (oblate configuration) 
the relative distance vector $\vec r$. The function $f(r)$ is
negative, favouring a prolate shape for the deuteron 
\cite{BW}. 
}
\end{figure}

One of the first experimental facts that suggested the presence of
a tensor component in the NN interaction is the quadrupole moment of
the deuteron.
More generally, 
it is known that the tensor force plays an essential
role to produce the binding for systems like the deuteron.
As is shown schematically in Fig. \ref{fig:a1}, the tensor
interaction acts on the spin triplet state ($S=1$) of a two nucleon system.
When the relative distance vector of a proton and a neutron is
aligned in the direction of the two spins, 
the deuteron gets an extra binding energy
through the action of the tensor interaction,
\be
V_T=f(r)S_{12}, 
\label{eq:tensor1} 
\ee
since $f(r)$ is negative and $S_{12}=3(\vec{\sigma}_1\cdot\hat{r})
(\vec{\sigma}_2\cdot\hat{r})-\vec{\sigma}_1\cdot\vec{\sigma}_2$ = 2.
On the other hand, if the relative distance vector 
of a proton and a neutron is perpendicular
to
the spins direction, the deuteron will loose its binding energy since
$S_{12}=-1$.
Thus, the deuteron assumes a deformed
prolate shape and makes a bound
system only when the tensor correlations are taken into account.

The
main origin of tensor interactions stems from the $\pi-$nucleon coupling 
and the tensor part of the $\rho-$nucleon couplings: these
are spin and isospin dependent. The isoscalar tensor coupling 
arises from the tensor $\omega-$nucleon coupling.  
In momentum space, the one-pion exchange reads \cite{Yukawa1935}
\begin{eqnarray}
V_{OPEP}({\vec k})=- \frac{4\pi f_{\pi}^2}{m_{\pi}^2}{\vec \tau_1}
\cdot{\vec \tau_2} \frac{{(\vec \sigma_1 \cdot \vec k)(\vec \sigma_2 
\cdot \vec k)}}
{k^2+m_{\pi}^2}.
\label{eq:OPEP}
\end{eqnarray}
Eq. (\ref{eq:OPEP}) is decomposed into a central and a tensor part as
\begin{eqnarray}
V_{OPEP}({\vec k})=- \frac{4\pi f_{\pi}^2}{3m_{\pi}^2}{\vec \tau_1}
\cdot {\vec \tau_2} \left[\frac{3{(\vec \sigma_1 \cdot \vec k)
\  (\vec\sigma_2 \cdot \vec k) - 
\vec \sigma_1 \cdot \vec \sigma_2} k^2}{k^2+m_{\pi}^2} + 
{\vec\sigma_1 \cdot \vec\sigma_2 }
\left(1-\frac{ m_{\pi}^2}{k^2+m_{\pi}^2} \right)   \right] ,
\label{eq:OPEPT}
\end{eqnarray}
where the pseudo-vector $\pi-$nucleon coupling constant is 
$f_{\pi}^2$=0.08, the pion mass is $m_{\pi}$=138 MeV 
and ${\vec k}$ is the momentum transfer. The first term in the bracket 
of Eq. (\ref{eq:OPEPT}) is a tensor interaction 
and the second term is a central interaction.  
The momentum-indpendent part of the ${\vec\sigma_1 \cdot \vec \sigma_2}$ 
term of the central part (the term 1 in the round bracket) will 
become $\delta({\vec r_1-\vec r_2})$ in the Fourier transform to the 
coordinate space, and can be dropped because it is
overcome by the short-range NN repulsion. 
$V_{OPEP}$ is transformed to the coordinate space as follows,
\begin{eqnarray}
V_{OPEP}({\vec r})=f_{\pi}^2m_{\pi}{\vec \tau_1}\cdot 
{\vec \tau_2}\left[\left(\frac{1}{3m_{\pi}r}+\frac{1}
{(m_{\pi}r)^2}+\frac{1}{(m_{\pi}r)^3} \right) e^{-m_{\pi}r}S_{12} 
+\frac{1}{3}{\vec \sigma_1 \cdot \vec\sigma_2 } 
\frac{e^{-m_{\pi}r}}{m_{\pi}r} \right].
\label{eq:OPEPr}
\end{eqnarray}

The $\rho-$meson exchange potential resulting from the 
tensor $\rho-$nucleon coupling reads in the momentum space
\begin{eqnarray}
V_{\rho}({\vec k})=- \frac{4\pi f_{\rho}^2}{m_{\rho}^2}
{\vec \tau_1}\cdot{\vec \tau_2} \frac{{(\vec \sigma_1 \times \vec k) 
(\vec \sigma_2 \times \vec k)}}{k^2+m_{\rho}^2}.
\label{eq:Vrho}
\end{eqnarray}
This potential has a very similar structure to the $\pi-$exchange one,
\begin{eqnarray}
V_{\rho}({\vec r})=f_{\rho}^2m_{\rho}{\vec \tau_1}\cdot {\vec \tau_2}
\left[-\left(\frac{1}{3m_{\rho}r}+\frac{1}{(m_{\rho}r)^2}
+\frac{1}{(m_{\rho}r)^3} \right) e^{-m_{\rho}r}S_{12} 
+\frac{2}{3}{\vec \sigma_1 \cdot \vec \sigma_2 } 
\left( \frac{e^{-m_{\rho}r}}{m_{\rho}r}- \frac{4\pi}{m_{\rho}^3} 
\delta({\vec r})\right) \right],
\label{eq:Vrhor}
\end{eqnarray}
where $m_{\rho}$=770 MeV and $f_{\rho}^2$=4.86. One should 
notice that the tensor 
part of the potential has the opposite sign compared to the $\pi-$exchange 
one. 
In general, the tensor part of 
the $\rho-$exchange potential has a much shorter range character and gives a
smaller contribution to the matrix elements than the $\pi-$ exchange one, 
because of the much larger $\rho$ meson mass compared with the 
pion mass.  

The $\omega-$tensor exchange potential is written as
\begin{eqnarray}
V_{\omega}({\vec k})=- \frac{4\pi f_{\omega}^2}{m_{\omega}^2} 
\frac{{ (\vec \sigma_1 \times \vec k) (\vec \sigma_2 
\times \vec k)}}{k^2+m_{\rho}^2}
\label{eq:Vomega}
\end{eqnarray}
in the momentum space. 
This tensor potential has no isospin dependence. 
The $\omega-$tensor potential in the coordinate space
becomes
\begin{eqnarray}
V_{\omega}({\vec r})=f_{\omega}^2m_{\omega}
\left[-\left(\frac{1}{3m_{\omega}r}+\frac{1}{(m_{\omega}r)^2}
+\frac{1}{(m_{\omega}r)^3} \right) e^{-m_{\omega}r}S_{12}  
+\frac{2}{3}{\vec \sigma_1 \cdot \vec \sigma_2 } 
\left( \frac{e^{-m_{\omega}r}}{m_{\omega}r}- 
\frac{4\pi}{m_{\omega}^3}\delta({\vec r}) \right) \right],
\label{eq:Vomegar}
\end{eqnarray}
where the mass of the omega meson is $m_{\omega}$ = 783 MeV.

\subsection{Tensor interactions in relativistic Hartree-Fock model}
\label{tensorRHF}
The relativistic mean field (RMF) model was originally based on the 
Hartree theory and involved only the scalar-isoscalar and vector-isoscalar 
mesons, namely the $\sigma$ and $\omega$ mesons with 
($J^{\pi},T$) = (0$^{+}$,0) and (1$^{-}$,0) respectively. 
Then the vector-isovector $\rho$ meson with ($J^{\pi},T$) = 
(1$^{-}$,1) was 
also introduced to get reasonable agreement with the experimental 
systematics of masses and radii.
Within the Hartree model, the $\pi$ meson$-$ and $\rho-$ and $\omega-$ 
tensor couplings do not give any contributions in the static 
mean field calculations.  
In the 1980s, there were several attempts to extend the 
RMF model including the Fock term, which can accommodate the 
$\pi -$N coupling, as well as the $\rho$-tensor and also $\omega$-tensor  
interactions. A. Bouyssy {\it et al.} \cite{Bouyssy1987}  
did the first attempt to perform relativistic Hartree-Fock (RHF) calculations.   
The effective Hamiltonian density 
in the covariant relativistic model can be obtained
from the Lagrangian density $\cal L$ through the general Legendre transformation
\be
H=\frac{\partial {\cal L}}{\partial \dot\phi_i} \dot\phi_i - {\cal L}, 
\ee
where $\phi_i$ is the field operator.
This leads to the general form for the effective Hamiltonian in the 
nucleon space:
\be
{\it H}={\bar \psi}(x_1)(-i{\gamma} \cdot {\partial }+M) 
\psi(x_1) +\frac{1}{2} \int d^4 x_2 \sum_{i=\sigma,\omega,\rho,\pi, A }
{\bar \psi}(x_1){\bar \psi}(x_2) \Gamma_iD_i(x_1,x_2)\psi(x_2)\psi(x_1), 
\ee
where $\psi(x)$ is the nucleon field operator and  
$D_i(x_1,x_2)$ represents the corresponding meson propagators. 
The interaction vertices
$\Gamma_i$ for mesons are defined as 
\bea
\Gamma_{\sigma}(1,2)&=&-g_{\sigma}(1) g_{\sigma}(2),            \\
\Gamma_{\omega}(1,2)&=&+g_{\omega}(1) \gamma_{\mu}(1) g_{\omega}(2) 
\gamma^{\mu}(2), \\
\Gamma_{\rho} (1,2)&=& +g_{\rho}(1) \gamma_{\mu}(1) {\vec \tau}(1) 
\cdot g_{\omega}(2) \gamma^{\mu}(2) {\vec \tau}(2),         \\
\Gamma_{\pi} (1,2)&=&-\left[\frac{f_{\pi}}{m_{\pi}}  {\vec \tau}
\gamma_{5}\gamma_{\mu} \partial^{\mu}\right]_1\cdot  
\left[\frac{f_{\pi}}{m_{\pi}}  {\vec \tau}\gamma_{5}
\gamma_{\mu} \partial^{\mu}\right]_2.    
\eea
In the point coupling models, the meson-nucleon couplings 
$g_{\sigma}, g_{\omega}, g_{\rho}$ and $f_{\pi}$ are taken 
to be constants. In the density dependent coupling model,
these couplings are assumed to be a function of the baryonic density.
In the RHF model of  
Ref. \cite{Bouyssy1987}, the authors took the bare point coupling for
the $\pi-$N vertex, and also for
the $\rho-$N and $\omega-$N tensor couplings. Their coupling constants  
were adjusted to obtain good fits of both nuclear matter and 
finite nuclear properties.
While the charge distributions are described well by this pioneering 
RHF method, nuclei were underbound 
due to the missing meson self-interactions.  　
Some improvements were obtained by taking into account the 
non-linear $\sigma-$couplings, but the RHF model based on the 
point couplings was not comparable with the RMF model 
in the quantitative description of nuclear observables.     

\begin{figure}[ht]
\centering
\includegraphics[width=8cm,clip]{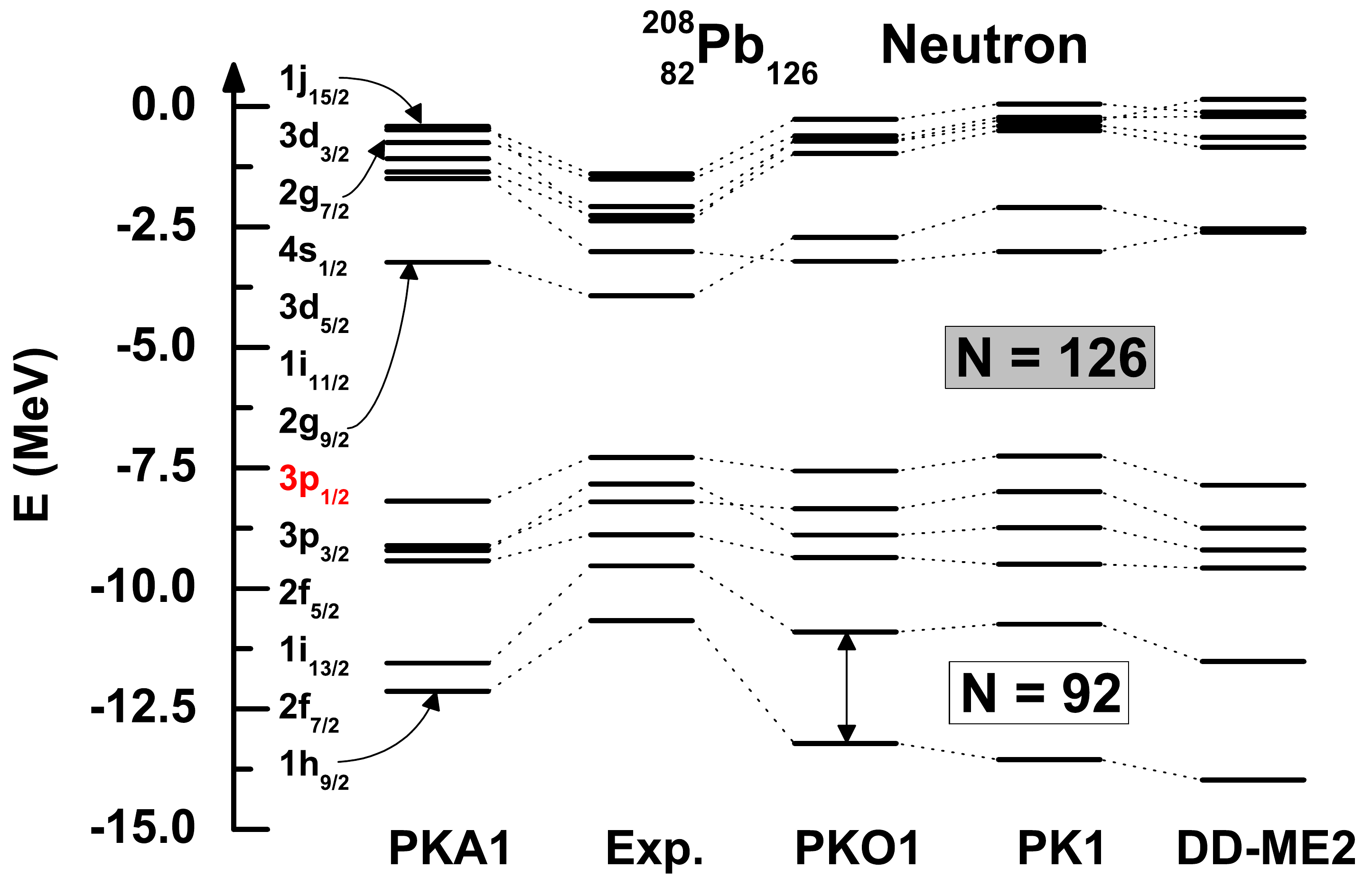}
\includegraphics[width=8cm,clip]{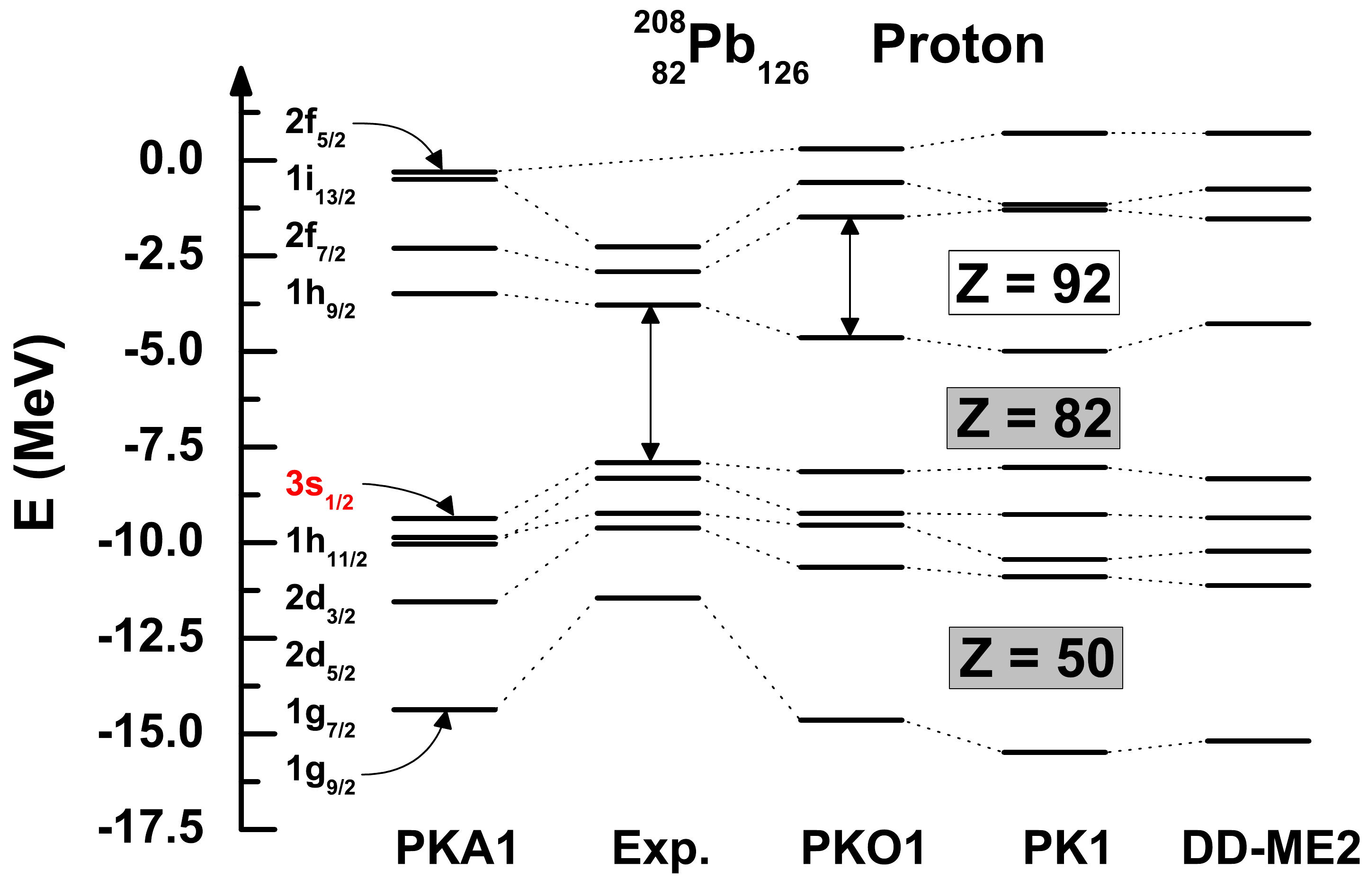}
\vspace*{0.5cm}
\caption{\label{fig:RHF-Pb}
Single-particle energies of $^{208}$Pb \cite{Long-priv}. 
The calculations are performed 
using the RHF model with the Lagrangians PKA1 \cite{Long2007} 
and PKO1 \cite{PKO1}, and 
using the RMF model with PK1 \cite{PK1} and DD-ME2 \cite{DDME2}.
The experimental data are taken from Ref. \cite{Oros1996}.
}
\end{figure}

Twenty years after the introduction of the point-coupling 
RHF model, W.H. Long {\it 
et al.} \cite{Long2006} introduced a RHF model with density 
dependent couplings between mesons and nucleons. The 
tensor coupling due to the pion exchange, and the vector coupling 
associated with the $\rho -$exchange, were introduced 
in the effective Lagrangian. The  $\pi-$ and $\rho-$ couplings in the model 
are the bare couplings in the free space, but they are 
very much quenched in the nuclear medium. 
The authors of Ref. \cite{Long2006} obtained good agreement 
with the experimental data for 
the ground state properties (masses and radii of several closed shell nuclei), 
at the same level or even better than 
the RMF models.
Moreover, the description of the nuclear effective mass and of 
its isospin and energy dependence have been improved by the RHF model.  
More recently \cite{Long2007}, 
it was shown that the RHF model can remove the problem of artificial 
shell closures at N, Z = 58 and 92 existing in some 
RMF calculations, by taking into account the 
Fock term associated with the pion and the $\rho-$tensor interactions 
(cf. Fig. \ref{fig:RHF-Pb}).  
The RHF+RPA model has also been developed for the change-exchange 
excitations, and it has provided a good description of Gamow-Teller 
and spin-dipole states  \cite{Liang2008}. 

\subsection{From bare tensor to Skyrme tensor}
\label{Skyrme}

In the first paper by T.H.R. Skyrme \cite{Skyrme1}, 
the central part of the effective interaction was introduced 
in a simplified form for the case of finite nuclei in the 
same spirit of Bruckner's self-consistent 
nuclear model. The simplified interaction has all the terms 
proportional to a zero-range $\delta-$function, but includes 
quadratic momentum-dependent terms that mimick 
a finite-range interaction. In the subsequent paper by J.S. Bell 
and T.H.R. Skyrme \cite{Skyrme2}, the spin-orbit coupling 
was introduced and the resultant 
potential has a radial dependence of the type $\sim r^{-1}d\rho/dr$,  
with a reasonable 
magnitude required to explain the empirical spin-orbit 
splitting of nuclei with mass A=16$\pm1$.   
The tensor terms were introduced 
in 1959 by  T.H.R. Skyrme \cite{Skyrme} 
under the requirement of making a complete parametrization of zero-range,   
momentum-dependent effective interactions. But 
nothing was mentioned about the role of these tensor terms in 
the mean field calculations.
These tensor terms have been neglected in the seminal
paper by D. Vautherin and D.M. Brink \cite{Vautherin-Brink}, and also
in most of the Skyrme parameter sets that have been fitted in the next
20 years. 

There have been a few exceptions. 
In fact, the effect of the tensor interactions was first 
discussed for the single-particle energies 
in the
Hartree-Fock (HF) calculations by Fl. Stancu {\it et al.} 
\cite{Stancu:1977}. We will discuss their findings in Sec. 3.2.
In a somewhat different context, the tensor effect on the spin-orbit splittings were discussed in the papers
by J. Dudek {\it et al.} \cite{Dudek}, and also by M. Ploszajczak 
and M.E. Faber \cite{Ploszajczak}. 
F. Tondeur \cite{Tondeur} included the spin-orbit densities $\vec J$ 
(see below) 
in his construction of the Skyrme energy density functional for 
HF+Bardeen-Cooper-Schrieffer (HF+BCS) calculations.  
He chose the value of the parameters 
giving the dependence of the energy density on $\vec J$
to fit the spin-orbit splittings of 
$^{16}$O, $^{48}$Ca and $^{208}$Pb, and   
implicitly included the tensor effects on the spin-orbit splitting.
K.-F. Liu {\it et al.} \cite{Liu:1991} worked out the energy-weighted 
and non-energy weighted sum rules
of the electromagnetic transitions, 
Fermi transitions and Gamow-Teller 
transitions including the tensor interactions. However, these
authors never 
performed any practical calculation 
of the sum rule values including the tensor correlations.
In summary, 
we can say that serious attempts to clarify the role 
of the tensor correlations  
have never been performed until very recently, when such
studies have been 
started with the aim of clarifying
the shell evolutions in the exotic nuclei (cf. Sec. \ref{spstates}). 

\begin{figure}[ht]
\centering
\includegraphics[width=10cm,clip]{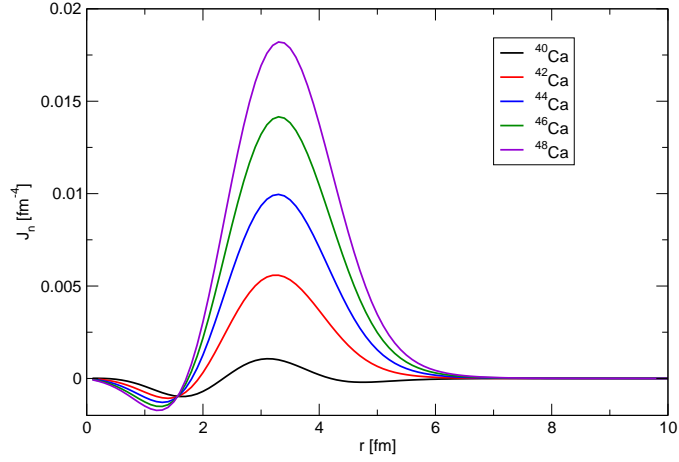}
\caption{\label{fig:spin-density}
Neutron spin-orbit density $J_n$ of the Ca 
isotopes. Since $^{40}$Ca is a $\vec l\cdot\vec s$-saturated nucleus 
the spin-orbit density is negligible. In the heavier isotopes, the
contribution of the f$_{7/2}$ orbit is making the spin-orbit
density non-negligible.}.
\end{figure}

We provide now the basic formulas for the tensor force within
the Skyrme framework.
The Skyrme tensor interaction is the sum of
the triplet-even and triplet-odd tensor zero-range tensor parts, namely
\begin{eqnarray}
v_T &=& {T\over 2} \left\{ \left[ (\vec \sigma_1\cdot {k^\prime})
(\vec \sigma_2\cdot {k^\prime}) - {1\over 3}(\vec \sigma_1\cdot\vec \sigma_2)
{k^\prime}^2 \right] \delta({\vec r_1}-{\vec r_2}) \right.
\nonumber \\
&+& \left. \delta({\vec r_1}-{\vec r_2})
\left[ (\vec \sigma_1\cdot {k})(\vec \sigma_2\cdot {k}) - {1\over 3}
(\vec\sigma_1\cdot\vec\sigma_2) {k}^2 \right] \right\} 
\nonumber\\
&+& U \left\{ (\vec \sigma_1\cdot {k^\prime}) \delta({\vec r_1}-{\vec r_2}) 
(\vec\sigma_2\cdot {k}) - {1\over 3} (\vec\sigma_1\cdot\vec\sigma_2) 
\left[ {k^\prime}\cdot \delta({\vec r_1}-{\vec r_2}) {k} 
\right] \right\},
\label{eq:tensor}
\end{eqnarray}
where the operator ${k}=(\overrightarrow{\nabla}_1-
\overrightarrow{\nabla}_2)/2i$ acts on the right and
${k'}=-(\overleftarrow{\nabla}_1-\overleftarrow{\nabla}_2)/2i$ on the left.
The coupling constants $T$ and $U$ denote the strength of the
triplet-even and triplet-odd tensor interactions, respectively.
The tensor terms (\ref{eq:tensor}) give contributions to the
binding energy and to the spin-orbit splitting 
that are proportional to the spin-orbit density $\vec J$. In spherical 
nuclei only the radial component of this vector does not
vanish and its expression reads \cite{Vautherin-Brink}
\begin{equation}
 J_q(r)=\frac{1}{4\pi r^3}\sum_{i}v_{i}^2(2j_{i}+1)\left[j_i(j_i+1)
           -l_i(l_i+1) -\frac{3}{4}\right]R_i^2(r),
\label{eq:sd}
\end{equation}
where $i=n,l,j$ runs over all  states and $q=0 (1)$ is the
quantum number 
$(1-t_z)/2$ ($t_z$ being the third isospin component) 
for neutrons (protons).
The quantity $v_{i}^2$ is the occupation probability
of each orbit determined by the BCS approximation and $R_i(r)$ is
the 
radial part of the 
HF single-particle wave function (cf. Fig. \ref{fig:spin-density}). 
In the HF-Bogolyubov (HFB) 
approximation, $v_{i}R_i(r)$ will be
replaced by the lower component of the HFB wave fucntion $V_i(r)$. 
It should be noticed that
the exchange part of the central Skyrme interaction gives
the same kind of contributions to the total energy density.
The central exchange and tensor
contributions give extra terms to the energy density that read
\begin{equation}
\delta E= {1\over 2}\alpha(J_n^2+J_p^2) + \beta J_n J_p.
\label{eq:dE}
\end{equation}

The spin-orbit potential is expressed as
\bea
V_{s.o.}^{(q)}(r)=U_{s.o.}^{(q)}(r){\vec l}\cdot{\vec s}
\eea  \label{eq:Vso}
with
\bea
U_{s.o.}^{(q)}(r) = {W_0\over 2r} \left( 2{d\rho_q\over dr} +
{d\rho_{1-q}\over dr} \right) + \left( \alpha
{J_q\over r} + \beta {J_{1-q}\over r} \right).
\label{eq:dW}
\eea
In Eq. (\ref{eq:dW}), 
the first term on the r.h.s comes from the Skyrme
spin-orbit interaction and the second term include contributions
both from the exchange part of the central force and from 
the tensor terms, that is, $\alpha= \alpha_C +\alpha_T$
and $\beta=\beta_C +\beta_T$.  
The central exchange contributions are
given by
\begin{eqnarray}
\alpha_C & = & {1\over 8}(t_1-t_2) - {1\over 8}
(t_1x_1+t_2x_2), \nonumber\\
\beta_C & = & -{1\over 8}(t_1x_1+t_2x_2), 
\label{eq:dWc}
\end{eqnarray}
in terms of the parameters of the Skyrme force 
as defined 
in Refs. \cite{Skyrme,Vautherin-Brink,Bender}.
The tensor contribution are expressed as
\begin{eqnarray}
\alpha_T & = & {5\over 12}U, \hfill\nonumber\\
\beta_T & = & {5\over 24}(T+U),
\label{eq:dWT}
\end{eqnarray}
in terms of the triplet-even and triplet-odd terms 
appearing in Eq. (\ref{eq:tensor}).

It is useful from the very beginning to stress that, at
variance with the case of the standard central Skyrme
force, the notation for the tensor parameters is not unique.
Other authors write the tensor force as
\begin{eqnarray}
v_T &=& {t_e\over 2} \left\{ \left[ 3(\vec\sigma_1\cdot {k^\prime})
(\vec\sigma_2\cdot {k^\prime}) - (\vec\sigma_1\cdot\vec\sigma_2)
{k^\prime}^2 \right] \delta({\vec r_1}-{\vec r_2}) \right.
\nonumber \\
&+& \left. \delta({\vec r_1}-{\vec r_2})
\left[ 3(\vec\sigma_1\cdot {k})(\vec\sigma_2\cdot {k}) - 
(\vec\sigma_1\cdot\vec\sigma_2) {k}^2 \right] \right\} 
\nonumber\\
&+& t_o \left\{ 3(\vec\sigma_1\cdot {k^\prime}) \delta({\vec r_1}-{\vec r_2}) 
(\vec\sigma_2\cdot {k}) - (\vec\sigma_1\cdot\vec\sigma_2) 
\left[ {k^\prime}\cdot \delta({\vec r_1}-{\vec r_2}) {k} 
\right] \right\},
\end{eqnarray}
and of course the following equations hold:
\begin{eqnarray}
t_e & = & \frac{U}{3}, \nonumber \\
t_o & = & \frac{T}{3}, \nonumber \\
\alpha & = & \frac{5}{4}t_o, \nonumber \\
\beta & = & \frac{5}{8}\left( t_e + t_o \right).
\end{eqnarray}

We will now discuss some qualitative arguments that are very 
important to compare tensor forces defined within the Skyrme
framework and in different contexts. 
In general, the tensor interaction can be divided into two parts, 
i.e., the isospin-independent and the isospin-dependent component,
\bea\label{eq:tensor-isospin}
V_T(r)=v^{IS}_T(r)S_{12}+ v^{IV}_T(r)\tau_1\cdot\tau_2S_{12}. 
\eea
We will preferably call these two components isoscalar (IS) and
isovector (IV) tensor force, respectively (other authors 
\cite{Moreno} use instead the terminology ``pure tensor'' 
and ``isospin tensor''). In the HF mean field of spin-saturated 
nuclei, the direct term of the tensor interaction does not 
contribute whereas the exchange term  
gives a finite contribution. In 
the spirit of the density matrix expansion theory by J.W. 
Negele and D. Vautherin \cite{Negele1972}, one can look at 
the expectation value $\frac{1}{2}\sum_{ij}\langle ij \vert 
V \vert ij \rangle$ of the tensor interaction
of Eq. (\ref{eq:tensor-isospin}). The isoscalar tensor $v^{IS}_T(r)S_{12}$ 
gives 
\bea\label{eq:IS-T}
\langle v^{IS}_T({\vec r}_1-{\vec r}_2)S_{12} \rangle 
=-\frac{1}{2}\int d^3r_1d^3r_2\ v^{IS}_T({\vec r}_1-{\vec r}_2)
\left(|{\vec s}_n({\vec r}_1,{\vec r}_2)|^2+
|{\vec s}_p({\vec r}_1,{\vec r}_2)|^2\right),
\eea
where the spin density matrix is defined as
\bea \label{eq:spin-den}
{\vec s}({\vec r}_1,{\vec r}_2) = \sum_{i,\sigma_1,\sigma_2}
\phi_i^*({\vec r}_1,\sigma_1)\langle\sigma_1|{\vec \sigma}|\sigma_2\rangle 
\phi_2({\vec r}_2,\sigma_2).
\eea
The isovector part of the tensor interaction has a different 
dependence on the spin density matrix, namely
\bea \label{eq:IV-T}
\langle v^{IV}_T({\vec r}_1-{\vec r}_2) \vec\tau_1\cdot\vec\tau_2 S_{12} \rangle 
=-\frac{1}{2}\int d^3r_1d^3r_2\ v^{IV}_T({\vec r}_1-{\vec r}_2) 
\left( |{\vec s}_n({\vec r}_1,{\vec r}_2)|^2+
|{\vec s}_p({\vec r}_1,{\vec r}_2)|^2    
+ 2|{\vec s}_n({\vec r}_1,{\vec r}_2)\cdot{\vec s}_p({\vec r}_1,{\vec r}_2)
|\right).  \nonumber \\
\eea
Let us look at this expectation value of the isovector tensor interaction 
in the neutron-proton (np) channel, that is, 
\bea \label{eq:tensor-hf}
\langle v_T^{np} ({\vec r}_1-{\vec r}_2) \vec\tau_1\cdot\vec\tau_2 S_{12} \rangle 
=-\int d^3r_1d^3r_2\ v_T({\vec r}_1-{\vec r}_2)
|{\vec s}_n({\vec r}_1,{\vec r}_2)\cdot{\vec s}_p({\vec r}_1,{\vec r}_2)
|.
\eea
The spin density matrix (\ref{eq:spin-den}) for spherical nuclei 
can be factorized as \cite{Negele1972} 
\bea \label{eq:spin-m}
{\vec s}({\vec r}_1,{\vec r}_2)=i({\vec r}_1 \times {\vec r}_2)
\rho_1({\vec r}_1,{\vec r}_2)
\eea
with
\bea \label{eq:spin-density}
\rho_1({\vec r}_1,{\vec r}_2) = \pm \sum_{nlj}\frac{1}{2\pi r_1^2 r_2^2 }
R_{nlj}(r_1)R_{nlj}(r_2)P_l'(cos\theta).
\eea
Here $P_l'(cos\theta)$ is the derivative of 
the Legendre polynomials $P_l(cos\theta)$, $\theta$
is the angle between $\vec r_1$ and $\vec r_2$
and the sign $\pm$ stands for $j=l\pm1/2$. For a short-range 
interaction with $\theta\sim0$,
$P_l'(cos\theta\sim 1)\sim l(l+1)/2.$  Then Eq. (\ref{eq:spin-density}) 
gives a factor
\bea
\pm l(l+1)=\frac{1}{2}(2j+1)[j(j+1)-l(l+1)-3/4] \,\,\,\,\,\,\,\, 
\mbox{for} \,\,\,\,\,\,\,\, j=l\pm 1/2, 
\eea
which is the same factor in the spin-orbit density $J_q(r)$. 
If the radial wave function is the same for
$j=l\pm 1/2$, the contribution to $\rho_1({\vec r}_1,{\vec r}_2)$ 
vanishes if both orbits $j=l\pm 1/2$ are either completely
occupied or completely empty. 
Accordingly, the tensor interaction can be represented by means of 
the spin-orbit densities
in the short-range limit.

Based on the above picture, D.M. 
Brink and Fl. Stancu \cite{Brink} have examined the validity of 
the Skyrme-type tensor interaction ansatz. 
They have shown that when the finite-range tensor force acts 
among Hartree-Fock wave functions, its matrix elements are to 
a good approximation proportional to those of the zero-range momentum-dependent
force (\ref{eq:tensor}). 
Both for Gaussian and one-pion exchange potentials, 
the proportionality constant is almost the same for nuclei in the 
medium-heavy mass region with different A and also when the
matrix elements involve different angular momenta $l$. In
other words, the action of the long-range tensor can be recast in the form 
(\ref{eq:tensor}) without any significant error because of
the momentum dependence and of the structure of HF wave functions.
Therefeore, the reader should be advised to abandon the point of view that the 
Skyrme tensor force is truly zero-range. 

Moreover, 
it is interesting to notice from Eqs. (\ref{eq:IS-T}) and (\ref{eq:IV-T}) 
that the IS tensor, if rewritten in the zero-range limit in terms of 
the spin-orbit densities, gives only a contribution 
to the total energy [cf. Eq (\ref{eq:dE})] proportional 
to $\alpha$ but no contribution proportional to $\beta$. On the other hand, 
the IV tensor gives two contributions such that $\beta=2\alpha$
(cf. also \cite{Brink}). 

The necessity of having both IS and IV tensor terms can be
seen through a comparison with bare tensor forces.
The Skyrme-type tensor interaction (\ref{eq:tensor}) is essentially 
based on the short-range nature of the nuclear force.  
The one-pion exchange tensor interaction can be written as
\begin{equation}
V_T(r)=v_T(r)\vec\tau_1\cdot\vec\tau_2[\frac{3}{r^2}(
\vec\sigma_1\cdot{\vec r})
(\vec\sigma_2\cdot{\vec r})-
\vec\sigma_1\cdot\vec\sigma_2]
\end{equation}
as was derived in Eq. (\ref{eq:OPEPr}). This one-pion exchange 
tensor interaction is attractive.
The $\rho$-meson exchange potential gives also rise to a 
tensor interaction as was shown in Eq. (\ref{eq:Vrhor}). Its range is 
much shorter than the pion-exchange one
and the sign is positive (i.e., repulsive).   
This is due to the nature of the $\rho$-meson which is characterized by 
a pseudo-vector 
coupling and a mass which is about 6 times heavier than the
pion mass. The tensor force due to the $\rho$-meson exchange 
tends to cancel that of the pion exchange. However, the net
result is expected to be attractive. Thus, the tensor force is 
expected to give an attractive interaction in the np 
channel [that is, a positive value of $\beta$ in keeping with the
minus sign of Eq. (\ref{eq:tensor-hf})] in the deuteron-like
configuration. 
As recalled above, the IS potential is instead coming from
$\omega-$meson exchange.

In the mean-field and density functional theory (DFT) approaches, 
the tensor force is introduced as a phenomenological one in the 
framework of a Slater determinant description of the many-body system. There 
are, however, alternative approaches adopting a bare tensor force. In 
the unitary correlation operator (UCOM) approaches
in Refs.~\cite{UCOM}, 
a careful attempt has been undertaken to study how tensor correlations evolve 
as the model space employed for the calculations is changed. 
In this study, it was pointed out that strong bare tensor correlations 
become much weaker in the transformed potential which should be 
used in a smaller model space such as shell model or HF, 
when the high momentum components 
of the tensor force are taken into account in the 
correlated wave functions.  
Another interesting approach is the one introduced by T. Myo and
co-workers~\cite{Myo}, 
in which the tensor force is considered 
in the framework of a charge- and parity-breaking shell model approach. 
This approach allows a strong mixing 
of odd and even parity states by the tensor correlations 
and nuclear states are eventually obtained by the 
projection techniques. We shall not discuss further these approaches 
that can not be easily related with 
the present mean field or DFT approaches.  


\subsection{Skyrme force and Skyrme energy functional}
\label{skyrmeEDF}

In this subsection we discuss an important aspect related to the introduction of
the tensor force in the Skyrme framework. Let us assume that we start from the 
standard
expression of the central Skyrme force. This expression does not 
include any tensor term: however, as discussed above, the central exchange terms
generate in the energy functional some terms that look the same as those
generated through a tensor force, although the coefficients are fixed 
[$\alpha_C$ and $\beta_C$, cf. Eqs. (\ref{eq:dWc})]. When tensor terms are introduced, 
they give the same kind of contribution to the energy functional, so 
that $\alpha_T$ and $\beta_T$ are summed, respectively, to 
$\alpha_C$ and $\beta_C$ [cf. Eq. (\ref{eq:dWT})]. In short, we
can say that the central and tensor terms are not decoupled 
at all in the Skyrme framework. Moreover, there 
have been attempts in the last decade to go beyond the picture of 
the Skyrme {\em force} 
and fit directly a Skyrme {\em functional}. When 
this is done, at least in the spherical 
nuclei, there is no way of discussing separately the tensor and central exchange 
terms. Therefore, when we discuss the Skyrme results in the following of our paper,
we will try to distinguish carefully the two cases (either force or functional) 
because the way in which the parameters are defined and affect the results is
different. We shall refer to self-consistent mean-field (SCMF) calculations in 
the case in which an effective force, that is, an Hamiltonian picture, is assumed 
as a starting point, and we shall refer instead to density functional 
theory (DFT) 
in the case in which the functional is taken as a starting point. 
In the following of the subsection we
shall define the quantities related to the 
energy density functional (EDF) picture and 
we shall discuss the difference between spherical and deformed case.

The Skyrme energy functional is discussed at length in many papers (see, e.g., 
Ref. \cite{Bender}). It is the prototype of any conceivable local energy
functional ({\em viz.}, a functional of local densities only). To define local densities,
we start from the (non-local) density matrix
\begin{equation}
\rho_q(\vec r \sigma, {\vec r'}\sigma') \equiv \langle \Phi \vert \psi_q^\dagger 
({\vec r'}\sigma') \psi_q(\vec r,\sigma) \vert \Phi \rangle
= \sum_i \phi_{i,q}^*({\vec r'}\sigma') \phi_{i,q}(\vec r,\sigma),
\end{equation}
where $\vert \Phi \rangle$ is the Slater determinant made up with the 
single-particle wave 
functions $\phi_{i,q}$ labelled by a set of quantum numbers $i$. 
From this density matrix we can extract the scalar
density matrix and the spin density matrix as
\begin{equation}
\rho_q(\vec r \sigma, {\vec r'}\sigma') = \frac{1}{2}\rho_q(\vec r, {\vec r'}) \delta_{\sigma\sigma'} +
\frac{1}{2} \vec s_q(\vec r, {\vec r'}) \langle \sigma' \vert \vec \sigma \vert \sigma \rangle.
\end{equation}
In turn, from these non-local densities we extract the local quantities
\begin{eqnarray}\label{loc_densities}
\rho_q(\vec r) & = & \rho_q(\vec r, {\vec r'}) \vert_{\vec r = {\vec r'}}, \nonumber \\
\tau_q(\vec r) & = & \vec \nabla \cdot \vec \nabla' \rho_q(\vec r, {\vec r'}) \vert_{\vec r = {\vec r'}}, \nonumber \\
J_{q,\mu\nu}(\vec r) & = & -\frac{i}{2} \left( \nabla_\mu - \nabla'_\mu \right) s_{q,\nu}(\vec r, {\vec r'}) \vert_{\vec r = {\vec r'}},
\end{eqnarray}
that are called, respectively, density, kinetic energy density and spin-current density. 
These densities are enough if we restrict ourselves to the time-even part of the energy 
functional. If one is concerned also with the time-odd part of the energy
functional, then the following densities need to be considered:
\begin{eqnarray}
\vec j_q(\vec r) & = & \frac{1}{2i} \left( \vec\nabla - \vec\nabla' \right) 
\rho_q(\vec r, {\vec r'}) \vert_{\vec r = {\vec r'}}, \nonumber \\
\vec s_q(\vec r) & = & \vec s_q(\vec r, {\vec r'}) \vert_{\vec r = {\vec r'}}, \nonumber \\
\vec T_q(\vec r) & = & \nabla\cdot\nabla' \vec s_q(\vec r, {\vec r'}) \vert_{\vec r = {\vec r'}}, \nonumber \\
F_{\mu,q} (\vec r) & = & \frac{1}{2} \sum_\nu \left( \nabla_\mu\nabla_\nu' + 
\nabla_\mu' \nabla_\nu \right) s_{q,\nu}
(\vec r, {\vec r'}) \vert_{\vec r = {\vec r'}}.
\end{eqnarray}
These quantities are called, respectively, current density, spin density, 
spin-kinetic density and tensor-kinetic density. 
All these proton/neutron densities can be re-expressed in terms of isoscalar ($t=0$) 
and isovector ($t=1$) densities in the standard way. We drop the label $q$ in what follows,
since the same formulas hold for the $t=0,1$ components as well.

We can decompose the (pseudo-tensor) spin-current density 
that appears in (\ref{loc_densities}) into
\begin{equation}
J_{\mu\nu}(\vec r) = \frac{1}{3}\delta_{\mu\nu}J^{(0)}(\vec r) + \frac{1}{2} \sum_\kappa
\epsilon_{\mu\nu\kappa}J^{(1)}_\kappa(\vec r) + J^{(2)}_{\mu\nu}(\vec r),
\end{equation}
where the three terms are pseudo-scalar, vector and pseudo-tensor, and read
\begin{eqnarray}
J^{(0)}(\vec r) & = & \sum_\mu J_{\mu\mu}(\vec r), \nonumber \\
J^{(1)}_\kappa(\vec r) & = & \sum_{\mu\nu} \epsilon_{\kappa\mu\nu}J_{\mu\nu}(\vec r), \nonumber \\
J^{(2)}_{\mu\nu}(\vec r) & = & \frac{1}{2} \left[ J_{\mu\nu}(\vec r) + J_{\nu\mu}(\vec r) \right]
- \frac{1}{3}\delta_{\mu\nu} \sum_\kappa J_{\kappa\kappa}(\vec r).
\end{eqnarray}

The time-even part of the energy density functional that is obtained when a tensor force of 
the type (\ref{eq:tensor}) is introduced is
\begin{equation}
{\cal E}_{\rm tensor} = \sum_{t=0,1} {\cal E}_t,
\end{equation}
where
\begin{eqnarray}\label{Eeven}
{\cal E}_t = & -C_t^T & \sum_{\mu\nu} J_{t,\mu\nu}J_{t,\mu\nu} \hfill \nonumber \\
& -C_t^F & \left[ \frac{1}{2} \left( \sum_\mu J_{t,\mu\mu} \right)^2 +
\frac{1}{2} \sum_{\mu\nu} J_{t,\mu\nu}J_{t,\mu\nu} \right].
\end{eqnarray}
We follow here the notation of \cite{Bender:2009,Hellemans}. 
The notation $T$ and $F$ will become more transparent by looking
at Eq. (\ref{Eodd}) below. 
To deal with the spherical limit more clearly, we can transform 
Eq. (\ref{Eeven}) into
\begin{equation}\label{Etransfomed}
{\cal E}_t = C_t^{(J0)} \left( J_t^{(0)} \right)^2 + C_t^{(J1)} {\vec J}_t^2 +
C_t^{(J2)} \sum_{\mu\nu} J_{t,\mu\nu}^{(2)}J_{t,\mu\nu}^{(2)}.
\end{equation}
As above, 
the coefficients $C$ are actually sums of terms coming from the exchange part of
the central Skyrme force and terms associated with the tensor force. In fact,
\begin{eqnarray}
C^T_t & = & A^T_t + B^T_t, \nonumber \\
C^F_t & = & A^F_t + B^T_f, \nonumber \\
A^T_0 & = & -\frac{1}{8}t_1\left(\frac{1}{2}-x_1\right) + \frac{1}{8}t_2\left(\frac{1}{2}+x_2\right), \nonumber \\
A^T_1 & = & -\frac{1}{16}t_1 + \frac{1}{16}t_2, \nonumber \\
A^F_0 & = & 0, \nonumber \\
A^F_1 & = & 0, \nonumber \\
B^T_0 & = & \frac{1}{8}\left( t_e + 3t_o \right), \nonumber \\
B^T_1 & = & \frac{1}{8}\left( t_e - t_o \right), \nonumber \\
B^F_0 & = & \frac{3}{8}\left( t_e + 3t_o \right), \nonumber \\
B^F_1 & = & \frac{3}{8}\left( t_e - t_o \right),
\end{eqnarray}
and the following relations hold
\begin{eqnarray}
C_t^{(J0)} & = & -\frac{1}{3} C_t^T - \frac{2}{3} C_t^F, \nonumber \\
C_t^{(J1)} & = & -\frac{1}{2}C_t^T + \frac{1}{4}C_t^F, \nonumber \\
C_t^{(J2)} & = & -C_t^T - \frac{1}{2} C_t^F.
\end{eqnarray}

From Eq. (\ref{Etransfomed}), one can deduce that in the EDF picture 
one could introduce in principle six independent coupling constants.
However, even in the deformed case, the pseudo-scalar density 
is zero if the parity is conserved and then only four 
independent coupling constants can be introduced in this
case for deformed nuclei. These reduce to two in the
spherical case as discussed above. The four constants $C^{(J1)}$ and $C^{(J2)}$ 
are non-zero in the static calculations of deformed nuclei and 
they reduce to the two constants $C^{(J1)}$ in
static calculations of spherical nuclei. As we did at the 
start of the subsection, we stress once more the difference 
between the Hamiltonian and EDF schemes: if the starting point 
is the Skyrme Hamiltonian with tensor-even and tensor-odd terms, 
the independent constants are in any case only two.
If the energy functional strategy is enforced, the independent 
constants can be
more. The same holds in case of non-static, e.g. 
RPA calculations. If one starts from the
Hamiltonian, automatically one sticks to two independent constants.

We now briefly discuss what happens when time-odd terms have to be taken into
account. We remind that our discussion takes care of 
Galilean invariance, which is needed since the results
ought not to depend on the reference frame, and that
time-odd terms are to be included in odd (or odd-odd) nuclei
as well as in dynamical calculations of even-even ones. 
Following Ref. \cite{Hellemans}, the terms generated by the 
inclusion of tensor terms in the Skyrme force in both the
time-even and time-odd sectors of the EDF are
\begin{equation}
{\cal E}_{\rm tensor} = \sum_{t=0,1} {\cal E}_t,
\end{equation}
where
\begin{eqnarray}\label{Eodd}
{\cal E}_t & = & B_t^T \left( \vec s_t \cdot \vec T_t - 
\sum_{\mu\nu} J_{t,\mu\nu}J_{t,\mu\nu} \right) \hfill \nonumber \\
& + & B_t^{\Delta s} \vec s_t \cdot \nabla \vec s_t + C_t^{\nabla s} 
\left( \vec \nabla \cdot \vec s_t \right)^2 \nonumber \\
& + & C_t^F \left[ 
\vec s_t \cdot \vec F_t -
\frac{1}{2} \left( \sum_\mu J_{t,\mu\mu} \right)^2 -
\frac{1}{2} \sum_{\mu\nu} J_{t,\mu\nu}J_{t,\mu\nu} \right].
\end{eqnarray}
The new constants that appear in this expression are
\begin{eqnarray}
C^{\nabla s}_0 & = & A^{\nabla s}_0 + B^{\nabla s}_0, \nonumber \\
C^{\nabla s}_1 & = & A^{\nabla s}_1 + B^{\nabla s}_1, \nonumber \\
A^{\nabla s}_0 & = & 0, \nonumber \\
A^{\nabla s}_1 & = & 0, \nonumber \\
B^{\nabla s}_0 & = & \frac{9}{8} \left( t_e - t_o \right), \nonumber \\
B^{\nabla s}_1 & = & -\frac{3}{8} \left( 3t_e + t_o \right), \nonumber \\
C^{\Delta s}_0 & = & A^{\Delta s}_0 + B^{\Delta s}_0, \nonumber \\
C^{\Delta s}_1 & = & A^{\nabla s}_1 + B^{\Delta s}_1, \nonumber \\
A^{\Delta s}_0 & = & \frac{3}{192}\left( 3t_1 - 6t_1x_1 + t_2 + 2t_2x_2 
\right), \nonumber \\
A^{\Delta s}_1 & = & \frac{3}{192}\left( 3t_1 + t_2
\right), \nonumber \\
B^{\Delta s}_0 & = & \frac{3}{32} \left( t_e - t_o \right), \nonumber \\
B^{\Delta s}_1 & = & \frac{1}{27} \left( 3t_e - t_o \right). \nonumber \\
\end{eqnarray}
We refer to \cite{Perlinska} for these and other general 
formulas related to the Skyrme energy functional. 

\subsection{Gogny plus tensor}
\label{gogny}

The standard calculations performed using the finite-range
Gogny force, starting from the paper by J. Decharg\`e and
D. Gogny \cite{Decharge}, do not include any tensor term.
It should be mentioned that most of these calculations
are performed in the harmonic oscillator basis, and the 
formulas to calculate the matrix elements of central, two-body
spin-orbit but also tensor forces on that basis can be actually 
found in the earlier paper by D. Gogny \cite{Gogny:1975}. 
Despite this, early attempts to complement the Gogny force with 
tensor terms are scarce. In Ref. \cite{Blumel}, one-pion and one-rho
tensor terms have been added pertubatively to the D1 
Gogny set. Unfortunately, this study is only devoted 
to some hypothetic states in N=Z nuclei where a 
non-zero expectation value of $\langle \sigma\sigma\tau\tau\rangle$
might show up. The authors are interested in finding a
sort of ferromagnetic structure (called ``spin-isospin lattice'') 
in which proton spins are aligned along one direction 
and neutron spins are aligned in opposite direction, 
at variance with the normal shell model configurations. 
This study, focused on low-lying states in $^{32}$S, remains to 
some extent speculative. More or less at the same time, 
another type of finite-range (Gaussian) force that includes 
tensor terms has been introduced in Ref. \cite{Onishi:1978}. 
The strengths of these terms have been determined through
a fitting procedure, but the results are very sensitive to
the choice of the observables that enter the fit (that 
can be either bulk nuclear properties, single-particle states
or shell-model p-h matrix elements). Surprisingly, in 
most of the cases the tensor strength seems to be dominated
by the isoscalar term at variance with almost all the tensor 
parametrizations. No clear conclusion can be
extracted from these studies.

However, these early attempts suggest that a tensor force of the
type (1), namely
\begin{equation}
V_T(r) = f_G(r) S_{12}
\end{equation}
with a Gaussian form factor $f_G$, would fit appropriately the Gogny
ansatz. Such kind of force has been introduced only much later in
Ref. \cite{Otsuka:Gogny}. Following the spirit of Ref. \cite{Otsuka},
the tensor term added to the Gogny force is written
\begin{equation}
V_T(r) = f_G(r) S_{12}\ \vec \tau_1 \cdot \vec \tau_2,
\end{equation}
(and, for practical reasons, turned into 
$V_T(r) = g_G(r) \left[ \left[ \vec r \otimes \vec r \right]^{(2)} 
\otimes \left[ \vec \sigma_1 \otimes \vec \sigma_2 \right]^{(2)} 
\right]^{(0)}$ by using standard recoupling). 
This Gaussian function has the largest among the ranges
associated with the Gogny terms, namely 1.2 fm (in agreement with
the long-range nature of the pion exchange). The strength is adjusted
so that the volume integral is the same as that of AV8' \cite{Pudliner}.
The other (central and spin-orbit terms) of the new force
introduced in \cite{Otsuka:Gogny} are the same as in D1S \cite{Berger} 
but their 
strengths have been refitted and the new force is called GT2.
This force has been constructed with the goal of including the
specific attraction between protons and neutrons respectively
in the $j_<$ and $j_>$ (or $j_>$ and $j_>$) orbits, that we 
will discuss in detail in \ref{spstates} below. However, GT2
has not been systematically applied to many other nuclear properties
later. 

\begin{figure}[htp*]
\centering
\includegraphics[width=8cm,clip]{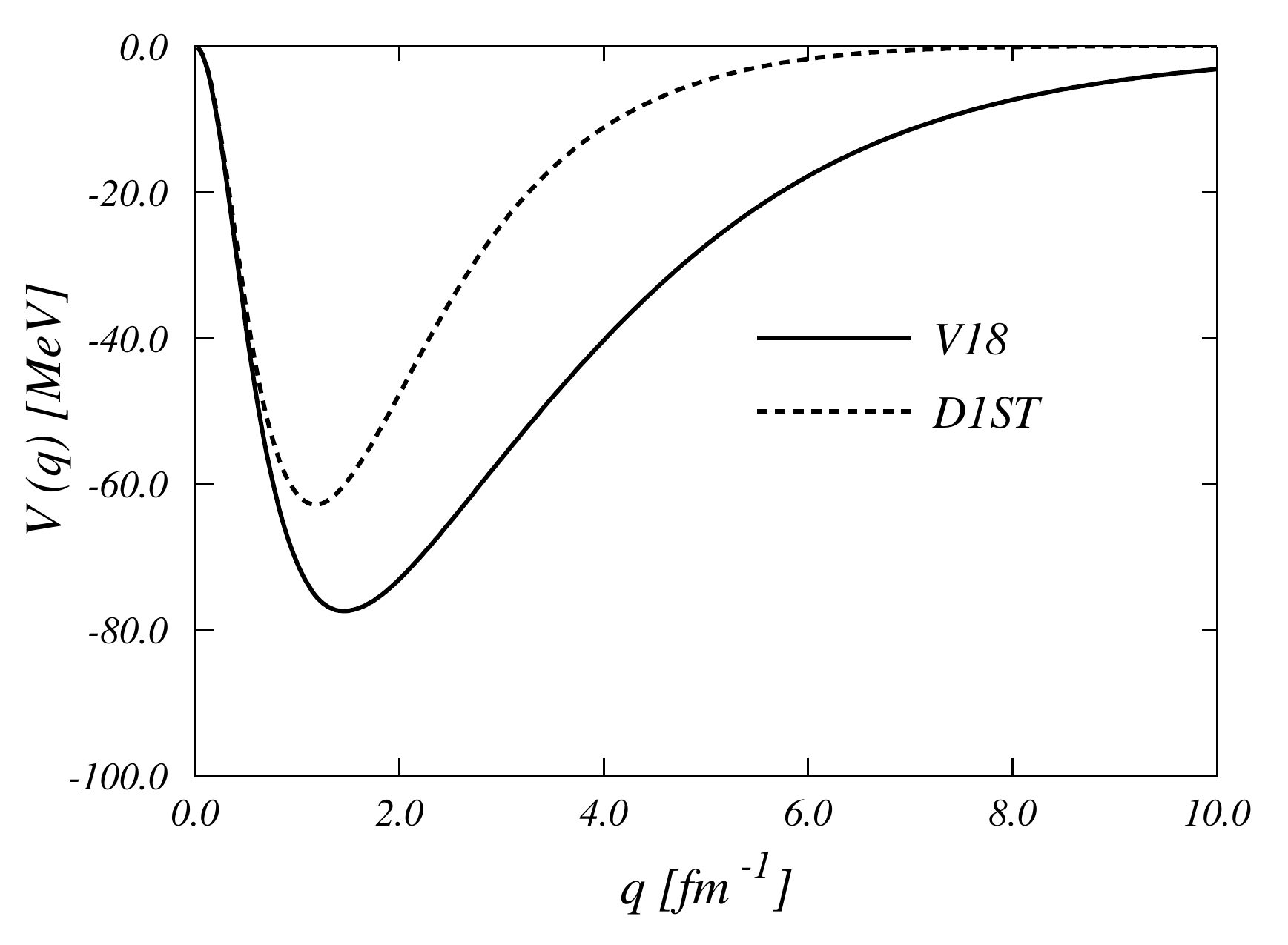}
\vspace*{1cm}
\caption{\label{fig:co}
Tensor component of the Argonne potential and screening
due to short-range correlations. \cite{DeDonno:2009,Co_priv}
}
\end{figure}

There have been many applications of Gogny plus tensor performed
by the groups of Granada and Lecce, as we shall discuss below.
However, they have at times employed hybrid approaches. 
A tensor force derived by the Argonne V18 potential and multiplied
by a correlation function that includes short-range effects 
\cite{Arias:2007},
\begin{equation}\label{tensor-co}
v_{\rm T}(r) = v_{\rm 6,AV18}(r) \left(1 - e^{-br^2} \right),
\end{equation}
where $v_{\rm 6,AV18}(r)$ is the radial function of the Argonne 
potential, has been introduced in \cite{DeDonno:2009}. 
We show in Fig. \ref{fig:co} the Fourier tranform of the bare
tensor component of the Argonne potential and of the IV tensor
force introduced in Ref. \cite{DeDonno:2009}. Although there
is a clear resemblance between the effective and bare tensor
forces, the quenching is not negligible.

We end this Section by mentioning another class of finite-range
tensor forces. In the papers by H. Nakada \cite{Nakada1,Nakada2,
Nakada3}, semi-realistic interactions derived from the
original M3Y interaction \cite{m3y} have been introduced. By
adding a density-dependent term to the M3Y ansatz, saturation
of uniform matter can be achieved and HF calculations in finite
nuclei can be performed with reasonable success. The parameters
have been refitted at different levels of approximation, and
the resulting sets are named M3Y-P$n$. They  
include tensor terms (although not big effort is
done to single out the specific effect of these terms).
Starting from the set M3Y-P2 \cite{Nakada1}, these sets
take care of the experimental systematics of single-particle
states: in fact, the quenching factor of the tensor force
with respect to the original M3Y interaction is obtained
by considering the ordering of states in $^{208}$Pb.

\section{Results}

\subsection{Masses}
\label{masses}

The question whether the tensor terms can improve the agreement of theoretical
binding energies with experiment has been analyzed mainly within
the Skyrme framework. It should be said that some accurately 
calibrated Skyrme functionals have achieved great success 
in reproducing experimental masses, yet with
a few {\it ad hoc} additional parameters associated with correlations. 
When the term $\delta E$ of Eq. (\ref{eq:dE}) coming from the tensor force 
is added to the binding energy, the effect is not huge. For instance, in
$^{208}$Pb, when the tensor force SLy5$_{\rm T}$ introduced in Ref. 
\cite{Colo} is employed the total binding energy
is changed only by about 0.5\%. However, recently there has 
been ample discussion
whether a fine agreement with the experimental 
masses can be reached at the level of 
less than $\approx$ 1 MeV, or several hundreds of keV. In this
respect, the tensor terms of the force can play a role. 

In particular, it can be seen from Fig. \ref{fig:spin-density} that the spin-orbit
densities $J$ have a specific isotope dependence: they are
negligible in $\vec l\cdot\vec s$ saturated nuclei, i.e., in the 
magic nuclei of the light or medium mass region (like in $^{40}$Ca), 
whereas they are 
large in the middle of the shell where $j_>$ orbits are
filled and $j_<$ orbits are empty (like in $^{48}$Ca). The
associated binding energy $\delta E$ of Eq. (\ref{eq:dE}), 
\begin{displaymath}
\delta E= {1\over 2}\alpha(J_n^2+J_p^2) + \beta J_n J_p,
\end{displaymath}
is also likely to display ``arches'', namely to be minimal
for magic nuclei and maximal in the middle of the shell.
This argument has been raised first in Ref. \cite{Dobaczewski}
(cf. also Fig. 4 of that work), where also the hope was
expressed that these arches can remedy some open theoretical
problem in the reproduction of experimental masses
by theoretical models. In fact, it is known that when 
a functional is designed with the goal of 
reproducing accurately the binding energies, often 
double-magic nuclei turn out to be underbound (see, e.g., the
review paper \cite{Lunney}). It should be added, though, that
this deficiency is more often ascribed either (i) to the
fact that a large-scale fit of any functional 
is dominated by open-shell nuclei,
and pairing gives a strong bias to the results of the fit so
that the errors are larger when pairing is absent, or (ii) to the fact that 
correlation energies are systematically different in
open-shell and closed-shell nuclei. 

The conclusion of the systematic study of Ref. 
\cite{Les07}, about the possible improvement of the 
agreement with the data
when tensor is included in a Skyrme functional, is 
rather on the negative side. As is well known, 
masses and charge radii are among the quantities that are fitted in the 
protocol to determine a Skyrme functional. In the fit of Ref. \cite{Les07}, the 
value of $\chi^2$ is minimum when $\beta$
is equal to zero, so that masses and charge radii seem to 
discard the proton-neutron tensor force to some extent. 
This is at variance with the conclusion obtained from 
the study of single-particle states (cf. Sec. \ref{spstates} below) 
and, more generally, with the overall conclusions of many
works and of the current review.
Within the specific ansatz made in \cite{Les07}, the analysis 
of the difference between theoretical and 
experimental binding energies would suggest 
that the  best
parameter sets for the binding energies give unrealistic single-particle spectra. 
Even the analysis of the 
charge radii brings in some contradiction: for instance, the difference between
the charge radii of $^{40}$Ca and $^{48}$Ca is controlled 
by the position of  proton 
1d$_{3/2}$ level, and the parameter sets
that would 
fit the radii at best make 
the overall single-particle spectra worse. 
In conclusion, there seems to
be little room to accommodate within the chosen framework 
an overall agreement between theory and experiment at the same time for
single-particle states, binding energies and radii.

Interestingly, the same conclusion can be drawn within the framework of
relativistic functionals. The authors of Ref. \cite{Lalazissis} have 
started from conventional RMF and extended it by including pion coupling 
in the RHF framework. They also point out that the pion tensor is important
for single-particle states but the fit of masses would instead ``prefer'' 
not to have the pion-coupling included.

It would be interesting to compare these conclusions 
with analogous ones coming from extensive studies performed with 
finite-range nonrelativistic interactions like the Gogny force. We do not
dispose yet of such studies in which we can compare e.g. the results
of D1M (the Gogny set which is the best suited for masses 
\cite{Goriely}), and an analogous
study by including the tensor. As we have discussed in Sec. \ref{gogny}, 
another type of finite-range force has been introduced by 
H. Nakada \cite{Nakada1,Nakada2}. In the case of his M3Y-P* forces, the
results for binding energies of specific nuclei do not seem to deteriorate
if the tensor force is included. 
At this stage, we may conclude that the deterioration of the results
for masses when some realistic tensor terms are included seems 
to be (only) the outcome
of large scale fits.

\subsection{Single-particle states}
\label{spstates}

\subsubsection{Early attempts and general considerations}

Even before the advent of modern self-consistent
calculations, there have been pioneering attempts 
to relate the behaviour of the spin-orbit splittings, 
in both $\vec l\cdot\vec s$ saturated and unsaturated 
nuclei, to some kind of effective tensor force (cf., e.g.,
\cite{Wong:1968,Scheerbaum:1976a,Scheerbaum:1976}). The mass
dependence of the $l$=5 spin-orbit splitting has been 
explained, within the HF-BCS framework, in terms of the 
tensor force in Ref. \cite{Goodman:1978}. The main
difference between non self-consistent calculations,
and self-consistent Hartree or Hartree-Fock ones, 
is that in the latter case the tensor terms
affect directly the spin-orbit splittings but 
affect indirectly other features of the spectrum as well. 

Within the Skyrme-HF framework, the role of the tensor 
interactions was firstly discussed in Ref. \cite{Stancu:1977}.
In this work, the authors added tensor terms on top of the 
SIII parameter set \cite{Beiner}. They favored 
values of $\alpha$ and $\beta$ in the ``triangle'' 
defined by -80 MeV$\cdot$fm$^5 < \beta < 0$, 
$0 < \alpha < 80$ MeV$\cdot$fm$^5$, and 
$\vert \alpha \vert < \vert \beta \vert$. Thus, they
have been the first to note that the signs of $\beta$ and
$\alpha$ agree and disagree, respectively, with what 
extracted from $G$-matrix calculations.
However, their conclusion was based on the analysis of
few examples of spin-orbit splittings. It did not
focus on the trend of spin-orbit splittings with
the mass number which, as discussed in Sec. \ref{Skyrme}, is the main
effect of tensor terms. Also the conclusion of Ref. 
\cite{Stancu:1977} did not point to a clear improvement
in the comparison of theory and experiment after the 
inclusion of the tensor force. Therefore, such kind
of research has been for some time abandoned. There were
some exceptions \cite{Tondeur, Liu:1991} that have 
been mentioned in Sec. \ref{Skyrme}, but still we can state that 
the tensor force was
essentially dropped in most Skyrme parameter sets which have been
used widely in nuclear structure calculations in the 
1980s and 1990s. Serious and systematic attempts to assess the impact of
the tensor force on single-particle states have been 
performed only in the last decade \cite{Dobaczewski,Brown,Colo,
Brink,Les07,Wei}.

This interest for the tensor force has been revived indeed 
by the paper by T. Otsuka and collaborators \cite{Otsuka}
(cf. also Ref. \cite{Otsuka-early}).  
In this work, a specific and strong effect induced by the 
proton-neutron tensor force on the evolution of the 
single-particle states as a function of the  neutron excess has been pointed
out. If a proton and a neutron lie in spin-orbit partner
orbitals, $j_<$ and $j_>$ respectively, the $S=1$ component
of the relative motion (the only one sensitive to the tensor
force) corresponds in a semiclassical picture to orbits 
having opposite direction and large relative momenta;
consequently, the wave function is spatially confined
and in this deuteron-like configuration the tensor force
is known to give attraction (cf. Fig. \ref{fig:a1} in this review and Fig. 5.1 of 
Ref. \cite{BW} in case of the deuteron). 
The complementary case is that in which 
both particles are in either ($j_<$, $j_<$) 
or in ($j_>$, $j_>$) configuration. Here, the orbits 
have aligned orbital angular momenta, small relative 
momentum, and spread wave function on which the tensor
force give a repulsive effect. These attractive 
or repulsive effects have been shown to be relevant when
looking at the evolution of s.p. states as a function of
the neutron number, going far from stability.

This effect can be easily rephrased in the EDF language.
So, before going to the discussion of the detailed studies
performed so far in this domain, let us illustrate how 
the tensor (and the central exchange) produce the attractive
and repulsive effects that have been just discussed. To this aim,
we analyze their contributions to the Eq. (\ref{eq:dW}) for the
spin-orbit splitting. The first important point concerns the
mass number dependence of the the first and second terms 
in Eq. (\ref{eq:dW}). The Skyrme spin-orbit force proportional 
to $W_0$ gives rise to the first term of Eq. (\ref{eq:dW}), that 
is, to a spin-orbit splitting which is proportional to the 
derivatives of the densities, whose mass number dependence
is very moderate in heavy nuclei. We should nonetheless notice
that this first term is negative because $W_0$ is positive and
the derivatives $\frac{d\rho}{dr}$ are negative functions. On 
the other hand, the second term in Eq. (\ref{eq:dW}) 
depends on the spin density 
$J_q$. This quantity is negligible for $\vec l\cdot\vec s$-saturated 
nuclei. Particles that occupy the $j_>$ orbit give a positive
contribution to $J_q$, while particles that occupy the $j_<$
orbit give a negative contribution to $J_q$. Thus, if $\beta$ is 
positive and the $j_>$ orbit for $q$ is occupied, the second term of
Eq. (\ref{eq:dW}) is opposite to the first term and the spin-orbit
splitting for $1-q$ is reduced; in other words, the occupation
of the $j_>$ orbit for $q$ pushes the $j_>$ orbit for $1-q$ up
and brings the $j_<$ orbit for $1-q$ down, exactly along the
line of the previous discussion. The occupation of the $j_<$ orbit
for $q$ changes the sign of $J_q$, and therefore enlarges
the spin-orbit splitting for $1-q$; this pushes the $j_<$ orbit 
for $1-q$ up and brings the $j_>$ orbit for $1-q$ down, again
following the same pattern. In conclusion a positive $\beta$
corresponds to the findings of Ref. \cite{Otsuka}.

\begin{figure}[ht]
\centering
\includegraphics[width=16cm,clip]{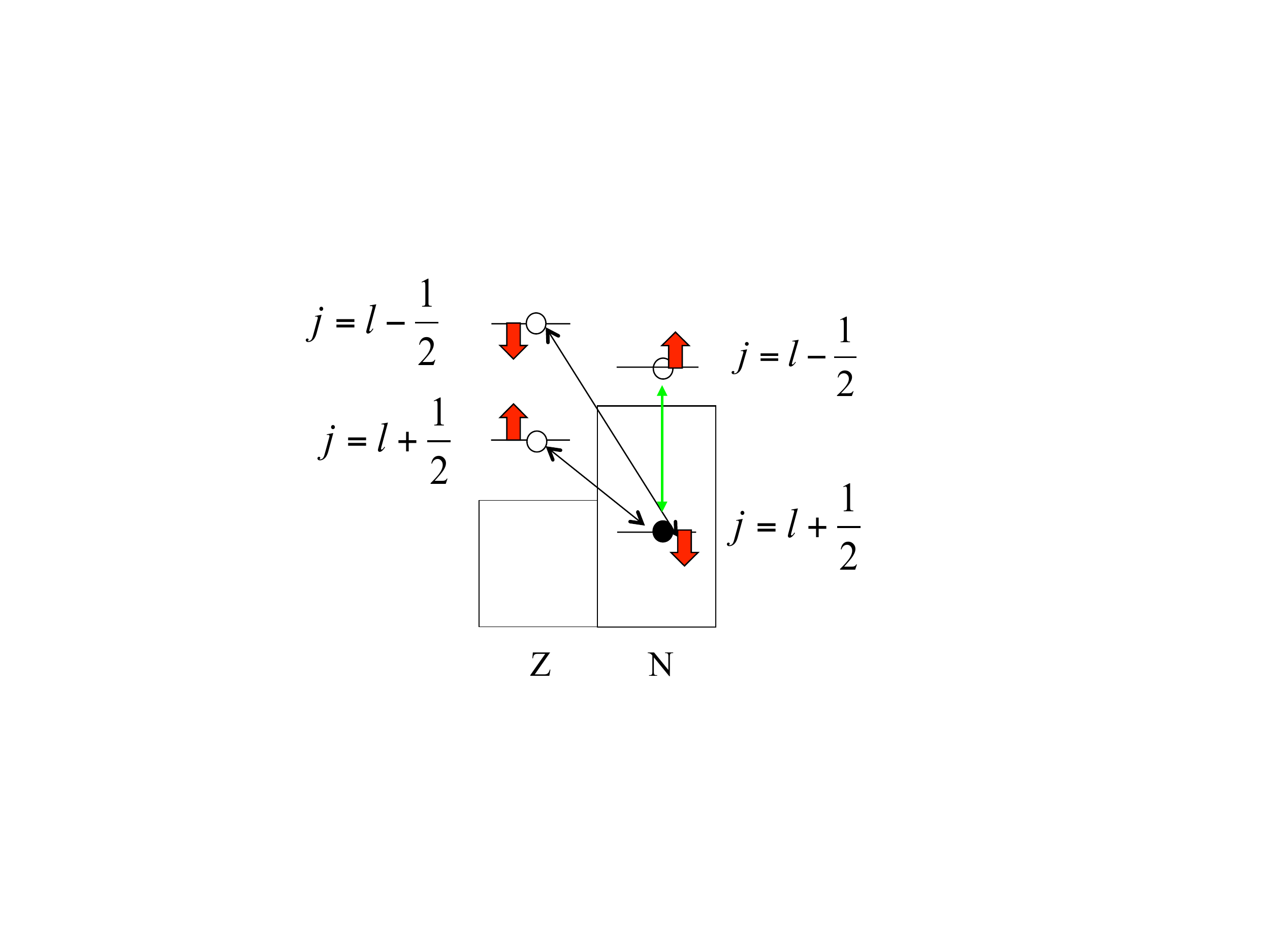}
\vspace*{-2.5cm}
\caption{\label{fig:a2}
Effect of a neutron in the $j_>$ orbit on the
proton and neutron spin-orbit splittings, in the
case in which $\beta$ is positive and $\alpha$ is
negative. See the text for the corresponding discussion.
}
\end{figure}

A positive (negative) value of $\alpha$ would produce the
same (opposite) result on the orbits for $q$ when $J_q$ is
varied. 
This is illustrated in Fig. \ref{fig:a2}, in which we consider 
a nucleus where the last occupied orbit is a neutron $j_>$ orbit
(like, for instance, $^{90}$Zr or $^{48}$Ca). A negative value 
for $\alpha$ (which may be preferable according to \cite{Stancu:1977} and 
to our discussion below, albeit with many warnings) increases the spin-orbit splitting for neutrons
because of the positive contribution to $J_{q=0}$ from the 
$j_>$ orbit, whereas a positive value for $\beta$ 
decreases the spin-orbit splitting for protons.


A first evidence for a
positive value of $\beta$ in modern Skyrme functional calculations
was obtained in Ref. \cite{Dobaczewski}. In this work, the study of 
single-particle states has been motivated by experiments in 
neutron-rich Ti isotopes \cite{Fornal,Dinca}, that indicate a 
sort of sub-shell closure at N=32 although in an indirect
way, that is, through the increase in energy and decrease in the
electromagnetic transition probability of the 2$^+$ state in 
$^{54}$Ti (similar to that observed in $^{50}$Ti and associated 
with the N=28 sub-shell closure). It has been found that
in the N=32 isotones the proton-neutron tensor force does not 
produce any effect on neutron levels at Z=20 ($^{52}$Ca) due to the 
$\vec l\cdot\vec s$ saturation and the vanishing of $J_p$. 
Then, as the proton f$_{7/2}$ orbital is filled, for positive 
values of $\beta$, the spin-orbit splittings of f and p orbit
is reduced. For Z=22 the p orbit splitting is still large enough
so that the N=32 sub-shell closure is visible, but this is not
the case when increasing Z so that, e.g., in $^{60}$Ni this
sub-shell closure is gone. 

The parameter $\alpha$, namely the tensor force between equal
particles, has not been considered in Refs. 
\cite{Otsuka,Dobaczewski}. This term  comes purely from the triplet-odd
tensor interaction [cf. Eq. (\ref{eq:dWT})].  
B.A. Brown and collaborators \cite{Brown} have been the first to study how
to complement a Skyrme type force with both isoscalar and
isovector tensor terms. Their study is based on the Skyrme 
set SkX \cite{SkX} and focused on single-particle states in
$^{132}$Sn and $^{114}$Sn. Their starting point is a $G$-matrix 
tensor force \cite{Gmatrix}. If the strengths of the tensor terms
added to SkX are calibrated on the $G$-matrix results, values
of $\alpha_T$ = 60 MeV$\cdot$fm$^5$ and $\beta_T$ = 110 
MeV$\cdot$fm$^5$ are obtained. The force
obtained by re-fitting the central and spin-orbit Skyrme parameters,
named SkXa, is not very satisfactory. A re-fitting of the
isoscalar tensor term, together with the central and spin-orbit 
parameters, leads to $\alpha_T$ = -118 
MeV$\cdot$fm$^5$
(the set is named SkXb).
Their conclusion is that, because of the Skyrme ansatz and the
resulting form of the one-body potential, there is a tendency
for a good fit to lead to $\alpha_T \approx -\beta_T$. This
conclusion is not based, however, on a very systematic
study including different mass ragions and deformed nuclei.

  
In subsequent works, it has been shown that the inclusion of tensor
terms in the Skyrme HF calculations (with positive values of
$\beta_T$ and negative values of $\alpha_T$) can bring 
considerable success in explaining some features of the 
evolution of single-particle states along isotopic or 
isotonic chains~\cite{Colo,Brink,Wei}. The tensor terms
were added perturbatively in Refs.~\cite{Colo,Wei} and~\cite{Brink}
to the existing standard parameterizations SLy5~\cite{Chabanat} and 
SIII~\cite{Beiner}, respectively.

\subsubsection{The Sn isotopes and the N=50 isotones}

\begin{figure}[htb]
\centering
\includegraphics[width=16cm,clip,bb=0.0 0.0 720.0 540.0]{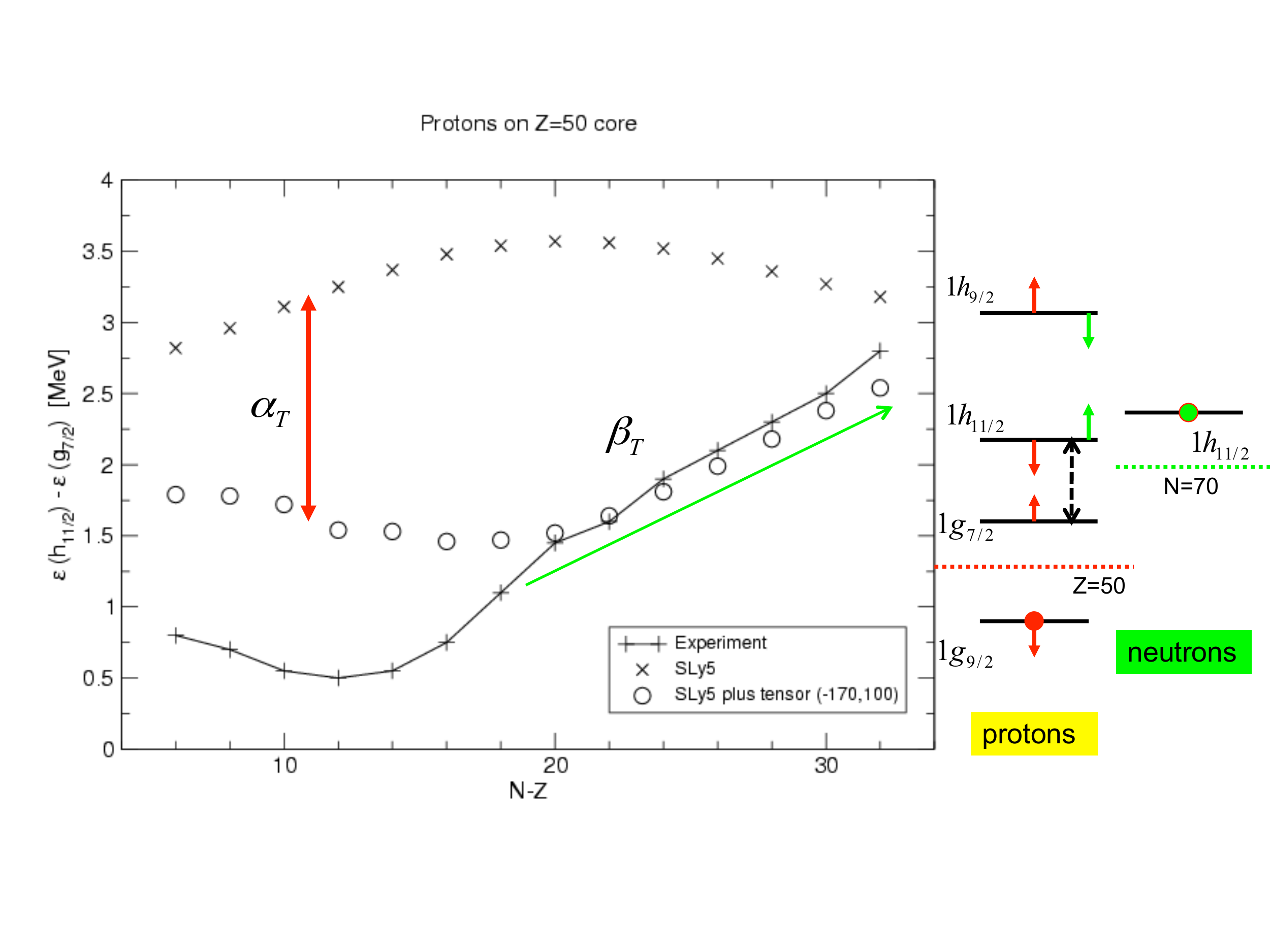}
\caption{\label{fig:Z50}
Energy difference between the 
single-particle 1g$_{7/2}$ and 1h$_{11/2}$ proton states along the Z=50 isotopes.
The calculations are performed with and without tensor 
terms in the spin-orbit potential (\ref{eq:dW}), on top of the 
SLy5 \cite{Chabanat} parameter set.
The experimental data are taken from Ref. \cite{Schiffer}.
See the text for details.
}
\end{figure}

One of the main experimental benchmark for these kinds
of studies is provided by the data on single-particle states
in $N=82$ isotones and $Z=50$ isotopes \cite{Schiffer}. 
In Ref. \cite{Colo},  
the optimal parameters $\alpha_T$ and $\beta_T$ are determined to be
($\alpha_T$, $\beta_T$) = (-170,100) MeV$\cdot$fm$^5$.
The qualitative reasons why these values provide a quite reasonable
fit to the experimental results can be understood by applying
the arguments discussed in this subsection. 
In Fig. \ref{fig:Z50}, the energy differences for
the proton single-particle states $\Delta e(h_{11/2}-g_{7/2})$ 
in the Z=50 isotopes are shown as a function of the neutron excess N-Z. 
The original SLy5 interaction fails to reproduce the experimental 
trend qualitatively and quantitatively. Firstly, the energy 
differences of the HF results are much larger than the empirical data. 
Secondly, the experimental data decrease as the neutron excess 
decreases and reach about 0.5 MeV at the minimum value. On 
the other hand, the energy differences obtained with the original 
SLy5 force increase as the neutron excess decreases and attain
the maximum at around N-Z=20 (several other Skyrme parameter 
sets show almost the same trends as 
those of SLy5). Then, when the tensor force is included, the results 
marked by open circles in Fig. \ref{fig:Z50} are obtained and
the substantial improvement is clear. The force SLy5 plus tensor
force parameters ($\alpha_T$, $\beta_T$) = (-170,100) MeV$\cdot$fm$^5$ 
will be denoted by SLy5$_{\rm T}$ in what follows. 

The results can be qualitatively understood by the above general 
arguments. Firstly, introducing $\alpha$ changes the 
strength of the proton spin-orbit potential. In the Z=50 core, 
only the proton g$_{9/2}$ orbit gives a significant positive contribution 
to the spin density $J_p$ in Eq. (\ref{eq:sd}): consequently, 
with a negative $\alpha_T$ value the spin-orbit splittings are increased,
in particular those associated with the 
g$_{9/2}$-g$_{7/2}$ orbits and h$_{13/2}$-h$_{11/2}$ orbits.
As a net effect, the proton energy difference $\Delta 
e(h_{11/2}-g_{7/2})$ decreases substantially.
Let us now discuss the N-Z dependence of this energy difference, 
for which the isovector parameter $\beta_T$ plays the essential role.
From N-Z=6 to 14, the g$_{7/2}$ neutron orbit is gradually filled.
Then the term associated with $\beta_T$ = 100 MeV$\cdot$fm$^5$ gives 
a negative contribution to the spin-orbit potential (\ref{eq:dW})
and increases the spin-orbit splitting. Therefore, the energy 
difference $\Delta e(h_{11/2}-g_{7/2})$ is decreasing. From N-Z = 
14 to 20, the s$_{1/2}$ and d$_{3/2}$ neutron orbits are occupied
and in this region the spin density is not so much changed. For 
N-Z = 20 to 32, the h$_{11/2}$ orbit is gradually filled: this 
gives a positive contribution to the spin-orbit potential (\ref{eq:dW}), 
that is, the the spin-orbit splitting is decreasing. This
explains why the value of the energy difference
$\Delta e(h_{11/2}-g_{7/2})$ increases. The magnitude of $\beta$ 
determines the slope of the N-Z dependence, so that a larger value of 
$\beta$ would give a steeper slope.

\begin{figure}[htb]
\centering
\includegraphics[width=16cm]{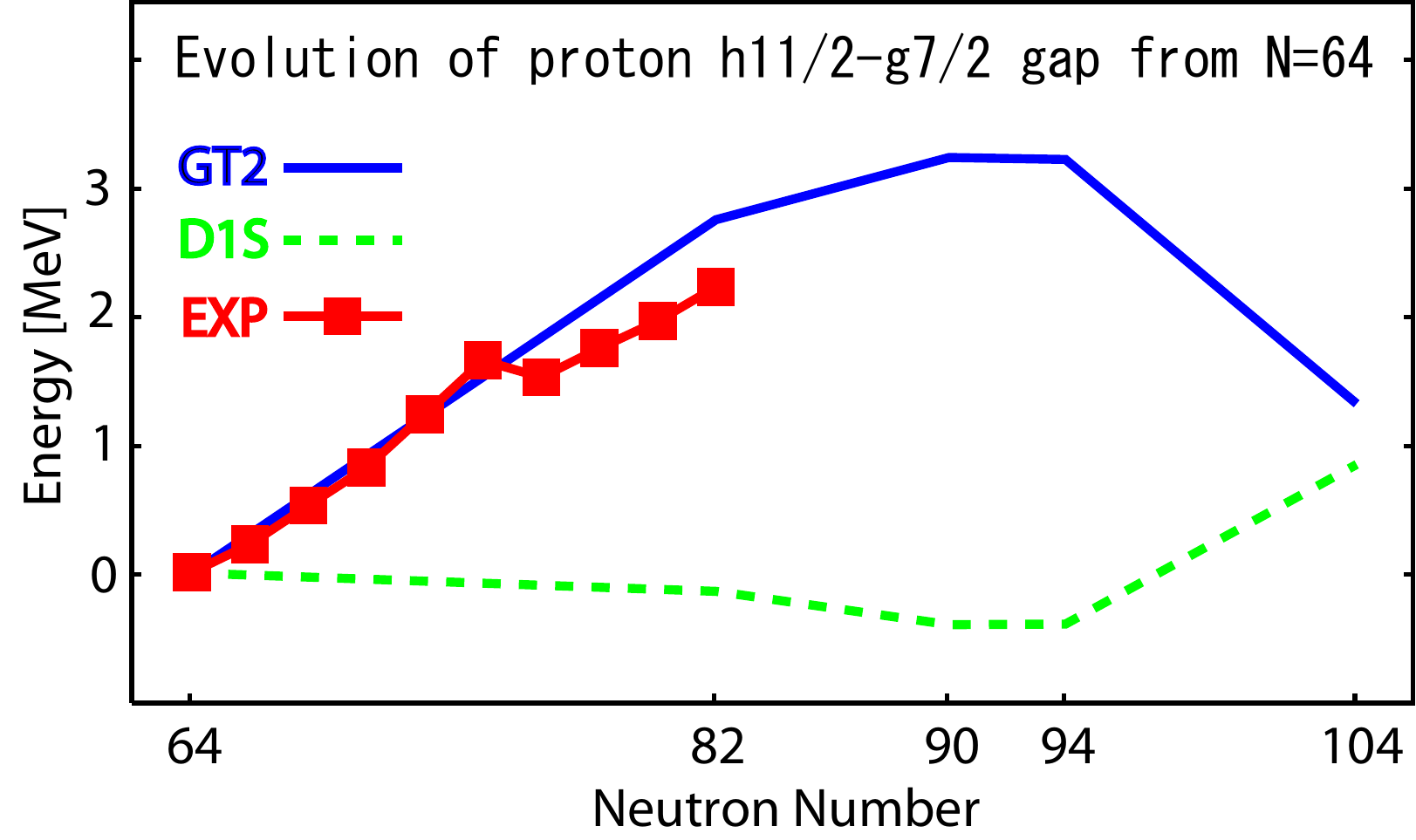}
\caption{\label{fig:GT2}
Energy difference between the 
single-particle 1g$_{7/2}$ and 1h$_{11/2}$ proton states along 
the Z=50 isotopes. The values are normalized to the difference in 
$^{114}$Sn. 
The calculations are performed with the Gogny force 
D1S (without tensor) and GT2 (with tensor).
The experimental data are taken from Ref. \cite{Schiffer}.
The figure is from Ref. \cite{Otsuka:Gogny}. 
See the text for details.
}
\end{figure}

In Ref. \cite{Otsuka:Gogny} the force GT2, that includes
an isovector tensor, produces the correct trend of the energy
difference $\Delta e(h_{11/2}-g_{7/2})$ as a function of the
neutron number at variance with D1S which is qualitatively
and quantitatively in conflict with experiment, as is
shown in Fig. \ref{fig:GT2}. 
In the calculations 
the isoscalar tensor associated with $\alpha_T$ is not introduced;
its effect is in fact not visible in Fig. \ref{fig:GT2} 
in which the energy difference is scaled so to set it at
zero in $^{114}$Sn. However, the fact that the trend as a function
of the neutron excess of the GT2 results is very similar to
that of the SLy5${_T}$ results shown in Fig. 
\ref{fig:Z50} is quite remarkable and shows that 
the effective p-n tensor force (in other words, the values
of $\beta$) is quantitatively rather close in the two cases.

\begin{figure}[htb]
\centering
\includegraphics[width=16cm]{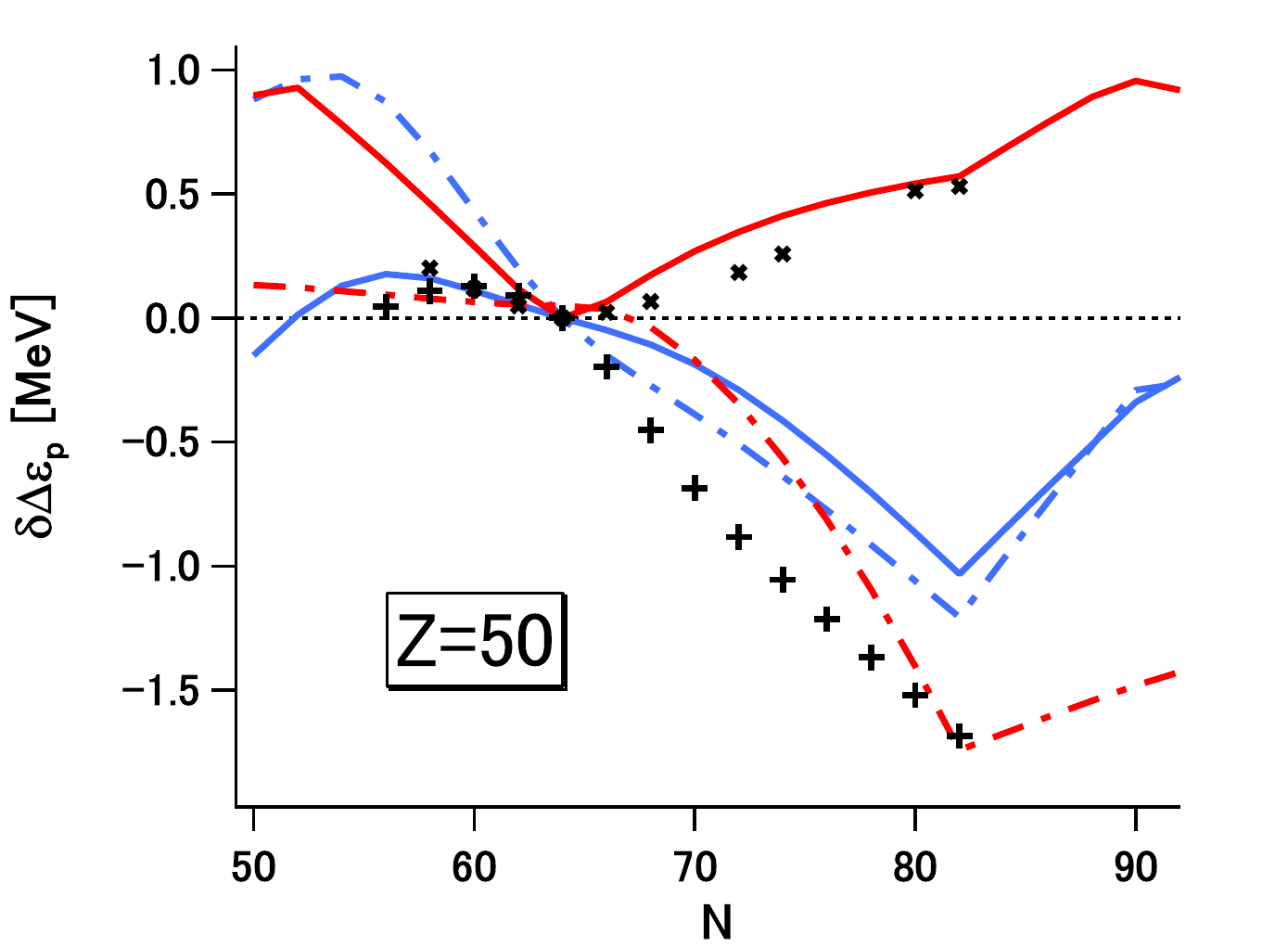}
\caption{\label{fig:Nakada}
$\delta\ \Delta e_p(j)$ for $j$ corresponding to 
the proton orbits 1g$_{7/2}$ (dot-dashed lines) and 
1h$_{11/2}$ (solid lines) in the Z=50 isotopes. 
The calculations are performed with tensor terms
(M3Y-P5, red lines) and without (D1S, blue lines).
The experimental data (pluses for 1g$_{7/2}$ and
crosses for 1h$_{11/2}$) are taken from Ref. \cite{Firestone}.
See the text for details, including the definition of
$\delta\ \Delta e_p(j)$.
}
\end{figure}

A similar argument can be applied to (at least some of)
the parameter sets M3Y-P$n$ that have been discussed at the
end of Sec. \ref{gogny}. In Fig. \ref{fig:Nakada} the
energies of the proton orbits 1g$_{7/2}$ and 1h$_{11/2}$ have
been considered with respect to 1d$_{5/2}$. The quantities 
$\Delta e_p(j) \equiv e_p(j) -
e_p(1{\rm d}_{5/2})$ have been calculated in 
Ref. \cite{Nakada1} and in Fig. \ref{fig:Nakada} they
are plotted after a scaling to the value in $^{114}$Sn
like in the previous Fig. \ref{fig:GT2}: in other
words, $\delta\ \Delta e_p(j)$ is 
$\Delta e_p(j)$ minus its value in $^{114}$Sn.
As above, the trends show that effectively the p-n tensor
force (i.e., the value of $\beta$) is not very different
from the cases that we have already discussed.

\begin{figure}[htb]
\centering
\includegraphics[width=16cm,clip,bb=0.0 0.0 720.0 540.0]{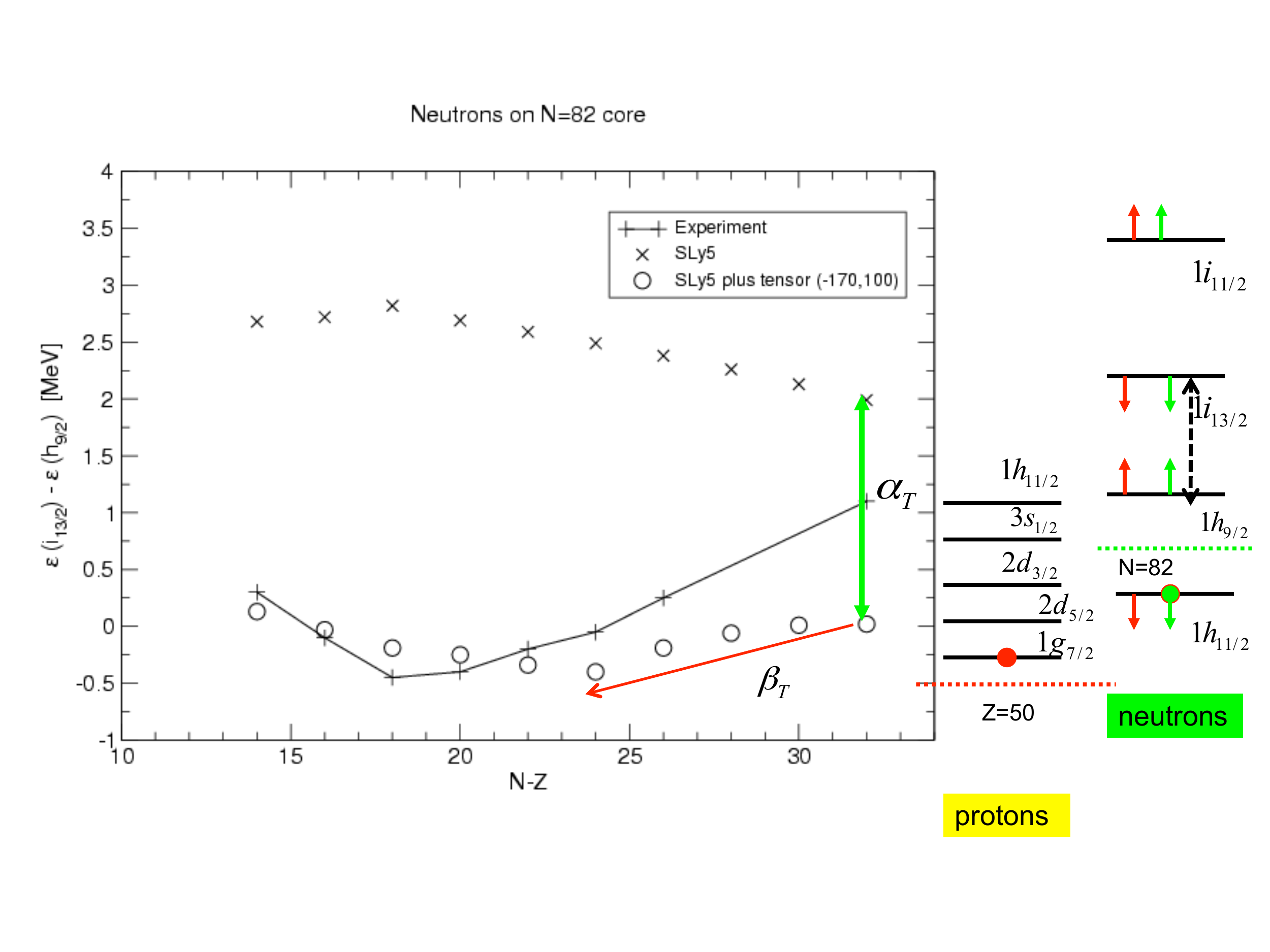}
\caption{\label{fig:N82}
Energy difference between the
single-particle 1i$_{13/2}$ and 1h$_{9/2}$
neutron states in the N=82 isotones.
The calculations are performed with and without tensor 
terms in the spin-orbit potential (\ref{eq:dW}), on top of the
SLy5 parameter set.
The experimental data are taken from Ref. \cite{Schiffer}.
See the text for details.
}
\end{figure}

We move back to the Skyrme case and analyze another
case that has been object of the experimental investigation
of Ref. \cite{Schiffer}. 
In Fig. \ref{fig:N82}, the values of the energy difference 
$\Delta e(i_{13/2}-h_{9/2})$ on the N=82 core are plotted 
as a function of the neutron excess. Similar arguments as
above can be applied. 
The last occupied g$_{7/2}$ level protons, because of the
negative value of $\alpha_T$, increase the neutron spin-orbit
as compared as in the original SLy5 calculation, so that 
the energy difference $\Delta e(i_{13/2}-h_{9/2})$ becomes 
substantially smaller. 
The isotope dependence in the figure can again be explained 
by the effect of the positive value of $\beta_T$.
The 1g$_{7/2}$ and 2d$_{5/2}$ are almost degenerate above the last
occupied proton orbit 1g$_{9/2}$ of the Z=50 core.
These two $j_<$ and $j_>$ proton orbits have opposite effects on 
the spin-orbit potential but the occupancy is larger for the larger 
$j$ orbit, so that the 1g$_{7/2}$ orbit plays a more important role 
in the nuclei with N-Z from 32 to 18. That is, the neutron spin-orbit 
splitting becomes larger for these isotones so that 
the energy gap $\Delta e(i_{13/2}-h_{9/2})$ becomes smaller
for the nuclei from N-Z = 32 ($^{132}$Sn) to N-Z = 18 ($^{146}$Gd).


\subsubsection{Results for the medium-mass region}

The work of Ref. \cite{Colo} has been extended to a different
mass region in \cite{Wei}, with a motivation coming from the
experimental results of Ref. \cite{Gaudefroy}. It has been concluded
that the same tensor force employed in \cite{Colo} can explain the 
trend of the proton single-particle states in Ca isotopes, and
the reduction of the spin-orbit splittings going from $^{48}$Ca to
$^{46}$Ar. The results of Ref. \cite{Brink} are essentially 
along the same line, and it is remarkable that although the 
tensor force is added on top of a different Skyrme force the
parameters are quite similar ($\alpha$ = -118.75 MeV$\cdot$fm$^5$
and $\beta$ = 120 MeV$\cdot$fm$^5$). The evolution of the 
single-particle states in the Ca isotopes has been
also analyzed in \cite{Grasso} and the role of the tensor force
has been emphasized. Interestingly, in this work there has 
been also the attempt to study separately the evolution
of all contributions (kinetic, central velocity-independent 
and velocity-dependent, and spin-orbit) to the s.p. energies.
This is important in view of the global refitting of Skyrme
forces including tensor that we discuss below.

\begin{figure}[htb]
\centering
\includegraphics[width=16cm]{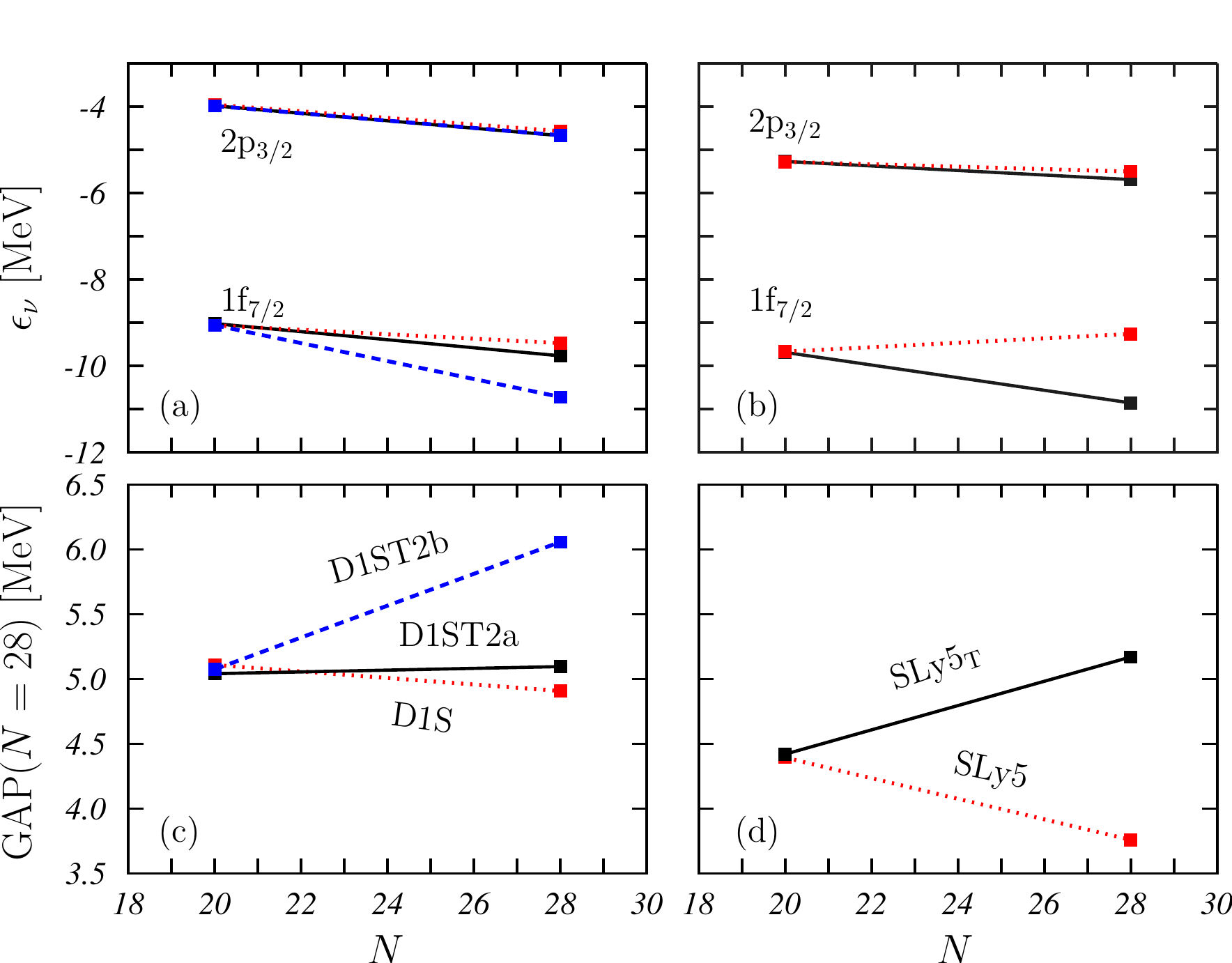}
\caption{\label{fig:MG}
(a), (b): Single-particle energies of 2p$_{3/2}$ and 1f$_{7/2}$ states 
in Z=20 isotopes with Gogny D1S and SLy5 interactions, respectively. 
(c), (d): Energy difference between the two 
single-particle 
neutron states in the Z=20 isotopes with Gogny and SLy5 interactions, respectively. 
The calculations are performed with and without tensor 
terms in the spin-orbit potential (\ref{eq:dW}), on top of the Gogny D1S and 
SLy5 parameter sets.
The experimental data are taken from Ref. \cite{Sorlin}.
See the text for details.}
\end{figure}

We focus now on some experimental fact that may point to 
a specific effect of the equal-particle tensor force
(i.e., non-vanishing value of $\alpha$). The experimental
evidence shows that the N=28 gap, or the difference between
the 1f$_{7/2}$ and 2p$_{3/2}$ single-particle states, increases
from $^{40}$Ca to $^{48}$Ca \cite{Sorlin}. This fact is not
reproduced by the Skyrme force SLy5 but it is reproduced
by SLy5$_{\rm T}$ in keeping with the negative value of
$\alpha$ [cf. panels (b) and (d) of Fig. \ref{fig:MG}]. 
The authors of Ref. \cite{Moreno} have used this and
a few other experimental facts to claim that the IS tensor
(or ``pure tensor'') should be included together with the IV
tensor (``isospin tensor'', present, e.g.,  in the GT2 set 
\cite{Otsuka:Gogny}) on top of the Gogny force. In
Fig. \ref{fig:MG} panels (a) and (c) display results
obtained with Gogny forces that include both kinds of tensor
terms. It can be seen that D1ST2b has the same trend as
SLy5$_{\rm T}$. These results should be considered as
a first step towards a fully refitted Gogny force with
tensor terms \cite{tbp}, which should overcome the  
limits of the calculations of Ref. \cite{Moreno} that
(in the case of the shell gaps of the Z = 8 isotopes, 
N = 8 isotones, Z = 20 isotopes, N = 20 isotones, Z = 28 isotopes
and N = 28 isotones) do not agree well with experiment.
One should also mention the work of Ref. \cite{Tarpanov}:
the reduction of the proton 1d$_{3/2}$-1d$_{5/2}$ spin-orbit
splitting going from $^{34}$Si to $^{42}$Si (that is,
filling the 1f$_{7/2}$ neutron orbital) is attributed to
the effect of a tensor force. This is clear in the nonrelativistic
approach but the authors also compare with the RHF case
(cf. Sec. \ref{RHFspstates}). 

In conclusion, in the medium-mass
region there are several examples of evolution of single-particle
energies that cannot be explained without the introduction of
a tensor force. We should keep in mind
that in self-consistent studies all terms of the functionals
are coupled and care must be taken before attributing effects
to a single term or to a group of terms. 

\subsubsection{Attempts of global fittings of single-particle
states}

In Ref. \cite{Zalewski:2008} the emphasis is put on a global
fitting of Skyrme functionals by considering also, or
mainly, single-particle states functionals. In this respect, 
tensor force plays a role. At variance with the works discussed
in the last paragraphs, the tensor force is fitted not on
isotopic or isotonic trends but rather on f$_{5/2}$-f$_{7/2}$ 
spin-orbit splittings in $^{40}$Ca, $^{56}$Ni and $^{48}$Ca.
This fit is performed on top of existing SkP \cite{SkP}, 
SLy4 \cite{Chabanat} and SkO \cite{SkO}, but the refit is
performed at the same time on the spin-orbit parameter and
on the tensor terms. For these latter terms, values of $\alpha$
around $\approx$ -100 MeV$\cdot$fm$^5$ and $\beta$ 
ranging from +15 to +55 MeV$\cdot$fm$^5$ are
obtained. The signs are consistent with our previous discussion,
but the magnitude is affected by the concurrent 20-40\% reduction
of the spin-orbit strength $W_0$.

\begin{figure}[htp*]
\centering
\includegraphics[width=8cm,clip]{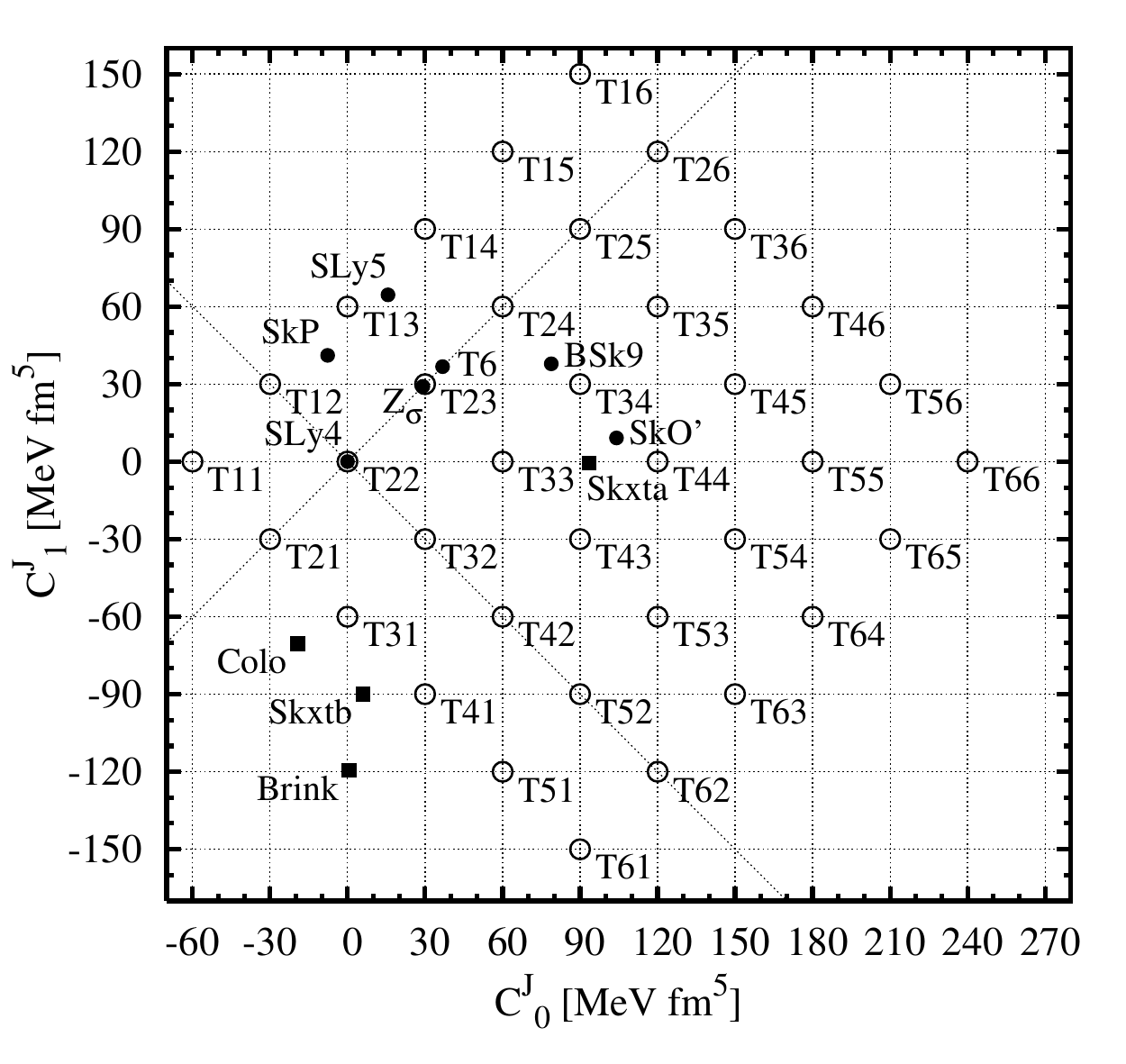}
\vspace*{1cm}
\caption{\label{fig:les1}
Values of the coupling constants $C^J_{t}$ defined
in Sec. \ref{skyrmeEDF} for the Skyrme sets introduced
in \cite{Les07} (circles). The diagonal lines
indicate the loci for $C^J_0+C^J_1=0$ (pure 
p-n tensor) and $C^J_0-C^J_1=0$ (pure
equal-particle tensor). Some other Skyrme sets are
shown, and the original references can be found in 
\cite{Les07}, from which the figure is taken.
}
\end{figure}

In Ref.~\cite{Les07} a very systematic study of the behaviour 
of the single-particle energies within the Skyrme HF framework, 
with and without the tensor interaction, has been performed. 
To this aim, new Skyrme sets have been built using basically 
the SLy protocol \cite{Chabanat} aside from small differences. 
The emphasis is set on the $J^2$-terms, whose coefficients 
have been varied to a quite large extent. These new sets 
are named T$IJ$, and the indices $I$ and $J$ correspond to 
given values of the constants $\alpha$ and $\beta$ that enter 
Eq. (\ref{eq:dWT}). In particular, the following relations hold
\begin{eqnarray}
\alpha & = & 60(J-2)\ {\rm MeV\ fm}^5, \nonumber \\
\beta & = & 60(I-2)\ {\rm MeV\ fm}^5.
\end{eqnarray}
This choice allows to span the range of coupling constants 
associated with the interactions that had been previously used in this section
and/or to scale their values up to a factor
$\approx$ 2. The values  of the coupling constants are
shown in Fig. \ref{fig:les1}.

As expected, the inclusion of tensor terms gives rise to 
specific oscillations in the trend of s.p. states as a function of 
the neutron or proton
numbers. The authors have looked at proton states as a function 
of the neutron number to fix the proton-neutron coupling 
constants, focusing in particular on the 1h$_{11/2}$, 
1g$_{7/2}$, and 2d$_{5/2}$ levels along the Sn isotopes 
and on the 1f$_{5/2}$ and 2p$_{3/2}$ levels along the Ni isotopes. 
In this way, they have extracted a positive value for $\beta$ 
around 120 MeV fm$^5$. This value  compares very well with
the values we have already discussed: 110 MeV fm$^5$ from \cite{Brown}, 
100 MeV fm$^5$ from \cite{Colo,Wei} and 120 MeV fm$^5$ from \cite{Brink}. 
This is quite comfortable albeit 
not surprising as similar data have been used as a reference. 
However, to fix the equal-particle coupling constant $\alpha$, 
T. Lesinski {\em et al.} have used the evolution of 2s$_{1/2}$ 
and 1d$_{3/2}$ states between $^{40}$Ca and $^{48}$Ca: in 
this way, they have extracted a positive value for
$\alpha$, around 120 MeV fm$^5$ at variance with Refs. 
\cite{Brown,Colo,Wei,Stancu:1977,Brink}.

\begin{figure}[htp*]
\centering
\includegraphics[width=9cm,clip]{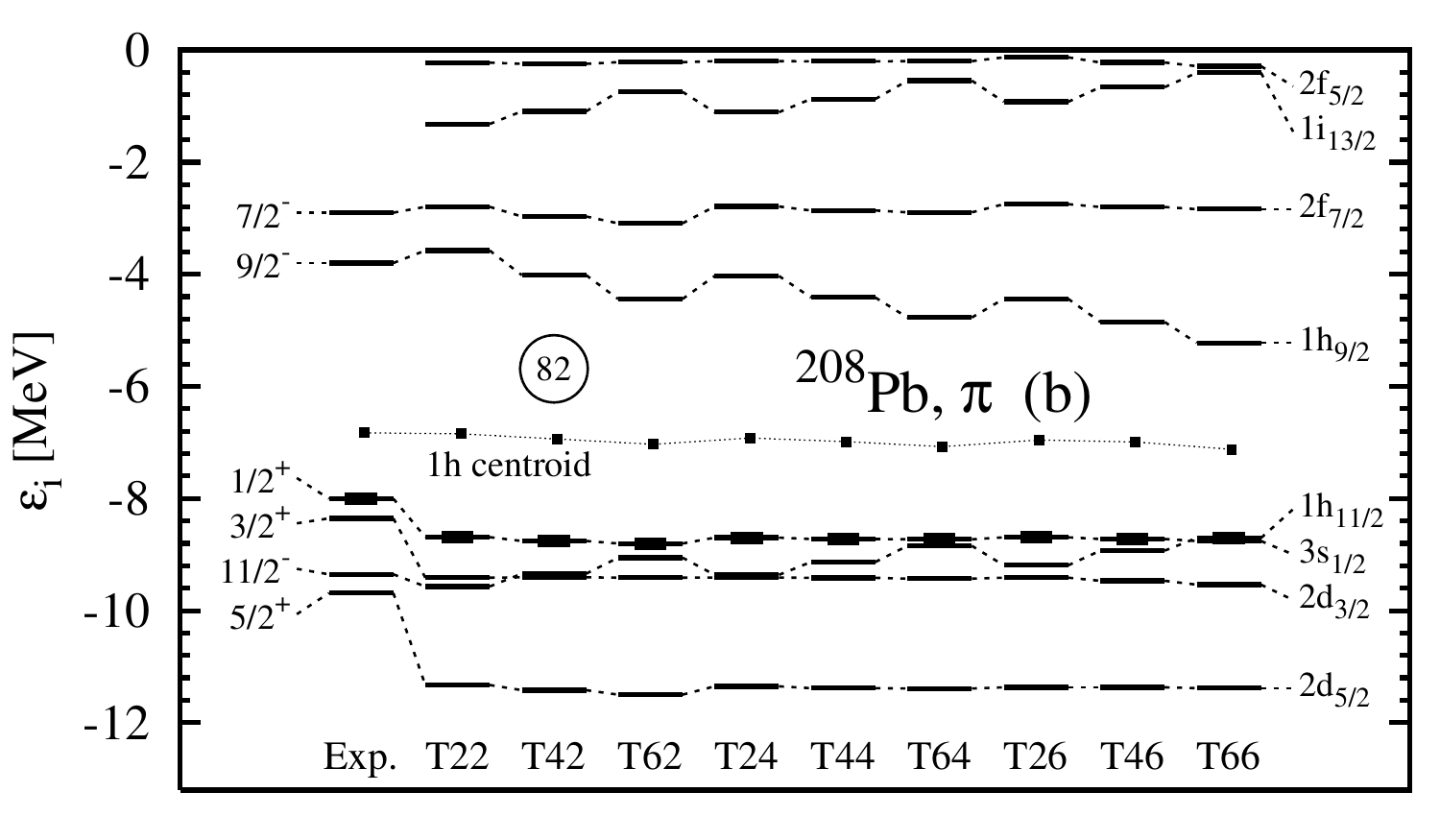}
\includegraphics[width=9cm,clip]{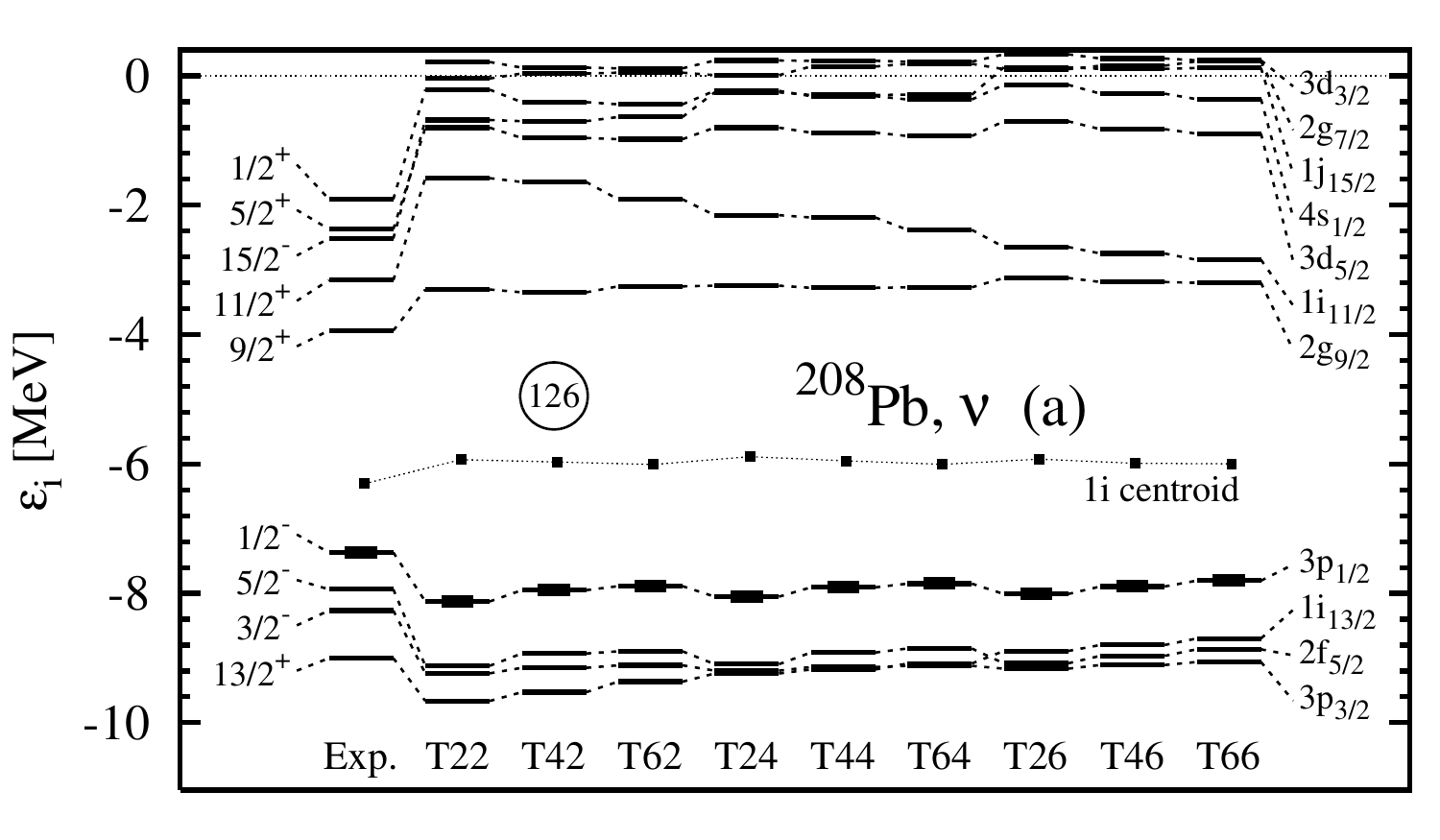}
\vspace*{1cm}
\caption{\label{fig:les2}
Single-particle energies in $^{208}$Pb for some of the 
sets T$IJ$. In panel (a) the neutron levels are displayed 
while in panel (b) the proton levels are displayed. A thick mark 
indicates the Fermi level. Figure taken from Ref. \cite{Les07}.
}
\end{figure}

\begin{figure}[htp*]
\centering
\includegraphics[width=9cm,clip]{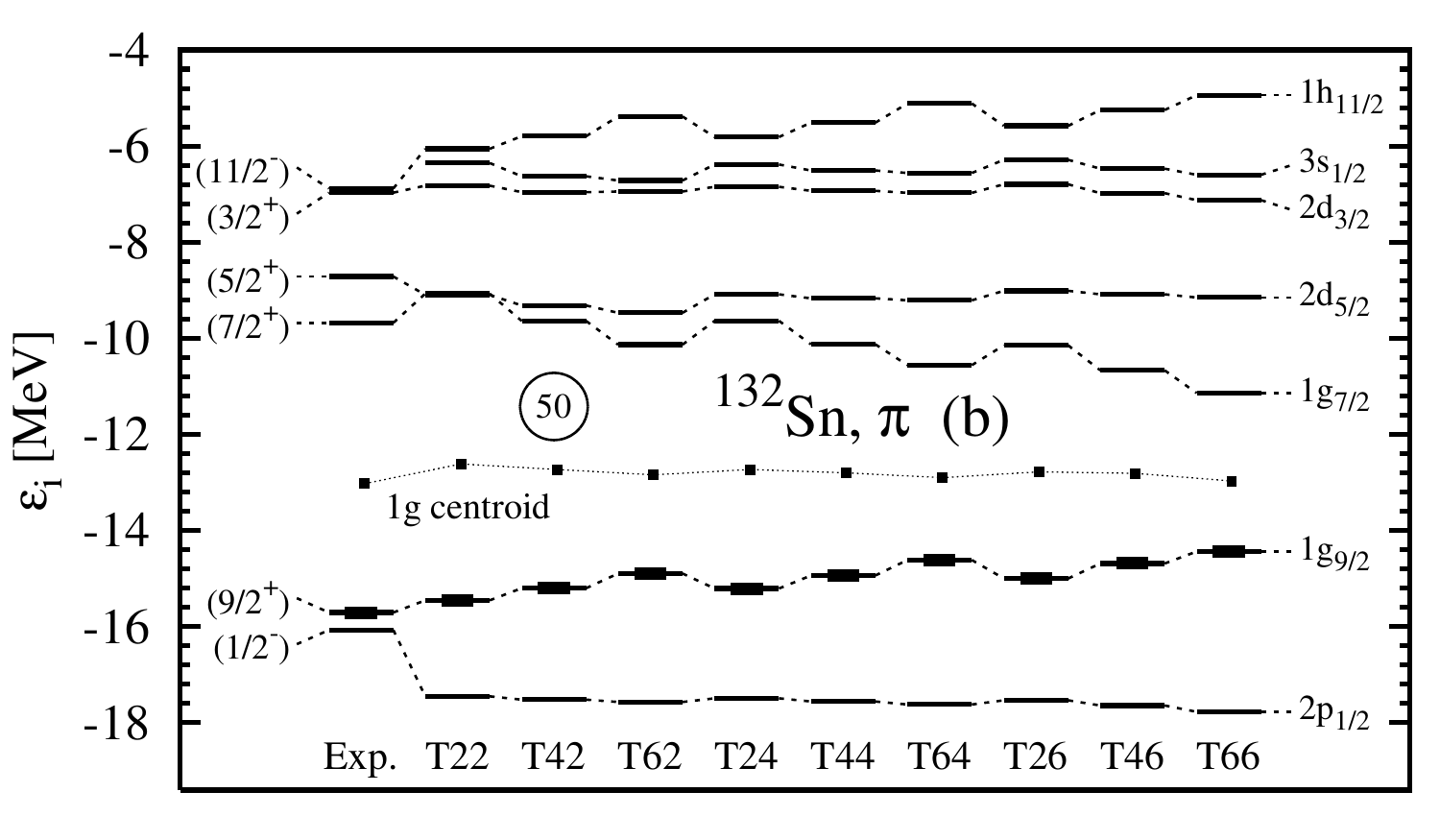}
\includegraphics[width=9cm,clip]{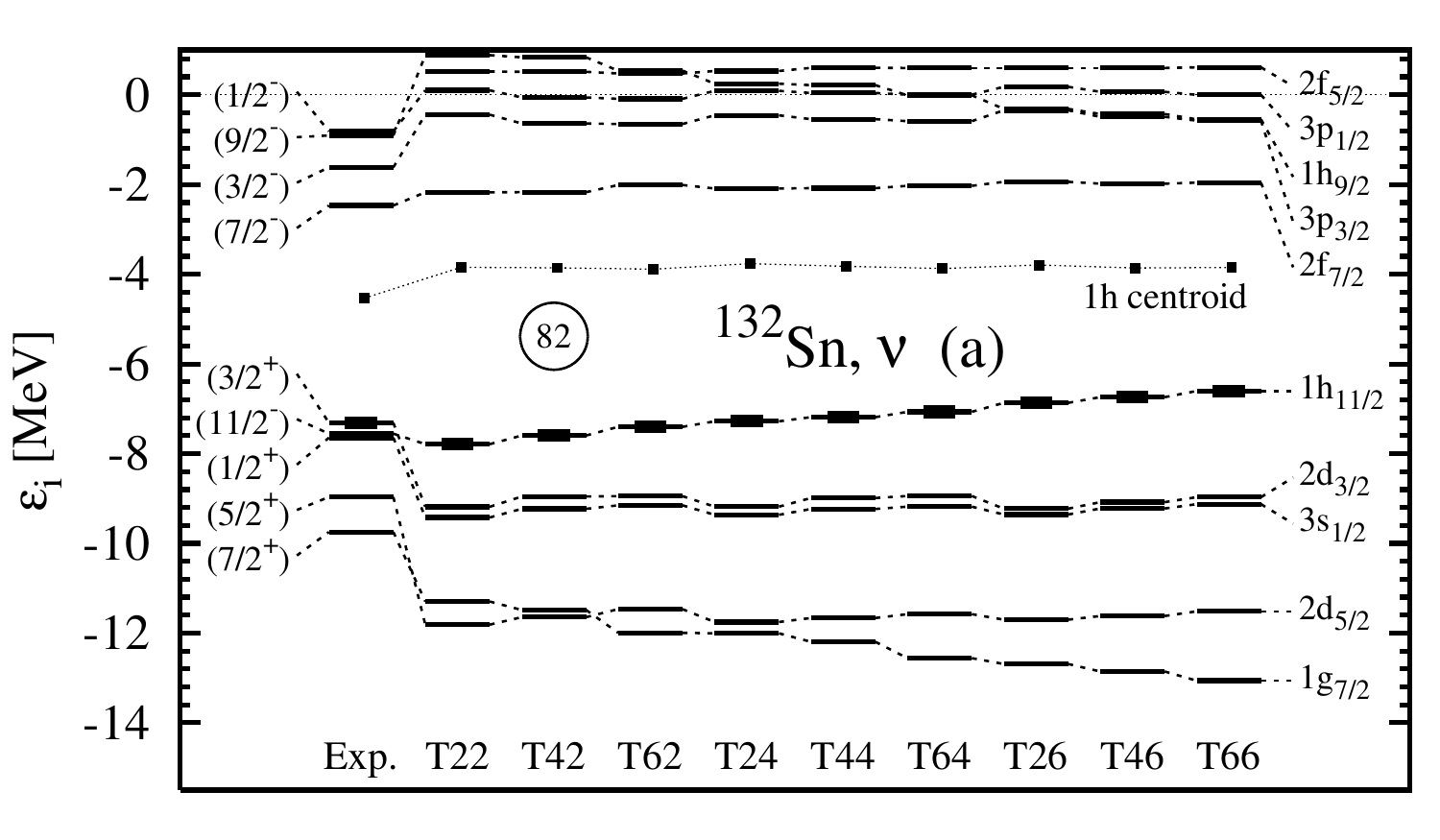}
\vspace*{1cm}
\caption{\label{fig:les3}
The same as Fig. \ref{fig:les2} in the case of $^{132}$Sn.
Figure taken from Ref. \cite{Les07}.
}
\end{figure}

When they look directly at the spin-orbit splittings 
in several magic nuclei they find that, no matter which 
parameter set is used, the splittings between states belonging 
to the same shell (i.e., in-shell splittings) tend to be larger 
than the empirical findings, whereas the splittings between 
states belonging to different shells (i.e, out-shell splittings 
in which one intruder state is involved) tend to be too small. 
Discrepancies with respect to experiments are of the order of 
40\% (in absolute value). If they consider single-particle
spectra as a whole, the systematics does not seem to single 
out any particular set that is satisfactory, in absolute or 
relative sense (cf. Figs. \ref{fig:les2} and \ref{fig:les3}). 

In this respect, the conclusions of the authors 
are pessimistic in an overall sense and do not put much emphasis 
on the aforementioned values of $\alpha$ and $\beta$. At the same 
time, it should be stressed that this conclusion seems to arise 
from a deficiency of the approach as a whole (i.e., the Skyrme 
ansatz plus a specific choice for the protocol fit), and not from 
a problem connected with the tensor in itself. The fact that 
in the fitting procedure the tensor and spin-orbit coupling 
constants are strongly dominated by the mass difference
among $^{40}$Ca, $^{48}$Ca and $^{56}$Ni is in this respect quite 
indicative.

\subsubsection{RHF and Single-particle states}
\label{RHFspstates}
\begin{figure}[htb]
\centering
\includegraphics[width=12cm,clip]{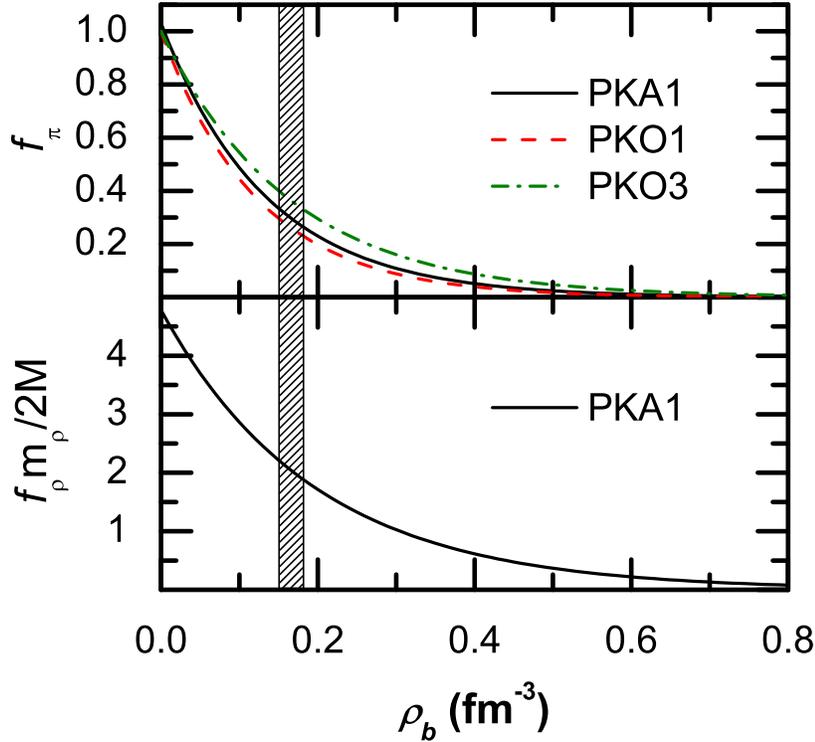}
\caption{\label{meson-coupling}
Density dependent meson-nucleon couplings in the isovector 
channels \cite{Long-priv} 
as a function of density for the RHF effective interactions 
PKA1 \cite{Long2007}
compared with PKO1 and PKO2 \cite{PKO1}. The shadowed area 
denotes the empirical saturation density region.   
See the text for details.
}
\end{figure}

The single-particle states are studied by a RHF model  
with density-dependent coupling constants between mesons and nucleons in 
Refs. \cite{Long2006,Long2007} (cf. Sec. \ref{tensorRHF}). The 
authors have included the 
tensor coupling due to the pion exchange and the vector coupling 
associated with the $\rho -$exchange 
in the effective Lagrangian.  
In order to obtain realistic values for the empirical observables 
in closed shell nuclei in their protocol, the $\pi-$ and $\rho-$ 
couplings in the model 
are the bare couplings in the free space, but they are 
very much quenched in the nuclear medium (cf. Fig. \ref{meson-coupling}).

As already discussed above, the authors of Ref. \cite{Long2007}
have shown that the RHF model can remove the problem of artificial shell 
closures at N,Z = 58 and 92 existing in some RMF calculations 
(cf. Fig. \ref{fig:RHF-Pb}).   
In particular, it was pointed out that the spurious shell closure 
at 92 is related to the pairs 
of high-$j$ states (2f$_{7/2}$, 1h$_{9/2}$). This pair is the 
pseudospin partner 1\~g state. The spurious shell closure 
is then related to the conservation 
of pseudospin symmetry (PSS), i.e., the existing artificial shell 
structure in RMF breaks largely the PSS. As is seen in Fig. 
\ref{fig:RHF-Pb}, PSS is successfully 
recovered for the 1\~ g state. To improve this shell structure, 
the $\rho-$ tensor correlations play a crucial role. This is the reason 
why the PKA1 parameter set 
with the $\rho-$ tensor conserves PSS better than PKO1 that does 
not include 
the $\rho-$ tensor, and thus improves the shell structure 
in the single-particle states around the proton and neutron 
Fermi energies.  
      
In Ref. \cite{Lalazissis},  
the role of the pion in covariant density functional theory was studied. 
Starting from conventional relativistic mean field (RMF) theory 
with a nonlinear coupling of the $\sigma$ meson and without 
exchange terms, the pion$-N$ coupling with a pseudovector type 
in relativistic Hartree-Fock approximation is added. In order to 
take into account the change of the pion field in the nuclear medium 
the effective coupling constant of the pion is treated as a free 
parameter and changed from 0 to the bare pion 
coupling in the free space. We have already discussed this
work in Sec. \ref{masses}. 

It is found 
that the non-central contribution 
of the pion (tensor coupling) does have effects on single-particle 
energies and on binding energies of certain nuclei. 
To obtain better agreement with the experimental data for the 
binding energies and single-particle energies, a weak pion coupling 
is necessary. 
Moreover, the shell structure of closed shell nuclei is 
improved by the introduction of the pion field in the model. 
These conclusions of Ref. \cite{Lalazissis} are consistent 
with those of the RHF models of Refs. \cite{Long2006,Long2007}

\subsubsection{Concluding remarks on single-particle states}

Concerning single-particle states, a fundamental remark is that
in principle one may argue whether they are indeed observable.
Information about hole (particle) states is usually obtained
by means of pickup (stripping) reactions on the corresponding
core nucleus. As a rule, one compares the measured cross sections 
with Distorted Wave Born Approximation (DWBA) calculations performed 
with conventional assumptions. In particular, one usually assumes 
that the wave function of the transferred nucleon can be taken 
as an eigenfunction of a static mean field potential, by adjusting 
the depth of that potential so that the binding energy becomes 
equal to the experimental separation energy and the correct asymptotic 
dependence is guaranteed. Such a procedure is reasonable for levels which are
concentrated in a single peak. This often happens around the 
Fermi energy. However, for states characterized by a broad 
distribution in energy, the comparison with a simple DWBA calculation 
is likely to be less reliable (cf., e.g., the discussion in 
Ref. \cite{Satchler}). These problems are still object of
strong debates.

Even when single-particle states are not broad or fragmented,
and their energies are associated with single peaks, they are
believed not to belong to the DFT framework 
and are known to be significantly affected by dynamical correlations such
as those associated with particle-vibration coupling (PVC)
\cite{Mahaux}. The authors of Ref. \cite{Les07} are well aware
of this fact, and try to argue that in some cases the PVC may
not affect too much their conclusions. One of their arguments is
that conventional wisdom may suggest that PVC just compresses
the spectrum in the sense that particle (hole) states are pushed
downward (upward) in energy, so that spin-orbit splittings between
states that are either occupied or unoccupied may not be affected
whereas spin-orbit splittings between states lying at opposite
sides of the Fermi energy should be systematically reduced.
This conventional wisdom may fail due to specific shell-effects,
in particular in light or medium-mass nuclei. Easily
the spin-orbit splittings can be changed by $\approx$ 500 keV
or more, as an effect of the PVC, when they are calculated
in a fully microscopic framework using Skyrme forces
(see Ref. \cite{colo:2010b}). In particular, the 2f$_{5/2}$-2f$_{7/2}$
spin-orbit splitting in $^{208}$Pb is reduced by
1.07 MeV when PVC is implemented of top of RMF, and by 0.62 MeV
if PVC is implemented on top of Skyrme-HF
(see Table I of Ref. \cite{colo:2010a}).
As a conclusion, one should not 
search for a fine-tuning agreement between theory and data
in such cases. Particle-vibration coupling calculations 
that include the tensor force are currently underway \cite{pvc-tensor}. 

\subsection{Deformation}

The subject of how tensor terms affect deformation properties has been thoroughly discussed in Ref. \cite{Bender:2009} with deformed HF calculations.
The main purpose of this work has been to test the forces introduced in \cite{Les07}. The strategy, in this case, has been to keep
the coupling constants $C^{(J1)}$ associated with the T$IJ$ sets and use them in deformed 
nuclei, by exploiting the
relation between $C^{(J2)}$ and $C^{(J1)}$ that is implied by their relations with the coefficients of the central and tensor
Skyrme forces. Other parameter sets have been considered, and also other ways to fix the coefficients
$C^{(J2)}$ that are in principle unconstrained by calculations in spherical symmetry if the EDF strategy is adopted.

The authors have looked systematically at the total energy as a function of the quadrupole deformation
parameter $\beta_2$. In general, for $\vec\ell\cdot\vec s$ saturated nuclei like $^{40}$Ca the contribution
of the tensor energy, which is close to zero at sphericity, increases with deformation, whereas in nuclei that
are not $\vec\ell\cdot\vec s$ saturated ($^{56,78}$Ni, $^{100,132}$Sn and $^{208}$Pb) the tensor energy
is largest at sphericity and decreases with deformation. The total energy, however, do not change as much
as the tensor energy if the sets T$IJ$ are employed. In other words, other terms of the energy functional
compensate the change induced by the $J^2$ terms. Also in most of the nuclei that have been studied the differences
among the predictions of the various sets are moderate ($\lesssim$ 2 MeV, yet with several exceptions).
These two latter statements become less true if the self-consistency of the functional is broken, e.g., if
one employs forces in which the tensor terms have been added pertubatively. This is a further argument, in
the authors' viewpoint, to advise for consistently fitted Skyrme sets. The situation becomes very complicated
in nuclei with open shells, where due to correlated effects from the various terms of the functional, no simple
conclusion can be drawn about the role played by the coupling constants on the total energy curve.
\begin{figure}
\centering
\includegraphics[width=12cm]{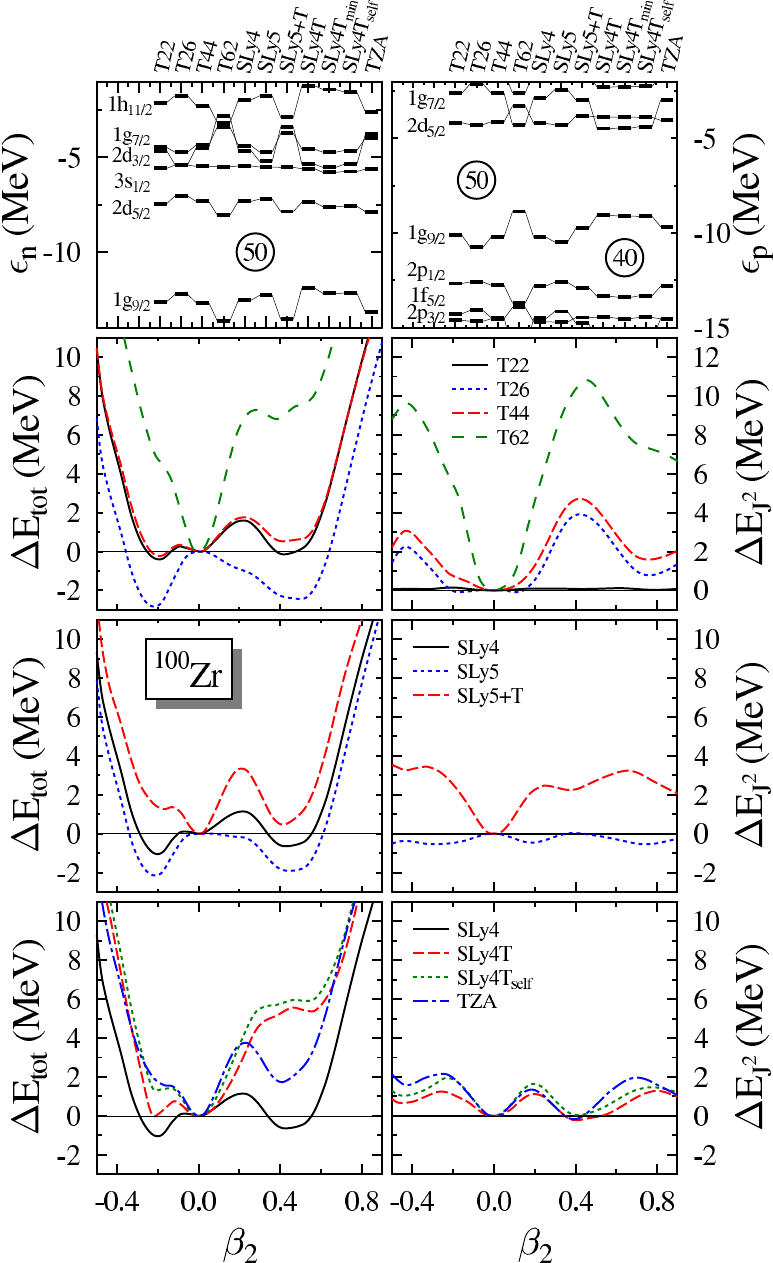}
\caption{The single-particle spectra at spherical
shape are shown in the upper panel for neutrons (left) and protons
(right) The deformation energy (left) and the variation of the total
tensor energy (right) are plotted on the lower panels for $^{100}$Zr and
different Skyrme parametrizations, as indicated.  
The protocols for the adopted Skyrme parameter sets can be found in Ref.   \cite{Bender:2009}. 
This figure is taken from Ref.   \cite{Bender:2009}.}
\label{fig:100Zr}
\end{figure}

In several cases, the authors have looked in detail at the Nilsson diagrams. 
Again, the conclusions are that the different tensor terms associated
with the T$IJ$ sets affect significantly the 
Nilsson levels around sphericity, 
while this is much less true at large deformations where the Nilsson levels
tend to lie on top of each other. If other Skyrme sets are employed, 
the level structure is affected in
a less obvious way.

Here we take one example, namely the nucleus $^{100}$Zr with 
neutron open-shells, from Ref. \cite{Bender:2009} (cf. Fig. \ref{fig:100Zr}).  
A large set of experimental data demonstrate that
$^{100}$Zr is located in a region of deformed nuclei with possible shape coexistence \cite{100Zr-exp}.
The single-particle spectra at the spherical shape, the deformation 
energy curve and the variation of the tensor energy
are plotted against quadrupole deformation in Fig. \ref{fig:100Zr}.  
The results obtained with the T44 and T62
parametrizations indicate a larger contribution from
the isovector tensor terms. 
The positions of the neutron s.p. energies 2d$_{3/2}$
and 
2g$_{7/2}$ are very much dependent on the sign and size 
of
the isovector coupling constant $C^J_
1$ . For a positive value 
as
in T26, the 2d$_{3/2}$ level is close to the Fermi level 
and is
occupied in such a way that it partially cancels 
the contribution
from the 2d$_{5/2}$  orbital. The isovector tensor 
terms are in
this case strongly reduced. In contrast, for a 
negative $C^J_
1$  
coefficient as in T62, the 2d$_{3/2}$  
level is pushed up and crosses
the 1h$_{11/2}$ level, increasing 
the neutron spin-current density.
Results obtained with the SLy5$_{\rm T}$ 
interaction, for which $C^J_
1$ 
is also negative, are similar, 
although less pronounced. For
even larger negative values of 
the tensor coupling constants,
this feedback mechanism
is suppressed 
by the reduced spin-orbit interaction.
Most total deformation energy curves in Fig. \ref{fig:100Zr} 
exhibit
spherical, prolate, and oblate minima. The inclusion of 
beyond
mean-field correlations should favor the prolate minima 
and
create a 0$^+$ excitation exhibiting some amount of 
configuration
mixing. Such results are consistent with experiment. 
The
spherical minimum is too much below the deformed one to
expect 
that additional correlations from the projection of $J = 0$ states 
will make $^{100}$Zr deformed in its ground state. 
For T62,
the deformation energy curve looks like that of a double 
magic
nucleus. 
For SLy4T and SLy4Tself \cite{Bender:2009}, it is the 
reduced spin-orbit
interaction that reinforces the proton Z = 40 
shell closure. The
prolate minimum becomes a shoulder around 5 MeV, 
leading
to the coexistence of spherical and oblate minima.

A different specific case that can illustrate the interplay of
tensor correlations and deformation has been thoroughly 
discussed in Ref. \cite{Li2013}, 
and concerns 
the tensor effect on the shape evolution of the Si isotopes. The 
basic questions in this case are: is the tensor-force-driven 
deformation present in other
neutron-rich Si isotopes, especially $^{30}$Si with a possible N =
16 subshell, since some models (e.g., the FRDM) predicted a spherical
shape for this nuclei~\cite{Wer96}  while the large B(E2) value
suggested its deformed nature~\cite{Ibb98}.  
For this purpose, the authors of   Ref. \cite{Li2013} used the 
deformed Skyrme-Hartree-Fock model
(DSHF)~\cite{Vau73} with the BCS approximation for the nucleon pairing.
For each nucleus the strength of the pairing force is
fitted by  taking care 
of the empirical data for the pairing gaps,
%
%
%
%
as done in Refs.~\cite{Sag04,Zho07,Li13}.
%
The parameter sets of Ref.~\cite{Les07} have been chosen, 
and they include: the Skyrme force T22 serving
as a reference as it has vanishing $J^2$ terms, T24 with a substantial
like-particle coupling constant $\alpha$ and a vanishing
proton-neutron coupling constant $\beta$; T44 with a mixture of
like-particle and proton-neutron tensor terms, T62 with a large
proton-neutron coupling constant $\beta$ and a vanishing
like-particle coupling constant $\alpha$; and T66 with large and
equal proton-neutron and like-particle tensor-term coupling
constants. The coupling strengths of various parameter sets used in
this study are collected in Tab.~1.

\begin{table}[!h]
\caption{Coupling strengths (in MeV$\cdot$fm$^5$) of various parameter sets used
in the work. $\alpha$ represents the strength of
like-particle coupling between neutron-neutron or proton-proton, and
$\beta$ is that of the neutron-proton coupling. The subscripts $T$  indicates the tensor
contribution. See text for details.} \label{tensor}
\vspace{5pt} \centering
\begin{tabular}{cccccc}
\hline
~~~&~~~~~T22~~~~~&~~~~~T24~~~~~&~~~~~T44~~~~~&~~~~~T64~~~~~&~~~~~T66~~~~~
\\ \hline
~~~$\alpha$   & 0 & 120  & 120 & 120 & 240
\\ \hline
~~~$\beta$    & 0 & 0    & 120 & 240 & 240
\\ \hline
~~~$\alpha_T$ & -90.6 & 24.7 &8.97&-0.246 & 113
\\ \hline
~~~$\beta_T$  & 73.9 & 19.4 &113 & 218 & 204
\\ \hline
\end{tabular}
\end{table}

\begin{figure}
\centering
\includegraphics[width=13.5cm,clip,bb=0 0 823 561]{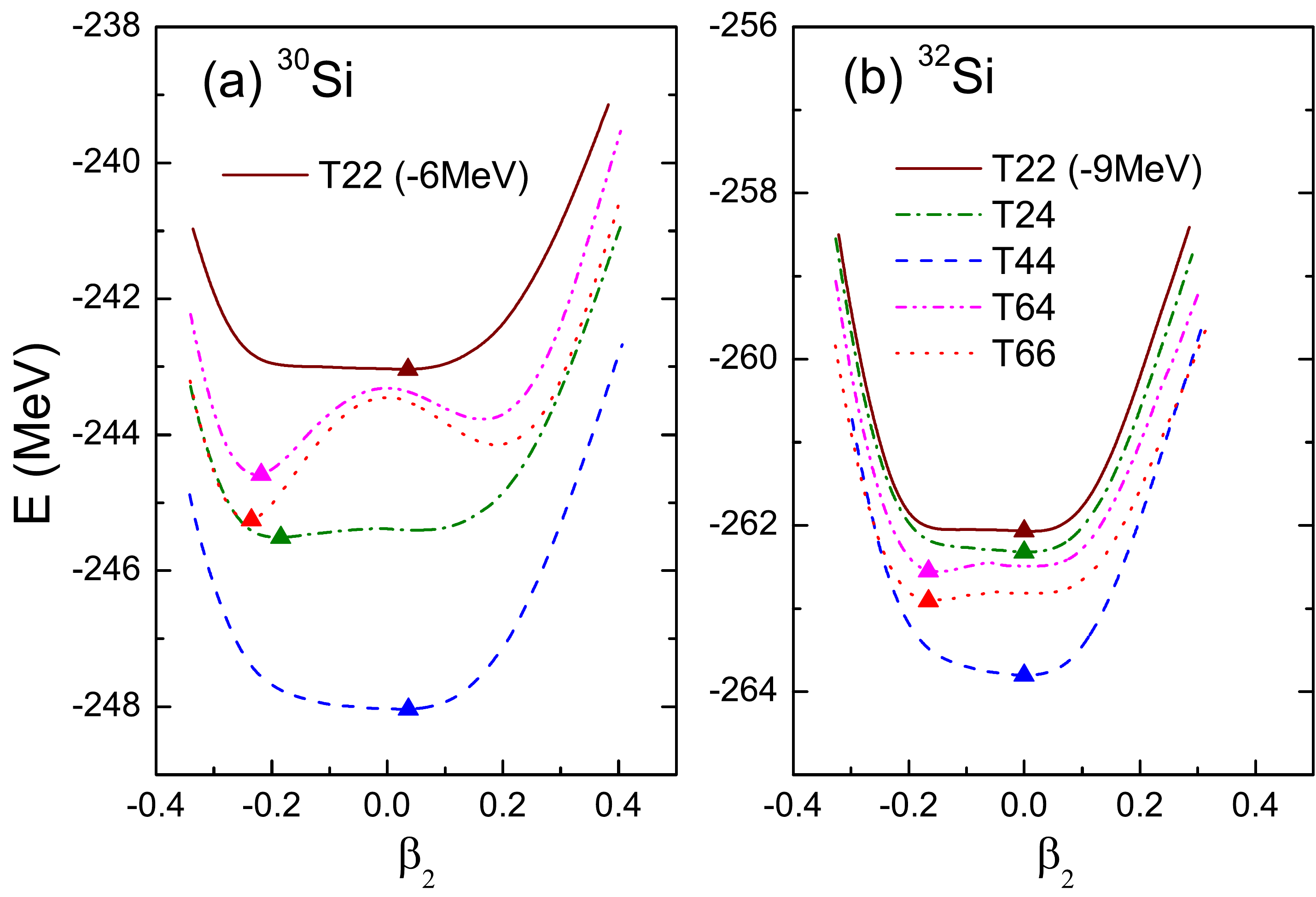}
\caption{Energy surfaces of $^{30}$Si (left panel)
and $^{32}$Si (right panel) as a function of the quadrupole
deformation parameter $\beta_2$ using T22, T24, T44, T64, and T66.
The results with T22 are shifted by $-$6 MeV and $-$9 MeV in
$^{30}$Si and $^{32}$Si, respectively. The energy minima are
indicated with triangles. This figure is taken from Ref. \cite{Li2013}.}\label{fig2-si}
\end{figure}

We begin by showing 
in Fig.~\ref{fig2-si} the energy surfaces of $^{30}$Si (left
panel) and $^{32}$Si (right panel) as a function of the quadrupole
deformation parameter $\beta_2$.   
The energy minima are indicated with triangles. 
The nucleus $^{30}$Si is suggested to be
deformed as mentioned before, but T22 and T44 with
relatively large pairing strengths ($\approx 1000$ MeV$\cdot$fm$^3$) fail to give
deformed energy minima. On the contrary, deformed ground states can
be achieved using the T24, T64 and T66 parametrizations with
associated small pairing strengths ($\approx 800$ MeV$\cdot$fm$^3$). The predicted
oblate shape of this nuclei is consistent with the recent RMF
result~\cite{Li11}. The interesting performance of a weak nucleon
pairing may stem from a well-known fact that the pairing interaction
forms the $J = 0^+$ pairs of identical particles which have
spherically symmetric wave functions, and as a result, nuclei tend
to be more spherical 
  with stronger
pairing couplings. 
We also
notice that large tensor terms present in T64 and T66 tend to make
the energy surface deep, namely the tensor force affects
dramatically to create a well-deformed ground state for $^{30}$Si.
The importance of tensor force will be seen also in $^{32}$Si as
shown in the right panel of the same figure. Its shape is predicted
to be spherical with T22, T24 and T44, but oblate with T64 and T66
with large tensor terms.

\begin{figure}
\centering
\includegraphics[width=13.5cm]{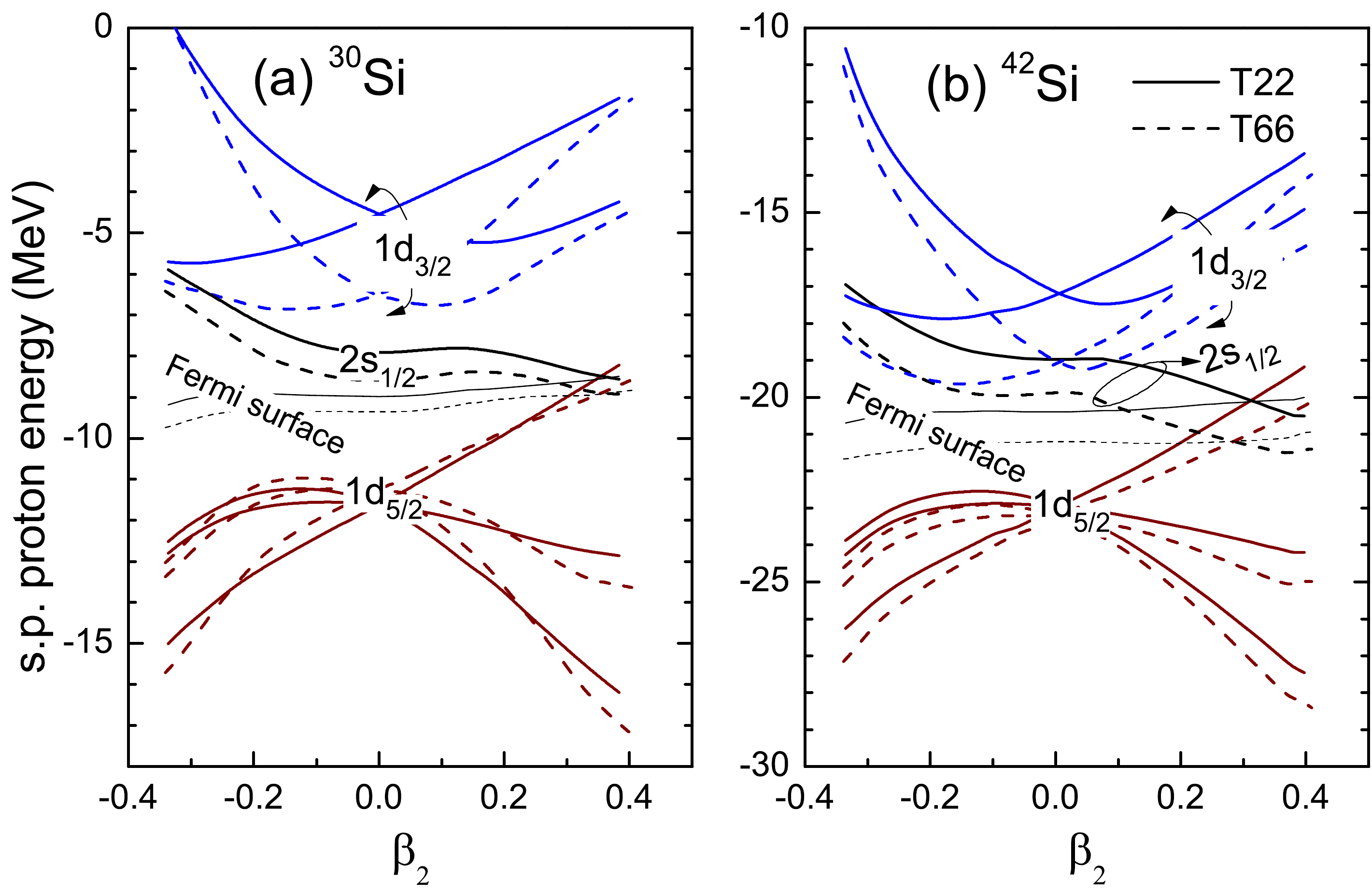}
\caption{The s.p. orbits of protons for $^{30}$Si
(left panel) and $^{42}$Si (right panel) are shown as a function of
the quadrupole deformation parameter $\beta_2$ using T22 (solid
lines) and T66 (dashed lines). The Fermi surface is also shown
between the 2s$_{1/2}$ and 1d$_{5/2}$ orbits in the spherical
limit. This figure is taken from Ref. \cite{Li2013}.}\label{fig4-si}
\end{figure}
To clarify the effect of tensor correlations on deformation,
through the HF fields, 
we plot in Fig.~\ref{fig4-si} the single proton levels in
$^{30}$Si (left panel) and $^{42}$Si (right panel) as a function of
the quadrupole deformation parameter $\beta_2$ using T22 (solid
lines) and T66 (dashed lines). The Fermi surface is also shown
between the 2s$_{1/2}$ and 1d$_{5/2}$ orbits in the spherical limit. One
can see that there are two main features indicating the tensor
effect to compare the results between T22 and T66 for $^{30}$Si: a
narrower 1d$_{3/2}$-1d$_{5/2}$ gap and a steeper 1d$_{5/2}$ orbits
downward in the oblate side. Then, in the case of T66, more mixing
between 1d$_{3/2}$ and 1d$_{5/2}$ results in an oblate shape for
this nuclei in comparison with the case of T22. The shell gap by T66
interaction is narrower also in $^{42}$Si, but both T22 and T66 give
oblate deformations, since the tensor effect in this nuclei is not
as evident as the case of $^{30}$Si regarding the 1d$_{5/2}$ orbits.

The effect of the tensor interaction on the stability of superheavy 
elements (SHEs) was studied in Ref.~\cite{Suckling2010} with the spherical 
Skyrme HF model. In Ref.~\cite{XRZ2012}, the deformed HF model with 
the tensor and the pairing correlations was adopted to study the 
stability of SHEs. It was pointed out in Ref.~\cite{XRZ2012} that the 
shell gaps at Z=114 and N=164 are more stabilized by the tensor 
correlations of the SLy5$_{\rm T}$ interaction.

\subsection{Vibrational states}
\label{vibr}

\begin{figure}[htp*]
\centering
\includegraphics[width=12cm,clip]{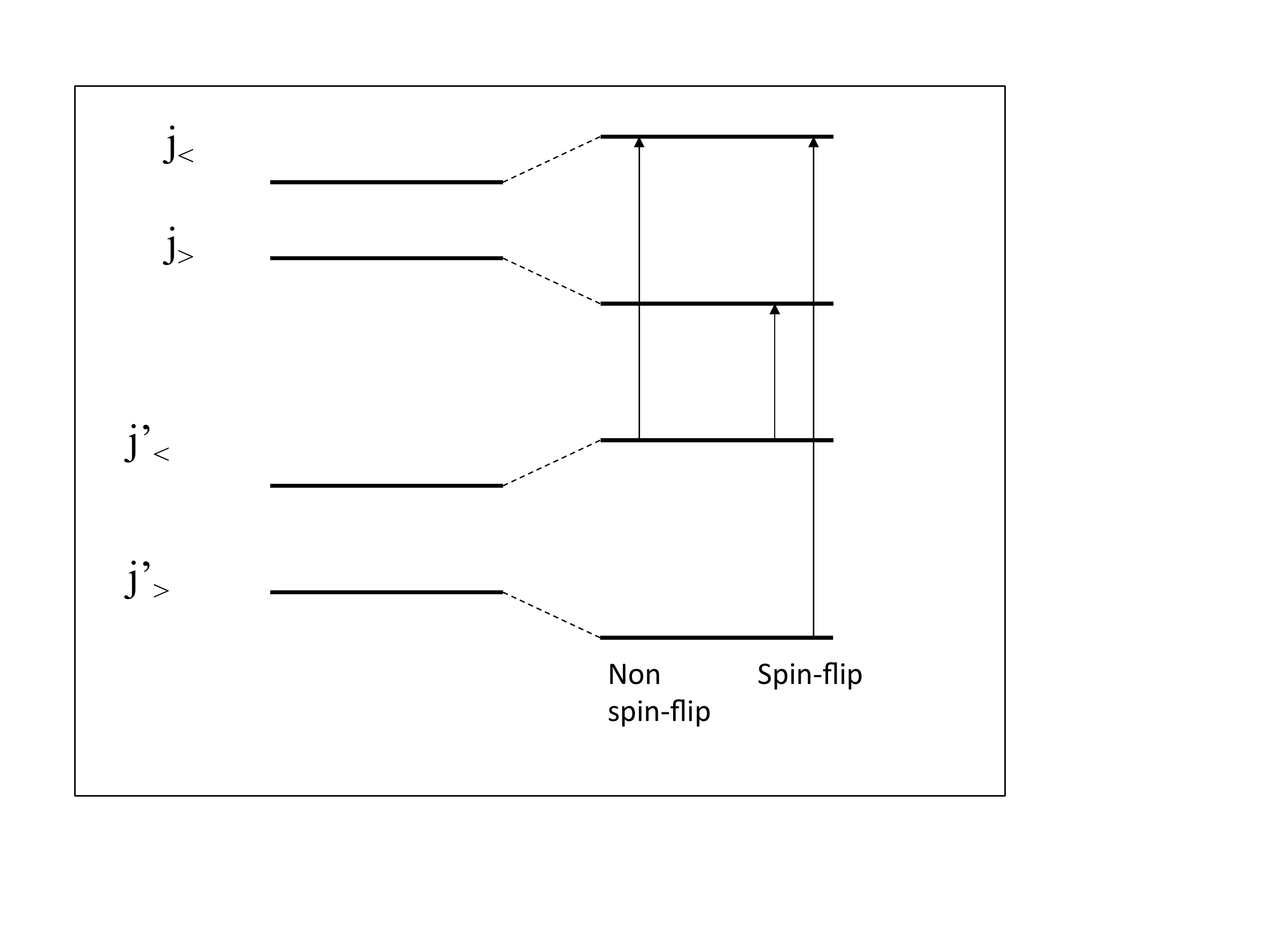}
\vspace*{-1cm}
\caption{\label{schem_rpa}
Schematic view of the effect of the tensor force on the mean-field 
responses to different
operators. The levels displayed on the left side evolve 
to those depicted on the right side due to
the effect of the tensor force. The arrows show 
the transitions that are excited mainly by either
non spin-flip (the single arrow in the left part) or spin-flip 
(the two arrows in the right part)
external fields.
}
\end{figure}

It can be expected that spin-flip excitations are quite
sensitive to the tensor force, while non spin-flip
excitations are less sensitive. At the unperturbed
level, this is schematically illustrated in Fig. \ref{schem_rpa}:
as the tensor force mainly changes the spin-orbit
splittings, particle-hole (p-h) transitions between
spin-orbit partners $j_>$ and $j_<$ are more
affected than transitions between $j_>$ and $j_>$
states (or between $j_<$ and $j_<$). This argument
remains true at the level of residual interaction
since the tensor force is only active between the
$S$ = 1 component of the p-h configurations.
In fact, in Ref. \cite{Davesne:2011} the response of nuclear matter
has been studied and the effect of the tensor force
is clearly much stronger in the $S$ = 1 than in the
$S$ = 0 channel.

Nevertheless, this $S$ = 1 component is not
completely absent in the natural parity states.
Therefore, a careful study of the effect of the
tensor force on the natural parity vibrational
states has been carried out in Ref. \cite{Sciacchitano}
by using self-consistent Skyrme-RPA \cite{CPC}
with and without the tensor force.
It has been found that the effects of the tensor force
are small, especially in the high-lying giant resonance
region. The low-lying 2$^+$ and 3$^-$ states, that
have been considered in the nuclei $^{40,48}$Ca
and $^{208}$Pb, are shifted in energy by the tensor force:
these shifts can range from very small values to several
hundreds of keV (reaching at most $\approx$ 700 keV in
the case of 2$^+$ and $\approx$ 1.3 MeV in the case
of 3$^-$). The electromagnetic transition probabilities
are affected in a less significant way. The energy
shifts induced by the tensor force can be written
approximately as
\begin{equation}\label{eq:RPAshift}
\Delta E_{\rm RPA} \approx \Delta E_{\rm HF} +
\langle V_{\rm tensor} \rangle,
\end{equation}
where the first term in the r.h.s. denotes the average
change of the p-h energies with and without the tensor
force and the second term is an average value of the
tensor force. The average value of the tensor force
can be numerically calculated, but also estimated with
within the framework of a separable approximation
(see Appendix B of \cite{Sciacchitano}). When forces
having negative values of $\alpha$ and positive values of
$\beta$ are employed, with $\alpha \approx -\beta$,
the tensor force is attractive in the natural parity
isoscalar channel and the second term  in the r.h.s. of Eq. (\ref{eq:RPAshift}) is as a rule negative; 
however, the sign of the first term 
turns out to depend on the nucleus and on the
specific states involved in the 2$^+$ and 3$^-$ transitions.
In summary, these low-lying states cannot be a good
constraint for the tensor force because of their dependence
on shell effects. Despite this, a more systematic study of the 
effects of the tensor force on the low-lying states in 
$^{40}$Ca and $^{208}$Pb has been carried out in Ref. 
\cite{Ligang11}. It has been found that these states
are reasonably described by the sets T44, T45 and SGII 
\cite{SGII} plus a perturbative tensor force.

There exist pioneering attempts to relate the low-lying
spin excitations to the tensor force.
One of the first papers to point out the sensitivity of the 0$^-$ states in
$^{16}$O to the tensor interaction is Ref. \cite{Blomqvist:1968}.
In the work of Ref. \cite{Zheng:1991} the authors had already noticed that the single-particle
states in a $\vec\ell\cdot\vec s$-saturated nucleus like $^{16}$O
are insensitive to the tensor force. On top of this, they have made attempts to understand
the effects of the spin-orbit and the tensor force on few excited states in light nuclei,
by using a simple interaction and a series of restricted Tamm-Dancoff approximation (TDA)
calculations. Looking at the negative parity excitations of $^{16}$O, described as
superpositions of 1p-1h configurations, they have found that the tensor force has a
large effect on the energies of the isoscalar states and this effect is attractive for the
non-natural parity states 0$^-$, 2$^-$ and 4$^-$, while it is repulsive for the natural
parity states 1$^-$ and 3$^-$. The absolute value of these energy shifts decreases
with $J$, being around $\approx$ 4 MeV for 0$^-$ and going down to 1 MeV or less
in the case of the high $J$ values. For the isovector states the energy shifts are significantly
smaller and they are of the order of 1 MeV or less. The 1$^+$ (M1) 
states in $^{12}$C have also
been studied. The tensor interaction has also a noticeable attractive effect in this
channel. The states come out too low with respect to experiment. However, it
has been noticed that in this case a full shell-model calculation differs from the
simple TDA result.

Looking at modern self-consistent calculations, one may try to
disentangle the effects of the tensor force on single-particle
states and on the residual interaction. The 1$^+$ state is, as we said, a possible benchmark
for the tensor force and the effect of this latter force could be
extracted from Eq. (\ref{eq:RPAshift}). 
Interactions like SLy5$_{\rm T}$ \cite{Colo} enlarge the spin-orbit splittings
in $^{48}$Ca, $^{90}$Zr and $^{208}$Pb (as compared with the original force SLy5) 
and push the unperturbed 1$^+$ states upwards; the effect of the residual
interaction is not easy to guess from simple arguments but
does not tend to alter the positive sign of the shift
induced by the change in spin-orbit splittings [for comparison, small
residual matrix elements can also be found by using the
phenomenological tensor forces that reproduce the
experimental findings and that have been introduced 
in \cite{DeDonno:2009} (cf. Table III of
this reference)]. 

The key point is that, anyway,  
simply adding a tunable tensor force does not automatically
ensure that one can reproduce the experimental findings.
In fact, one needs to start from an effective force in which 
the $\sigma\sigma$ and $\sigma\sigma\tau\tau$ 
terms are realistic. This is not the case for many
microscopic interactions. For instance, the Lyon forces SLy*
that are among the most modern and accurate Skyrme sets
\cite{Chabanat}, have an attractive (repulsive) isovector
(isoscalar) spin-spin residual force (these forces are
associated, respectively, with the Landau parameters $G_0$
and $G_0^\prime$; cf. Sec. \ref{instabilities} for a discussion
concerning these parameters). This is at variance with the empirical
findings. In fact, experimentally, the isoscalar M1 states are
found at lower energies than the isovector M1 states. It is
interesting to notice that the
same shortcoming is associated with the D1 Gogny force
\cite{Blaizot,Decharge}. The calculations of the
1$^+$ states performed, respectively, in \cite{Sciacchitano}
with the SLy5 force and in \cite{DeDonno:2009} with the D1
force show an inversion of the isospin doublet with respect
to experiment. By implicity commenting also the role of the tensor,
the authors of Ref. \cite{Vesely} did also discuss the
shortcomings of the current models as far as the prediction of M1
states are concerned. 
It would be interesting to check the performance
of the new Skyrme set SAMi proposed in Ref. \cite{roca-maza} for the properties of 1$^+$ states.
Once more, we may remark that forces in which all parameters 
are fit on equal footing may perform better. Along this line, 
it has been noted in \cite{Sciacchitano} that the force T44
introduced in \cite{Les07} produces reasonable results
for the 1$^+$ states.

In \cite{DeDonno:2009} other kind of excitations (2$^-$ and
4$^-$) in $^{12}$C, $^{16}$O, $^{40,48}$Ca, and $^{208}$Pb
have been considered in the framework of both phenomenological
and microscopic (i.e., Gogny-based) calculations. A step forward in the
microscopic approach by the same group is discussed in Ref. 
\cite{Anguiano:2011}. In this work, self-consistent RPA calculations have 
been performed. A tensor force of the type (\ref{tensor-co}) has 
been added to the forces D1S \cite{Berger} and D1M \cite{Goriely}.
In both cases, the parameter $b$ of Eq. (\ref{tensor-co}) has been
fitted (and the spin-orbit strength readjusted) so that
the 0$^-$ state in $^{16}$O is well reproduced. This state has been
identified as the most sensitive to the tensor force (not
surprisingly since it carries the same quantum numebers
of the pion). The new forces have been named D1ST and D1MT. 
By using them, the binding energies and radii are not affected in a significant manner, 
while the s.p. states are affected since this attractive bare pion-like
force has a positive value of $\beta$ and $\alpha\approx\beta/2$ as
recalled in Sec. \ref{Skyrme} (O, Ca, Ni, Zr, Sn and Pb isotopes have been
considered in \cite{Anguiano:2011}). In RPA, it has been found that the effect of the tensor
force follows the trends that we have previously discussed. The
natural-parity states are less affected than the spin-flip states,
and among those latter the IV states undergo smaller shifts than the IS states,
these shifts being also a decreasing function of $J$. The tensor force introduced in
Ref. \cite{Anguiano:2011} improves the agreement with experiment of
the 0$^-$ excited states in several nuclei and not only in $^{16}$O. 
For the states with different quantum numbers it is hard
to make such a statement. From Table II in the paper one notices
that, although the effect of the tensor force cannot be ignored, its
inclusion sometimes improves the agreement with experiment but
sometimes goes in the opposite direction. This is probably an indication
that for such observables a refit of the whole Gogny force would be
mandatory, including both IS and IV terms \cite{tbp}. The same group
considered also the magnetic states in nuclei with neutron excess
in Ref. \cite{Anguiano:2012}. 

\subsection{Rotation and tensor coupling constants}
 To study the high-spin properties of atomic nuclei is one the current 
subjects of mean field theories, especially because of recent advances 
in the experimental study of superdeformed (SD) rotational bands.  Recently,
 the influence of the tensor terms on the high-spin properties was 
studied using a cranked HFB approach for different 
parametrization of Skyrme tensor interactions  in Ref. \cite{Hellemans} 
  The authors separate the Skyrme EDF into two parts
  \be
  E=\int d^3r \sum_{t=0,1} \left[ {\cal E}_t^{\rm te}+{\cal E}_t^{\rm to} 
\right], 
  \label{eq:EDF2} 
\ee  
where the first (second) term in the bracket is the time-even (time-odd) part of the EDF and the suffix $t$ is the isospin index $(t=0,1)$.
All densities and currents are distinguished into isoscalar ($t=0$) and 
isovector ($t=1, t_z=0$).  Detailed expressions for 
${\cal E}_t^{te}$ and ${\cal E}_t^{to}$ can be found, e.g., in Ref. \cite{Perlinska} and
have been already discussed in Sec. \ref{skyrmeEDF}.  
Some arguments are repeated here from Sec. \ref{skyrmeEDF} for the
reader's convenience. 
The names ``time-odd'' 
and ``time-even'' associated to terms of the EDF 
refer to the properties under time reversal of the densities
and currents they are built from, but not to the properties of 
the EDF itself which is time-even by definition. 
For example,  
the part of the EDF constructed from the density $\rho(\br)$ 
or the kinetic density $\tau(\br)$ belongs to the time-even part, 
while that involving 
the spin density $\bf s (\br)$ belongs to the time-odd part. 

The SD rotational bands are calculated by the self-consistent cranked HFB approach.  The method can be seen as a semiclassical description of the collective rotation of a finite system with a constant angular velocity $\omega$.  In particular, the model takes into account 
the distortion of the nucleus intrinsic state by the centrifugal and Coriolis forces induced by the collective rotation. 
The variation of the EDF including the constraints on particle number and rotational frequency leads to the cranked HFB equations
\begin{eqnarray}
\left(\begin{array}{cc} h-\lambda -\omega J_z &  \Delta \\
  -\Delta &   -h^*+\lambda +\omega J_z^* \end{array} \right)
\left(\begin{array}{c} U_{\mu} \\
                      V_{\mu} \end{array} \right) = E_{\mu}
                      \left(\begin{array}{c} U_{\mu} \\
                      V_{\mu} \end{array} \right) , 
\label{eq:HFBC}
 \end{eqnarray}
 where $U_{\mu}$ and 
                      $V_{\mu}$ are the two components of the quasiparticle wave functions and  $E_{\mu}$ is the quasiparticle energy, often called 
                      Routhian in the rotating frame \cite{Ring-Schuck}.  The effective 
interaction in the HFB Hamiltonian manifests itself in the mean field Hamiltonian $h$ and the pairing field
                      $\Delta$.   The rotational frequency $\omega$ is self-consistently adjusted to fulfill the auxiliary condition for the mean value of the 
                      projected angular momentum $J_z$ on the rotating axis.  The component $J_z$ is chosen along the axis perpendicular to the axis of the longest  elongation.  At high spins and large deformations, the solution of Eq. (\ref{eq:HFBC}) can be seen as   an approximation of a variation after the projection of angular momentum and the model is particulaly adequate for the description of SD bands.
                      
                      In Ref. \cite{Hellemans}, the authors studied the dynamical moment of inertia defined by
\be
J^{(2)}=\frac{d<J_z>}{d\omega}
\ee
where $<J_z>$ is the average value of the projected  angular momentum 
on the rotation axis.   This quantity is the well-established 
observables of SD bands, and 
can be also expressed through the derivative 
of the energy with respect to the rotational frequency as
\be
J^{(2)}=\frac{1}{\omega}\frac{dE}{d\omega}, 
\ee
which makes it possible to separate the contributions of time-even 
and time-odd terms in the EDF.  

\begin{figure}
\centering
\includegraphics[width=13.5cm]{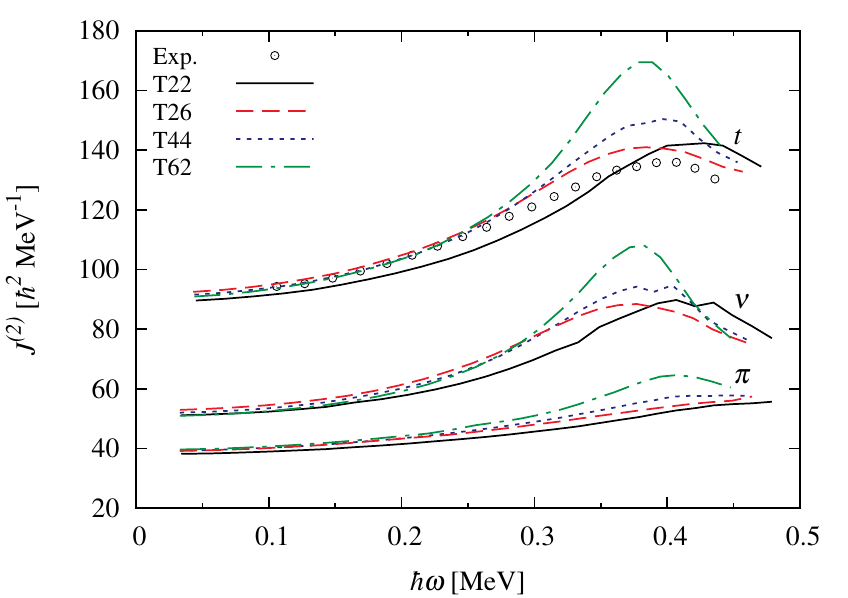}
\caption{Proton ($\pi$), neutron ($\nu$), and total ($t $)
dynamical 
moments of inertia in $^{194}$Hg as a function of the rotational
frequency 
for the SD band in  $^{194}$Hg calculated 
with the T22, T26, the T44, and 
the
T62 parametrization. Experimental data are taken from 
\cite{SD-exp}.
Figure adapted \cite{Hellemans-priv} from Ref. \cite{Hellemans}. 
}\label{fig:veerle}
\end{figure}

The dynamical moment of inertia of $^{194}$Hg was calculated by using 
Skyrme forces like SLy4 and members of the T$IJ$ family that include 
tensor interactions, and some results are shown in Fig. 
\ref{fig:veerle}.    In general, some of T$IJ$ interactions give equally good or somewhat better agreement with experimental data 
than  SLy4.  Especially T22, T44 and T26 give satisfactory results, but 
the T62 parametrization 
in which tensor 
acts only between protons and neutrons in spherical symmetry 
shows an artificial enhancement of $J^{(2)}$ in very high-spin states, 
and does not give a satisfactory description of experimental observations.  
 Concerning the tensor contributions, the time-even contribution to $J^{(2)}$ for T44 is small and varies between $-4$\% at low　 spin and　
  $2$\%  at high spin.  For T26, the contribution is larger 
at high spin going up to 4\%.   For the time-odd contributions, 
the spin dependent terms tend to cancel  largely 
 and the total contribution is less than 1\%.   As was pointed out in 
the previous studies using T$IJ$ interactions \cite{Bender:2009}, 
the presence 
 of the tensor terms has an impact on the $J^{(2)}$ value mainly 
through the rearrangement of other coupling constants in the fit 
and also through the self-consistency of the HFB approach.    
 

\subsection{Tensor correlations and spin-isospin 
charge-exchange excitations}

This section is devoted to the study of the impact of the tensor terms 
of the Skyrme
effective interaction in the self-consistent charge-exchange RPA
calculations. In particular, we will focus on 
the Gamow-Teller (GT) and Spin-Dipole (SD) transitions, which are 
expected to be significantly 
affected in keeping with the fact that the corresponding 
operator is spin-dependent.   
The results presented in this Section are mainly 
based on the recent studies of Refs. \cite{Bai1,Bai11,Bai_last} 
(see also \cite{Bai2,Bai3}). 

\subsubsection{Gamow-Teller states}
 In the
study of GT transitions, the quenching problem is of some
relevance. The experimentally observed strength from 10 to 20 MeV
excitation energy (with respect to the ground state of the target
nuclei) is about $50\%$ of the model-independent non-energy
weighted sum rule (NEWSR); this percentage becomes about $70\%$ if
the whole strength in the neighbouring energy region is
collected~\cite{Rapaport}. 
It is expected that the tensor force has an effect on the centroid
of the GT resonance (GTR), in analogy with what discussed for
the case of the M1 resonance in Sec.~\ref{vibr}. In addition, 
it is interesting to study if the
tensor force has an effect 
in producing some amount of quenching
by shifting strength in the high-energy region, already at one
particle-one hole (1p-1h) level. Coupling the GTR with 
two particle-two hole states is essential, e.g., to describe the 
resonance width (although it is not expected to affect strongly the 
position of the main GT peak) and will increase the quenching effect; 
the role of the tensor
force in connection with the 2p-2h coupling was studied in
Ref~\cite{Bertsch}.

As already discussed, the spin-orbit density $J_q$ 
gives essentially no contribution
in the $\vec l\cdot\vec s$-saturated cases. 
Therefore, we will discuss the example of the nuclei 
$^{90}$Zr and $^{208}$Pb. $^{90}$Zr is a proton 
$\vec l\cdot\vec s$-saturated nucleus, with a
neutron orbit 1g$_{9/2}$ contributing to $J_n$. $^{208}$Pb is chosen 
as it is not saturated either in protons or neutrons. The two
examples should allow elucidating separately the role of tensor-even
and tensor-odd terms in the results of self-consistent 
charge-exchange RPA.

The
operator for GT transitions is defined as
\begin{eqnarray}
O_{GT\pm}&=&\sum\limits_{im}t^i_{\pm}\sigma_m^i
\end{eqnarray}
in terms of the  isospin operators, 
$t_\pm=\frac{1}{2}(t_x{\pm}it_y)$.
In the charge-exchange RPA, the $t_-$ and $t_+$ channels are 
coupled and the corresponding eigenstates emerge from 
a single diagonalization of the RPA
matrix. 
In the self-consistent charge-exchange HF+RPA calculations, the 
NEWSRs $m_\pm(0)$ and the Energy-Weighted Sum Rules (EWSRs) 
$m_\pm(1)$ (associated with the two different isospin 
channels) satisfy the following relations:  
\begin{equation}
\begin{array}{lll}
m_-(0)-m_+(0)&=&\sum\limits_\nu\left( \vert\langle\nu\vert 
O_-\vert 0\rangle\vert^2-\vert\langle\nu\vert 
O_+\vert 0\rangle\vert^2 \right)\\
&=& \langle 0\vert \left[ O_-, O_+ \right] \vert 0\rangle, 
\end{array}
\label{diffNEWSR}
\end{equation}
\begin{equation}
\begin{array}{lll}
m_-(1)+m_+(1)&=&\sum\limits_\nu \left(  E_\nu\vert\langle\nu\vert 
O_-\vert 0\rangle\vert+ E_\nu\vert\langle\nu\vert 
O_+\vert 0\rangle\vert^2 \right)\\
&=& \langle 0\vert \left[ O_+, \left[ H, O_- \right] \right] 
\vert 0\rangle,
\end{array}
\label{sumEWSR}
\end{equation}
where $O_+$ ($O_-$) is a generic charge-changing operator 
proportional to $t_+$ ($t_-$). 
In the GT case, the difference of NEWSRs (\ref{diffNEWSR}) 
is model-independent and turns out to be
\begin{eqnarray}
m_-(0)-m_+(0)=3(N-Z),
\end{eqnarray}
whereas the sum of the EWSRs is model-dependent. This latter
receives a contribution from the
tensor interaction, which is obtained by replacing the total 
Hamiltonian $H$ in the double commutator of 
(\ref{sumEWSR}) with $V_T$. If there is enough neutron excess, and
the contributions from the $t_+$ channel to the sum rules, 
$m_+(0)$ and $m_+(1)$, are small, then we can estimate the 
effect of the tensor
interaction on the GT centroid in the $t_-$ channel 
by writing
\bea
\Delta E_{GT}&=&\frac{m_-(1)}{m_-(0)}\nonumber\\
&\sim&\frac{m_-(1)+m_+(1)}{m_-(0)-m_+(0)}\nonumber\\
&=&\frac{4\pi}{3(N-Z)}\int
drr^2\ [-(\frac{5}{2}U+\frac{5}{2}T)J_n J_p-\frac{5}{3}U(J_n^2+J_p^2)],
\label{shift}
\eea
where the last line comes from a lengthy but straightforward
evaluation of the double commutator \cite{Colo}

As far as the calculations shown in this Section are
concerned, the two-body spin-orbit 
residual interaction is not included in the 
RPA. 
However, this
term of the residual interaction has been shown to be very
small~\cite{Fracasso} in the case of the GTR;  
no further approximation is involved. 
The values reported in Table~\ref{Table1} are, however, 
calculated
by dropping completely the spin-orbit contribution, 
both at HF and RPA level. This calculation (with the
Skyrme parameter $W_0$ set at 0) is not supposed to be
compared with the experimental findings but respects
self-consistency in a strict sense. 
The shift in the GT centroid
caused by the inclusion of tensor terms, 
[calculated by using either RPA or the analytical formula 
(\ref{shift})], 
and the EWSR 
$m_-(1)$ 
obtained from RPA, 
are listed in Table~\ref{Table1} for the two nuclei $^{90}$Zr 
and $^{208}$Pb. One should notice the good agreement between 
the RPA results and the analytical results for the shift.  

\begin{table}[hbt] \centering
\caption{Values of the EWSR $m_-(1)$ obtained from self-consistent 
HF plus RPA calculations, with
and without the tensor terms. 
$\delta E_{RPA}$ and $\delta E_{DC}$ are the
contributions of the tensor terms to the GT centroid energy 
calculated, respectively, by using RPA and the analytical formula 
(\ref{shift}). In the case of the numbers reported here (not for
the other results in this paper), the spin-orbit term is dropped
both at HF and RPA level. See also the text for details. 
\label{Table1}}
\vspace{0.1cm}
\begin{tabular}{ccccc}
\hline\hline
   & $m_-(1;{\rm no\ tensor})$ & $m_-(1;{\rm with\ tensor})$ 
   & $\delta E_{RPA}$ & $\delta E_{DC}$\\
   & MeV & MeV & MeV & MeV \\
\hline \hline
 $^{90}$Zr & 271.45 & 338.68 & 2.241 & 2.276 \\
 $^{208}$Pb & 1854.12 & 2000.76 & 1.111 & 1.118\\
\hline
\end{tabular} \thinspace
\end{table}

\begin{figure}[tbp]
\includegraphics[scale=0.85,clip]{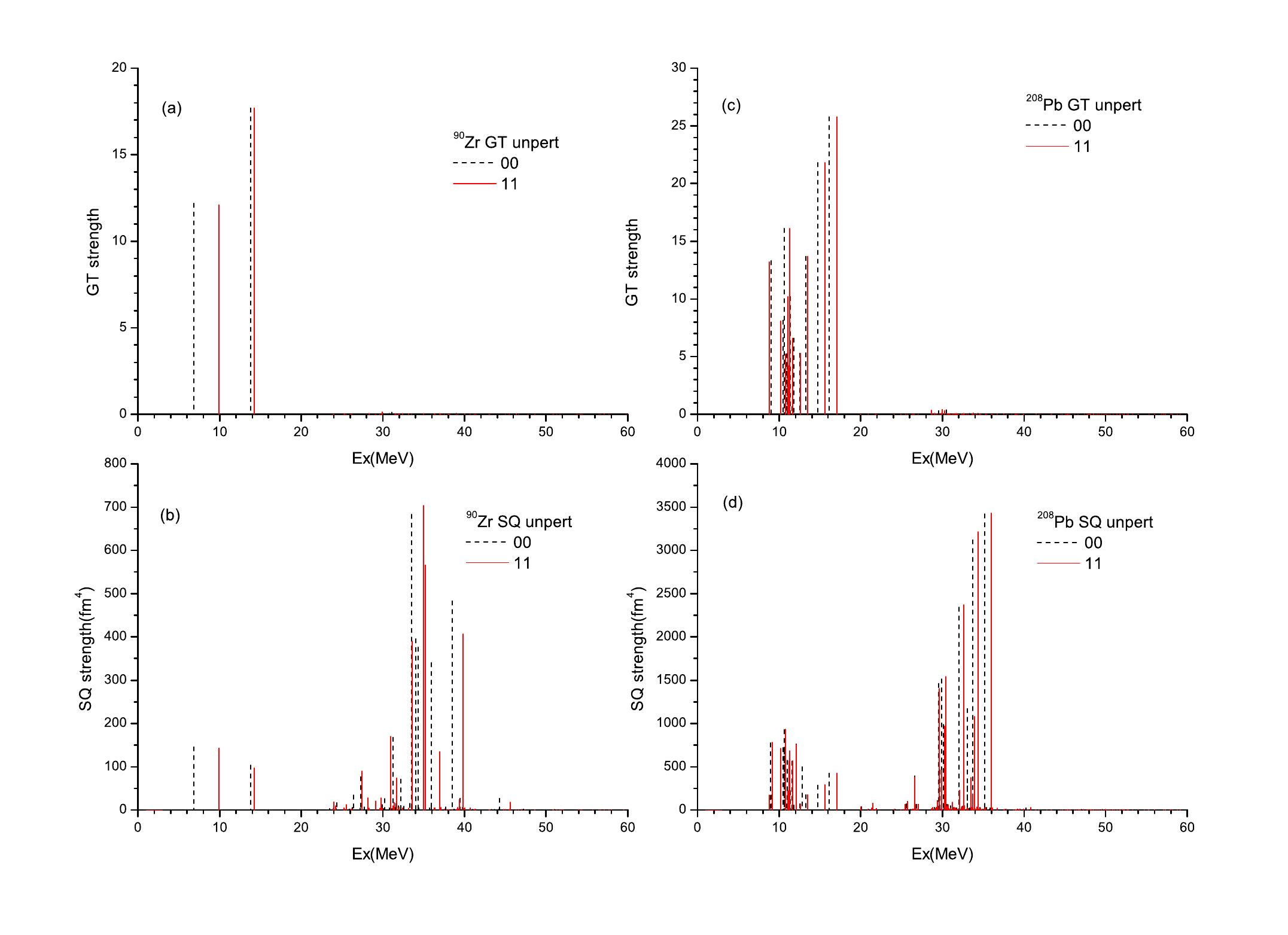} 
\caption{The unperturbed GT$_-$ and SQ$_-$ (spin-quadrupole) 
strength in $^{90}$Zr and $^{208}$Pb.
The results labeled by 00 correspond to neglecting the 
tensor terms both in HF and RPA, while 11 
corresponds to including the tensor terms both in HF and RPA. 
Figure taken from Ref. \cite{Bai1}.
}
\label{GT-fig1}
\end{figure}
\begin{figure}[tbp]
\includegraphics[width=18.cm,clip,bb=0. 0. 630. 450.]{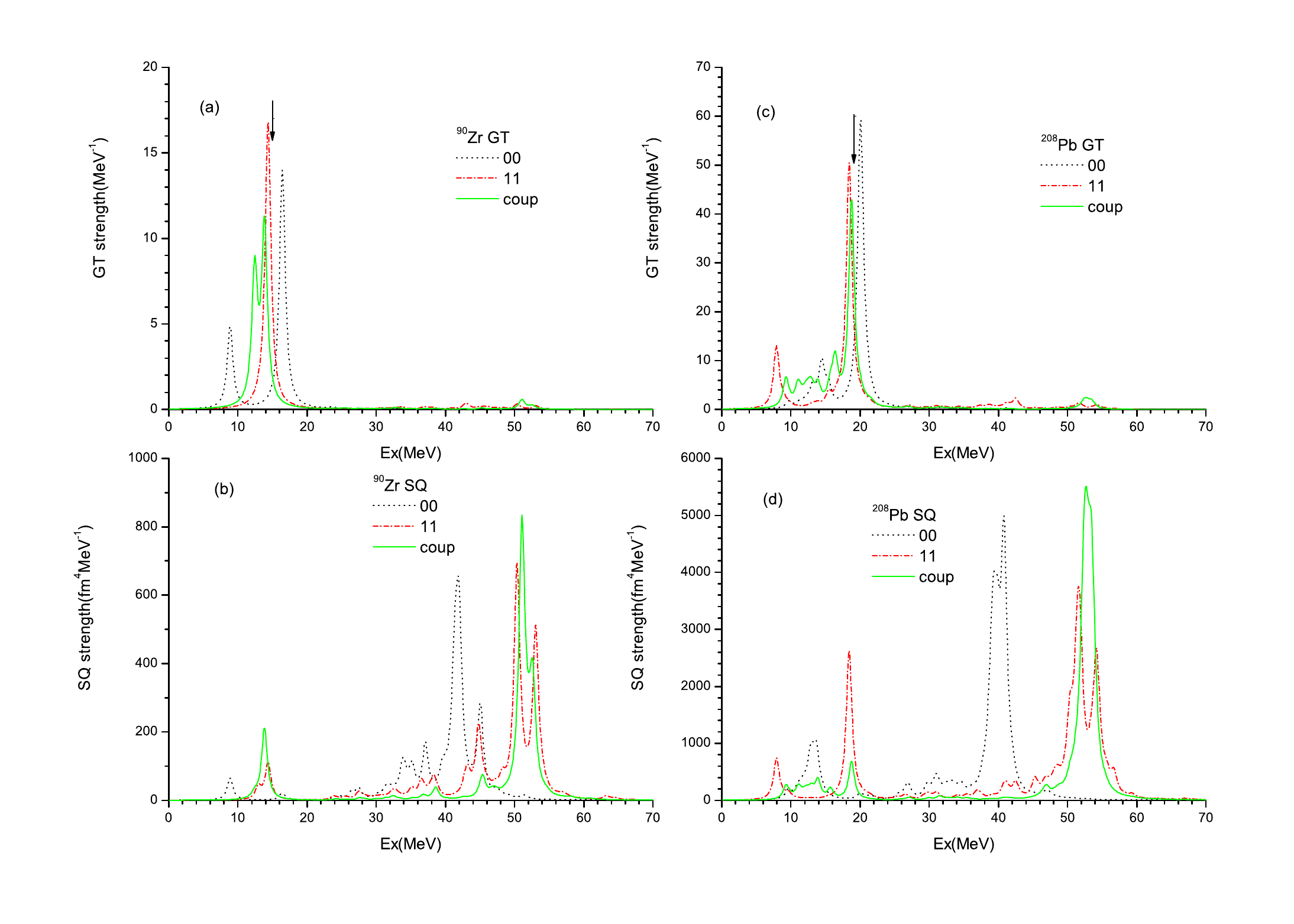} 
\caption{RPA results for the GT$_-$ and SQ$_-$ strength in $^{90}$Zr and $^{208}$Pb.
The results labeled by 00 correspond to neglecting the 
tensor terms both in HF and RPA, while 11 
corresponds to including the tensor terms both in HF and RPA without 
the coupling between GT and SQ states.
The solid line includes full tensor correlations, with 
the coupling between GT and SQ states.  
The RPA results are displayed by smoothing them 
with Lorentzian
weighting function
having 1 MeV width. Figure taken from Ref. \cite{Bai1}.}
\label{GT-fig2}
\end{figure}

The GT$_-$ strength distributions in $^{90}$Zr and $^{208}$Pb are shown in
Figs.~\ref{GT-fig1} and~\ref{GT-fig2}, respectively. 
The 
tensor force affects these results in two ways. 
Firstly, it changes 
the single-particle energies in the HF
calculation; 
secondly, it contributes to the RPA residual force. We display
the results of 
three different kinds of calculations to analyze separately these 
effects. In the first one, the tensor terms are not included at all. 
In the second one, the tensor terms are included 
in HF but dropped in RPA. This calculation is not self-consistent,
but it demonstrates the effects of the changes 
in the single-particle energies 
on the strength distribution. In the last one, the tensor terms are
included both in HF and RPA calculations. For simplicity, results
of the three categories of calculations are labeled by 00, 10 and
11, respectively.

\begin{table}[hbt] \centering
\caption{Values of the EWSRs $m_-(1)$ for $^{90}$Zr 
and $^{208}$Pb in different excitation energy regions. 
 The results labeled by 00 correspond to neglecting the 
tensor terms both in HF and RPA; 10 corresponds to including the 
tensor terms in HF but neglecting them in RPA;  11 
corresponds to including the tensor terms both in HF and RPA.
See the text for a discussion of the effects of the tensor terms. 
\label{Table2}}
\vspace{0.1cm}
\begin{tabular}{c|ccccc}
\hline\hline
& type of RPA     & $m_-(1)$ & $m_-(1)$  & $m_-(1)$ & $m_+(1)$ \\
& calculation & 0-30 MeV & 30-60 MeV & total    & total \\
\hline \hline
              & 00 & 395 & 26.2 & 421.8 & 10.1\\
$^{90}$Zr     & 10 & 444 & 22 & 466 & 11.1\\
              & 11 & 366.9 & 122 & 493.2 & 10.3\\
\hline
              & 00 & 2080 & 124.5 & 2212.8 & 18.8\\
$^{208}$Pb    & 10 & 2176 & 93 & 2269 & 21\\
              & 11 & 1658 & 694 & 2370 & 19.3\\
\hline
\end{tabular}\thinspace
\end{table}

The exhaustion of the EWSR $m_-(1)$ in the different excitation energy
regions is shown in Table~\ref{Table2}.
The EWSR in the energy region below 
30 MeV (where the 1p-1h transitions are located) decreases 
after the inclusion of the tensor term.
From Table~\ref{Table2}, we also see that an appreciable amount 
of EWSR is shifted
from the lower energy region (0-30 MeV) to the higher energy
region (30-60 MeV) by including tensor terms in HF plus RPA 
calculations.
We can also consider the values of the NEWSR in the 0-30 MeV and 30-60 MeV
energy regions for $^{90}$Zr and $^{208}$Pb. 
When the tensor is not included in the residual interaction 
(i.e., the calculations labeled by 00 and
10), the values of the NEWSR in the energy region between 30-60 MeV for
both $^{90}$Zr and $^{208}$Pb are
 small, only few \% of EWSR
  (see the
Figs.  \ref{GT-fig1}(a) and \ref{GT-fig2}(a)). But in the case 11, 
about $10\%$ of the NEWSR is
shifted from the lower energy region to the higher energy region
(corresponding to 25\% and 29\%  of the EWSR in  $^{90}$Zr and $^{208}$Pb, 
respectively). 
Moreover, we can see that
essentially no unperturbed strength appears in this region (see the
Figs. \ref{GT-fig1}(a) and \ref{GT-fig2}(a)). This means 
that including tensor terms in
  the RPA calculations shifts about $10\%$ of the GT strength to
the energy region 30-60 MeV. While 
2p-2h couplings will increase further these high energy strength, 
we can stress 
  that the tensor correlations move substantial GT strength from the low 
energy region  0-30 MeV to the high energy region 30-60 MeV
 even within 
the 1p-1h model space.  
 The coupling between GT and SQ states is responsible for this quenching.
 Without the tensor interaction, the coupling is almost nothing, but it becomes substantial when the tensor interaction is switched on.
 This strong coupling can be seen clearly by a decomposition of the tensor interaction into the spin-multipole couplings as will be given in Eq. (\ref{eq:V-te}).  

In $^{90}$Zr, from Fig.  \ref{GT-fig1}(a) one can notice that the GT strength is
concentrated in two peaks in the region below 30 MeV. There
there are only two important configurations involved which are
$(\pi1{\rm g}_{9/2}-\nu1{\rm g}^{-1}_{9/2})$ and
$(\pi1{\rm g}_{7/2}-\nu1{\rm g}^{-1}_{9/2})$ (see Fig.  \ref{GT-fig1}(b)). When  the tensor term
is included only in HF and neglected in RPA,  the
centroid in the energy region 0-30 MeV is moved
upwards by about 1.5 MeV, and the higher energy peak at 
 E$_x \approx$ 16 MeV is  moved upwards 
by only 
0.5 MeV, as compared with the results without tensor term. When  the
tensor term is included both in HF and RPA,  the
centroid of the GT strength in the energy region 0-30 MeV is moved
downwards by about 1 MeV, and the higher energy peak is also moved downwards
by about 2 MeV, as compared with the results obtained without tensor
term. Including tensor terms in RPA makes the two main separated 
peaks closer (this situation
also happens for $^{48}$Ca).
 This can be understood as a typical effect of the tensor correlations 
on the single-particle states (cf. Sec. \ref{spstates}).  
When the $\nu1{\rm g}_{9/2}$ orbit is filled by neutrons, the tensor 
 correlations provide some quenching of the spin$-$orbit splitting between 
 $\pi1{\rm g}_{9/2}$ and  $\pi1{\rm g}_{7/2}$ orbits 
  so that the unperturbed energies of the two 
main p-h configurations ~$(\pi1{\rm g}_{7/2}-\nu1{\rm g}^{-1}_{9/2})$ and
 $(\pi1{\rm g}_{9/2}-\nu1{\rm g}^{-1}_{9/2})$ are closer in energy. 
The RPA results in Fig. \ref{GT-fig1}(a) labelled by (10) and (11) reflect 
these changes of the HF single particle energies due the tensor correlations 
and the energy difference between two peaks is narrower.


In $^{208}$Pb, from Fig. \ref{GT-fig2}(a) we see that  the GT strength 
  is concentrated in two peaks in the low energy
region 0-30 MeV for all cases 00, 10 and
11. There are eleven important configurations
which contribute to these peaks. When  the tensor terms are 
only included in HF
and neglected in RPA, the centroid of these peaks is moved upwards
by about 0.5 MeV, and the higher energy peak 
at E$_x \approx$ 18 MeV
 is also raised by about 0.8 MeV.
When  the tensor terms are included in both HF and RPA, the
centroid of the peaks in the energy region 0-30 MeV 
 moves downward by about 1.5 MeV, and the
higher energy peak  moves also downwards by about 3.3 MeV,
compared  with
the result obtained without tensor terms. By
including tensor terms in the RPA calculation, the GT strengths in the
energy region 30-60 MeV are increased  substantially 
by the shift of
the strength  in the energy region of 0-30 MeV
 through the tensor force.

The tensor interaction is spin-dependent,
so we expect that it can have important
effects not only on the GT transitions, but also on spin-dipole 
and other spin-dependent  excitation modes as well. These issues will
be discussed in the next subsection.

\subsubsection{Spin-Dipole states}

The study of the charge-exchange spin-dipole (SD)
excitations of $^{208}$Pb (inspired by recent accurate measurements
\cite{Wakasa}) 
and of $^{90}$Zr will be shown to elucidate in a quite specific way
the effect of tensor correlations.
To get an unambiguous signature of the effect of the tensor
force, which is strongly spin-dependent, 
one can expect that the separation of the
strength distributions of the $\lambda^\pi=0^-$, $1^-$ and $2^-$ components
is of great relevance.

We will discuss here calculations that employ 
two different Skyrme parameter sets, 
namely the set T43 of
Ref.~\cite{Les07} 
the set SLy5+T$_w$ that is a set in which the tensor 
terms are added on top of the existing force SLy5, 
in a perturbative way \cite{Bai11}. 
One should notice that the tensor part of SLy5+T$_w$ 
is different from that of SLy5+T
which was introduced in Ref.~\cite{Colo} and discussed in the 
previous Sections. In Ref. \cite{Bai_last}, results obtained 
with the other forces T11, T22, T33, and T44 were also studied.  
The values of $T$, $U$, $\alpha$ and $\beta$ for the
adopted interactions are listed in Table \ref{Para}.
\begin{table}[hbt] \centering
\caption{The tensor strength parameters $T$ and $U$ of 
Eq. (\ref{eq:tensor}) as
well as
the $\alpha$ and $\beta$ values of Eq. (\ref{eq:dW}). All values are
in MeV$\cdot$fm$^5$.
\label{Para}}
\vspace{0.2cm}
\begin{tabular}{ccccccc}
\hline\hline Force & T & U & $\alpha_T$ & $\beta_T$ & $\alpha_C$ &
$\beta_C$ \\
\hline \hline
SLy5+T$_w$ & 820.0 & 323.4 & 134.76 & 238.2 & 80.2 & -48.9\\
T43 & 590.6 & -147.5 & -61.5 & 92.3 & 121.5 & 27.7\\
\hline\hline
\end{tabular}
\end{table}

\begin{table}[ht] \centering
\caption{The SD sum rules $m_-$(0) and $m_-$(1) for $^{208}$Pb
with and without the tensor terms.
$\Delta E$ is the difference
between the values of $m_-$(1)/$m_-$(0) calculated with and without tensor.
\label{Table1-1}}
\begin{tabular}{c|c|ccc|ccc|r}
\hline\hline
   &  &  \multicolumn{3}{c|} {without tensor} & \multicolumn{3}{c|} {with tensor} & \\ \hline
  force & $\lambda^{\pi}$ &   $m_-$(0)  &  $m_-$(1) & $m_-$(1)
/$m_-$(0) &  $m_-$(0)  &  $m_-$(1) & $m_-$(1)/$m_-$(0)
& $\Delta E$ \\\hline \hline
     &  $0^-$   & 158.6 & 4718 & 29.7  & 171.9   & 5138   & 29.9 & 0.2\\
  SLy5  &  $1^-$ & 432.0 & 11746 & 27.2 & 440.0 & 10111 & 23.0 & $-$4.2\\
  +T$_w$   &      $2^-$   & 646.0 & 13742 & 21.3 & 657.4 & 14008 & 21.3 & 0\\ \cline{2-9}
     &  sum &   1236.6 & 30206 &  24.4 & 1269.3 & 29256 &23.0  & $-$1.4 \\
\hline
     &  $0^-$   & 154.8 & 4693 & 30.3 & 164.0   & 6170   & 37.5 & 7.2 \\
   T43   &  $1^-$    &  440.3 & 12138 & 27.6 & 444.1 & 10366 & 23.3 & $-$4.3\\
      &     $2^-$    & 645.5 & 14067 & 21.8 & 649.4 & 14675 & 22.6 & 0.8 \\\cline{2-9}
    &  sum &  1240.6  & 30898  & 24.9  &  1257.5 & 31211  & 24.8 & $-$0.1\\
  \hline\hline
\end{tabular}\thinspace
\end{table}

The charge-exchange SD operator is defined as
\begin{eqnarray}
O_{\pm
}^{\lambda }=\sum\limits_{i}t_{\pm
}^{i}r_{i} \left[ Y_{1}(\hat r_i) 
\otimes \sigma^{i} \right]_{\lambda}. \label{multipole}
\end{eqnarray}
The $n-$th energy weighted sum rules $m_n$ for the
$\lambda$-pole SD operator are defined as
\begin{eqnarray}
m_n^{\lambda}(t_{\pm})=\sum\limits_{i}E_i^n|\langle i|
O_{\pm}^\lambda|0 \rangle|^2,
\end{eqnarray}
and 
the model-independent sum rule which is known to hold is
\be
m_0^{\lambda}(t_{-})
-m_0^{\lambda}(t_{+})=
\frac{2\lambda+1}{4\pi} (N \langle r^2 \rangle_n-Z \langle r^2
\rangle_p) , 
\ee 
where $\langle r^2 \rangle_n ( \langle r^2
\rangle_p)$ is the mean square radius of neutrons (protons).  

As above, 
we will discuss two kinds of calculations.
In the first one, the tensor
terms are neither included in HF nor in RPA.
In the second one, the tensor terms are included both in HF and
in RPA.
In Section 3.6.1, it has been found that the effect of tensor correlations in
HF is large for the Gamow-Teller mode. 
This is largely due to the fact that the unperturbed GT transitions are
exactly those among spin-orbit partners. 
On the other hand, this is not the case for the
SD transitions:
the average unperturbed energies are not much  affected by the spin-orbit
splittings since they are 1$\hbar \omega$-type excitations.
\begin{figure}[tbp]
\includegraphics[width=18.cm,scale=1.,clip]{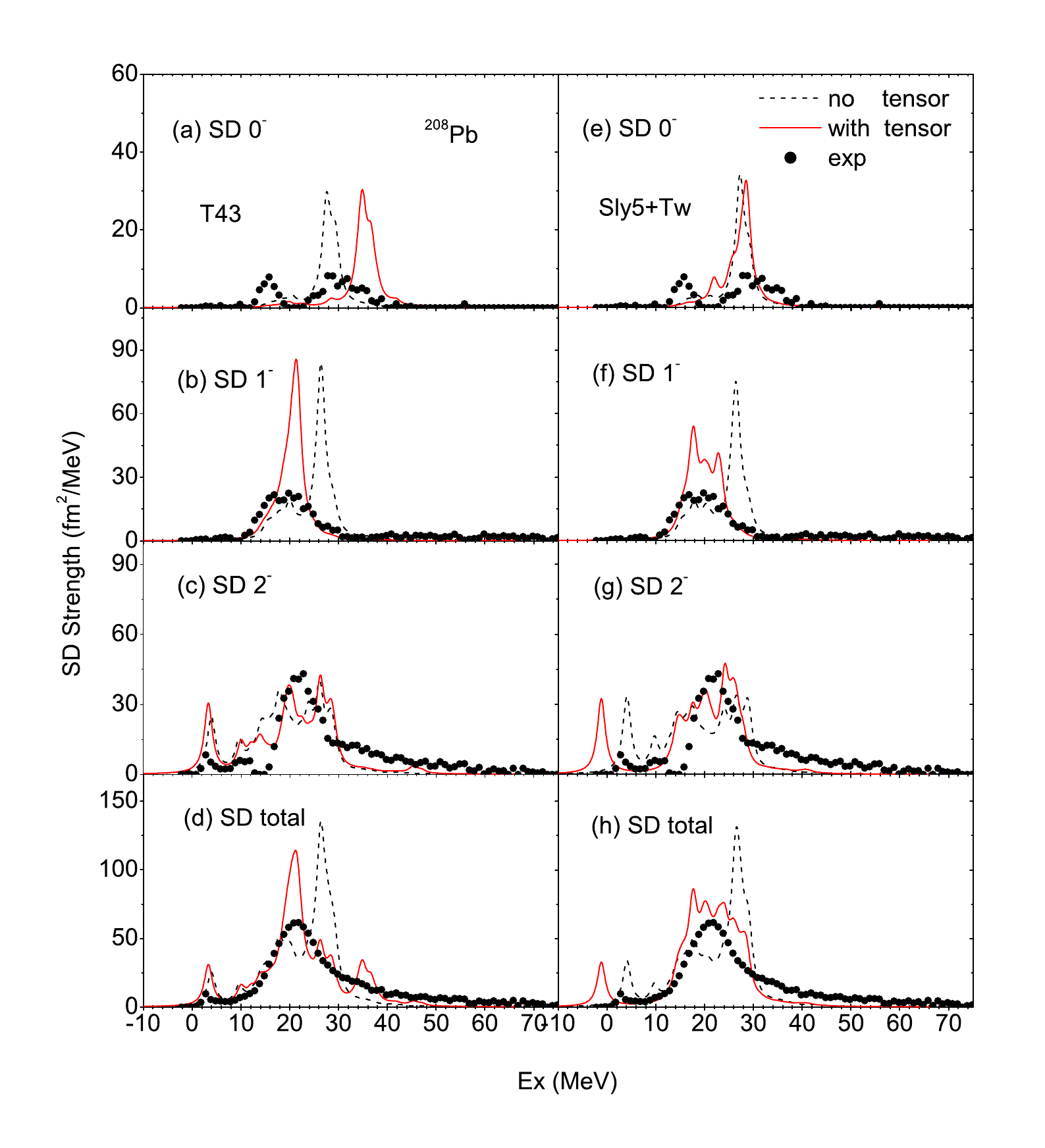}
\caption{Charge-exchange SD$_-$ strength
distributions in $^{208}$Pb. In the panels (a), (b), and (c) the
RPA results obtained by employing the interaction
SLy5+T$_w$ for the multipoles 0$^-$, 1$^-$,  2$^-$ are displayed.
In panel (d) we show the total strength
distribution. Panels (e), (f), (g) and (h) correspond
to similar results when the parameter set T43 is employed.
All these discrete RPA results have been smoothed by using a Lorentzian
averaging with a width of 2 MeV and compared with experimental
findings. The excitation energy is with respect to the ground state
of $^{208}$Bi. The experimental data are taken from Ref.~\cite{Wakasa}.
See the text for details and discussion. Figure taken from Ref. \cite{Bai11}.
}
\label{fig1-1}
\end{figure}
The numerical
results of the HF+RPA calculations with the forces T43 and SLy5+T$_w$
are shown in Fig. \ref{fig1-1}.
They are compared with experimental
data obtained by 
multipole decomposition analysis of the (p,n) reaction data
and extraction of the strength functions by means of
Distorted Wave Impulse Approximation (DWIA) calculations 
\cite{Wakasa}. From Fig. \ref{fig1-1}(a) and (b) one can see that
in the case of the T43 interaction the main peaks of the 0$^-$ and
1$^-$ strength distributions are shifted upwards by about 7.5 MeV
and downwards by about 5 MeV, respectively,
due to the tensor correlations.
There are several 2$^-$ peaks (cf. Fig.\ref{fig1-1}(c)).
The peak at excitation energy $E_x\approx$ 17.7 MeV is moved upwards by about 2 MeV by including
tensor forces, and comes close to an experimental peak, while another peak at
$E_x\approx$ 3.9 MeV is shifted downwards by about 0.6 MeV and is also
eventually close to the observed low energy peak.
For the total SD strength in Fig. \ref{fig1-1}(d), it is remarkable that
the main peak at 26 MeV is shifted to 21 MeV
when tensor is included, and this provides good agreement with the
experimental data.

In the same figure, the SD strength distributions in $^{208}$Pb
calculated by using the set SLy5+T$_w$ are also shown.
From Fig. \ref{fig1-1}(e), we see that the calculated 0$^-$ strength
is concentrated in one peak which is shifted upwards by about 1.3 MeV
by the tensor correlations. In Fig. \ref{fig1-1}(f), the RPA tensor
correlations move the 1$^-$ peak downwards and split it into three
peaks, in qualitative agreement with the bump-like experimental strength.
In the case of the 2$^-$ component [Fig. \ref{fig1-1}(g)], the main peaks
in the high energy region are rather near to the experimental main
peak. Therefore, as shown in Fig. \ref{fig1-1}(h), the inclusion of
the tensor terms in HF+RPA can make the calculated main peak of
the total SD strength coincide with the main measured peak.
However, in the low energy region the agreement is not good compared with
the experimental result in the case of SLy5+T$_w$.  
The sum rule analysis of the SD excitations can be seen 
in Table \ref{Table1-1}.  The general trends of the excitation
energies of the SD peaks are also reflected in the average 
energies $m_-$(1)/$m_-$(0) as one can deduce from 
the values of $\Delta E$ in the last column of the Table. 
The absolute values $m_-$(0)
and $m_-$(1) are also affected by the tensor interaction, but at 
most at the one or few \% level.    


We would like at this stage to obtain a better understanding of this
peculiar role of tensor interactions.
The diagonal matrix element of their triplet-even (TE) term on a state
with multipolarity $\lambda$ can be expressed as \cite{Gang10}
\begin{eqnarray}
V^{(\lambda)}_{TE} & = &
\frac{5T}{4}
\sum_{\ell,k,k'}
\frac{(-)^{k+k'+\lambda+\ell+1}\hat{k} \hat{k'}}{2\lambda +1}
\left\{ \begin{array}{ccc} k & k' & 2 \\
1 & 1 & \ell \end{array} \right\} \nonumber  \\
& \times &
\left\{ \begin{array}{ccc} 1 & 1 & 2  \\
k' & k & \lambda \end{array} \right\}
\langle p  \vert\vert \hat O_{k',\lambda} \vert\vert h \rangle
\langle p  \vert\vert \hat O_{k,\lambda}  \vert\vert h  \rangle ^*,
\label{eq:V-te}
\end{eqnarray}
in terms of the reduced matrix elements of the operator
$\hat O_{k,\lambda}= \sum_i [ \sigma_i \otimes (
{\bf \nabla}_i \otimes Y_\ell(i) )^{(k)}
]^{(\lambda)}$ and $6j$ symbols. In Eq. (\ref{eq:V-te}), the notation
$\hat k\equiv\sqrt{2k+1}$ is used.
For GT case with $\lambda=1$, there are the $ph$ matrix elements not only with $k$=$k'$=0, but also with $k$=0 (2) and $k'$=2(0).  The latter makes possible the 
coupling between GT and SQ states which are not connected directly by other terms of the Skyrme interactions.
For the SD excitations, taking $\ell=0$ and $k$=$k'$=1,
Eq. (\ref{eq:V-te})  gives
\begin{equation}
V^{(\lambda)}_{TE} = -
\frac{5}{12}T \left\{ \begin{array}{c}
 1 \\
-1/6 \\
  1/50 \end{array} \right\}
|\langle p  \vert\vert \hat O_{1,\lambda} \vert\vert h \rangle|^2 
\,\,\, {\rm for} \,\,\,  \lambda=\left\{ \begin{array}{c}
                                                    0^- \\
                                                    1^- \\
                                                    2^- \end{array} \right\}.
\label{eq:V-TE}
\end{equation}
The TO tensor part is also expressed in a similar way as
\begin{equation}
V^{(\lambda)}_{TO} =
\frac{5}{12}U \left\{ \begin{array}{c}
1 \\
-1/6 \\
1/50   \end{array} \right\}
|\langle p  \vert \vert \hat O_{1,\lambda} \vert \vert  h \rangle|^2 
\,\,\, {\rm for} \,\,\,  \lambda=\left\{ \begin{array}{c}
                                                    0^- \\
                                                    1^- \\
                                                    2^- \end{array} \right\}.
\label{eq:V-TO}
\end{equation}
We can see in Eqs. (\ref{eq:V-TE}) and  (\ref{eq:V-TO})
 that the diagonal
  p-h matrix element in the 0$^-$ case is the largest, and
that for 1$^-$ is the next. The effect on $2^-$ is rather small.
It should be noticed  that
these relative strengths of the Skyrme tensor interactions on each multipole are similar to
those obtained from the finite-range tensor interactions both
 in magnitude and in sign \cite{SB84}. We can sum the TE and TO direct matrix elements as
\begin{equation}
V^{(\lambda)}_{T}=V^{(\lambda)}_{TE}+V^{(\lambda)}_{TO}\equiv a_{\lambda}T+b_{\lambda}U.
\end{equation}
The proper antisymmetrization is easy to obtain for contact interactions
and gives, in the isovector channel,
\begin{equation}
V^{(\lambda)}_{T,AS}=[-\frac{1}{2}a_{\lambda}T+\frac{1}{2}b_{\lambda}U ]
\langle\vec\tau_1\cdot\vec\tau_2 \rangle.
\label{eq:tensor-a}
\end{equation}
Since the coupling constant $T$
is
positive for the interactions we considered, $V^{(\lambda)}_{TE}$
is repulsive for the $0^-$ and $2^-$ case, while it is attractive for $1^-$ .
The $V^{(\lambda)}_{TO}$ part may contribute with
  the same sign as the $V^{(\lambda)}_{TE}$ one if the value of $U$ is negative.
For the T$IJ$ family, the value of $U$ is negative or small
positive,
so that the
$V^{(\lambda)}_{TO}$ contributions have the same multipole dependence
  or almost negligible.
All together, the tensor correlations are strongly repulsive
  for 0$^-$ and
weakly repulsive for 2$^-$  in general.
For
$1^-$, the net effect will be  attractive.
For SLy5+T$_w$, the value of $U$ is positive and will give opposite
contributions to those of $T$.  However, the $T$ value is much larger
than the value of $U$ so that the same argument given for the T$IJ$ family
will hold.
One can see from Table \ref{Table1-1} that Eq. (\ref{eq:tensor-a})  provides
a very effective guideline for interpreting the numerical results of
microscopic RPA.



In conclusion, 
in the case of the charge-exchange
$t_-$ SD excitations of $^{208}$Pb 
it has been clearly demonstrated 
that tensor correlations
have a specific multipole dependence,
that is, they produce a strong hardening effect on the 0$^-$mode and a
softening effect on the 1$^-$ mode. A weak hardening effect is also
observed on the 2$^-$ mode.
These characteristic effects of the tensor force 
can be understood
by using analytic formulas based on the multipole expansion of
the contact tensor interaction, and 
are confirmed
by the favourable comparison with the experimental data. 

In Ref. \cite{Bai_last}, the constraints set on the tensor
force by both GT and SD excitations are summarized. It is
found that $T$ and $U$ can be restricted in a range that has
a reasonable overlap with the values discussed in Sec. 
\ref{spstates}, namely $\beta$ positive and $\alpha$ negative
(cf. Fig. 4 of Ref. \cite{Bai_last} and corresponding discussion). 

\subsection{$\beta$-decay}

Quite recently, the impact of the tensor correlations on
$\beta$-decay has been discussed in Ref. \cite{Minato}.
In keeping with our previous discussion about Gamow-Teller
transitions, a significant effect can be expected. However, 
no study has been performed earlier despite the difficulty
that mean-field calculations have in reproducing the
experimental data for $\beta$-decay. Using Skyrme interactions,
the work of Ref. \cite{Engel} was the first to demonstrate
that a fully self-consistent Quasiparticle RPA (QRPA) 
calculation of $\beta$-decay tends to overestimate the
experimental half-lives. The proton-neutron ($T=0$) pairing
residual force can produce an additional attractive effect
and bring the half-lives in agreement with data. In fact,
the proton-neutron $T=0$ pairing cannot be constrained
by ground-state data and escapes, strictly speaking, the
philosophy of self-consistent mean-field: its strength
is a free parameter. It is of obvious interest to assess if the
agreement with $\beta$-decay data can also be improved
by including tensor correlations, whose parameters can
be checked by using all observables that we have discussed
so far in the present review paper. Moreover, one should
add that the effect of proton-neutron pairing is expected
to weaken going towards drip-line nuclei, and among 
the nuclei that undergo $\beta$-decay 
there are magic nuclei in which pairing is negligible
anyway.

The authors of Ref. \cite{Minato} have studied the
effect of tensor on $\beta$-decay based on forces
with reasonable values of the Landau parameters, that is,
of the spin and spin-isospin residual forces, in order
to avoid the shortcomings that we have described in
connection with the M1 resonance in Sec. \ref{vibr}.
However, they have used tensor forces that are added
perturbatively on top, respectively, of the Skyrme sets
SkO \cite{SkO} and SkX \cite{SkX}. 

\begin{figure}[ht]
\centering
\includegraphics[width=6cm,clip]{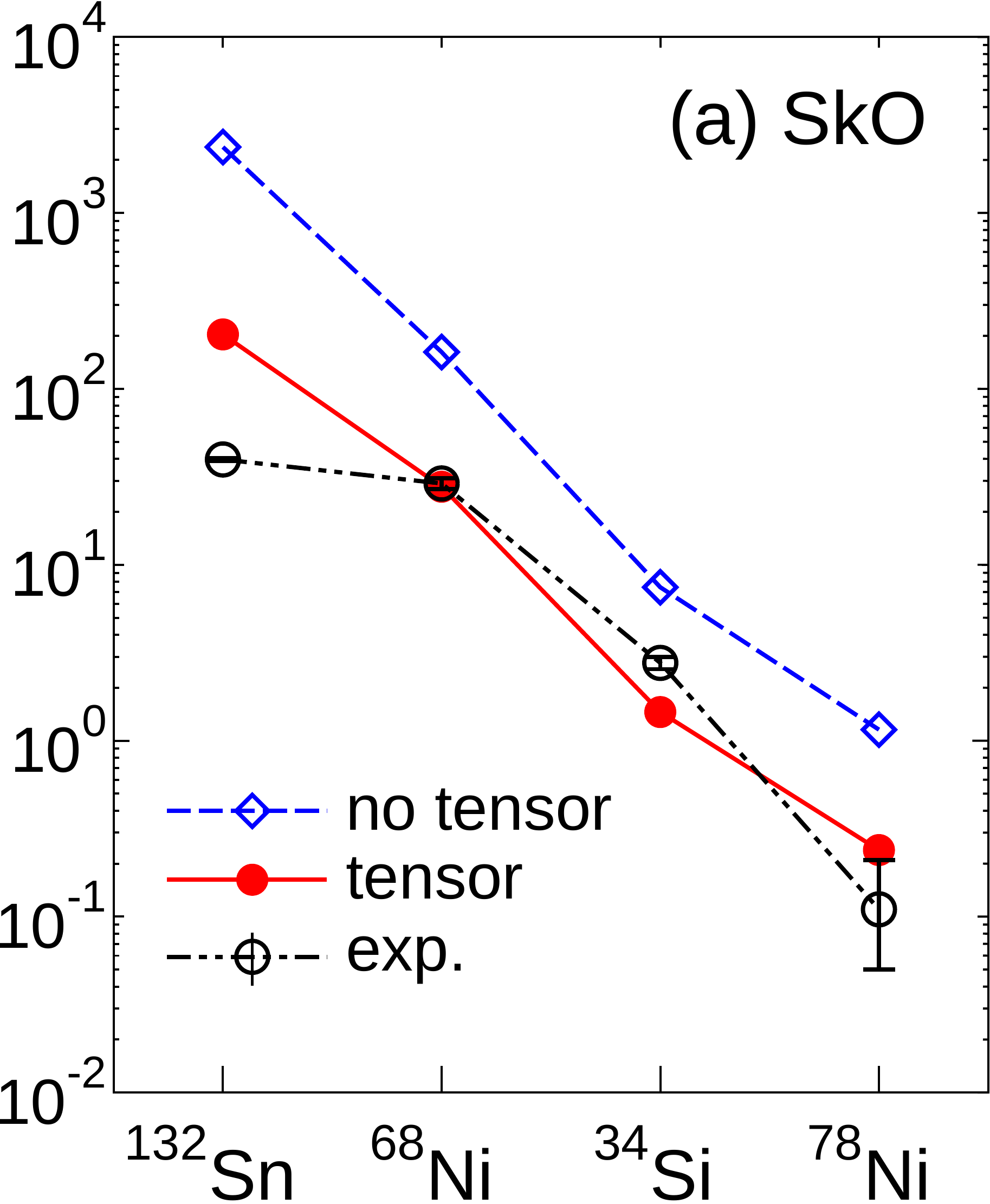}
\includegraphics[width=6cm,clip]{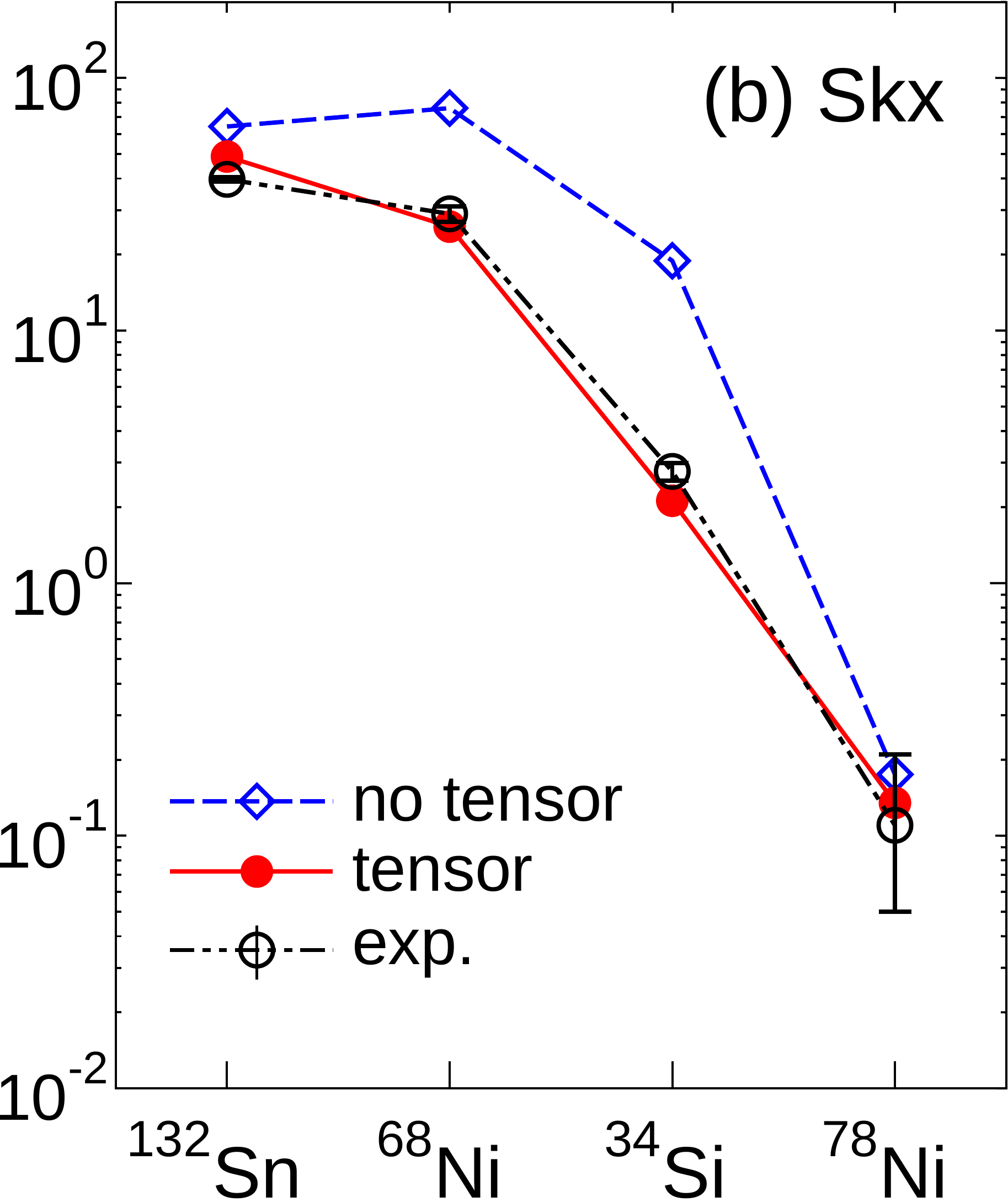}
\caption{\label{fig:half_life}
Half-lives of several nuclei obtained in QRPA with the 
SkO and SkX interactions, with and without tensor added. 
Figure taken from \cite{Minato}.}
\end{figure}
\begin{figure}[ht]
\centering
\includegraphics[width=6cm,clip]{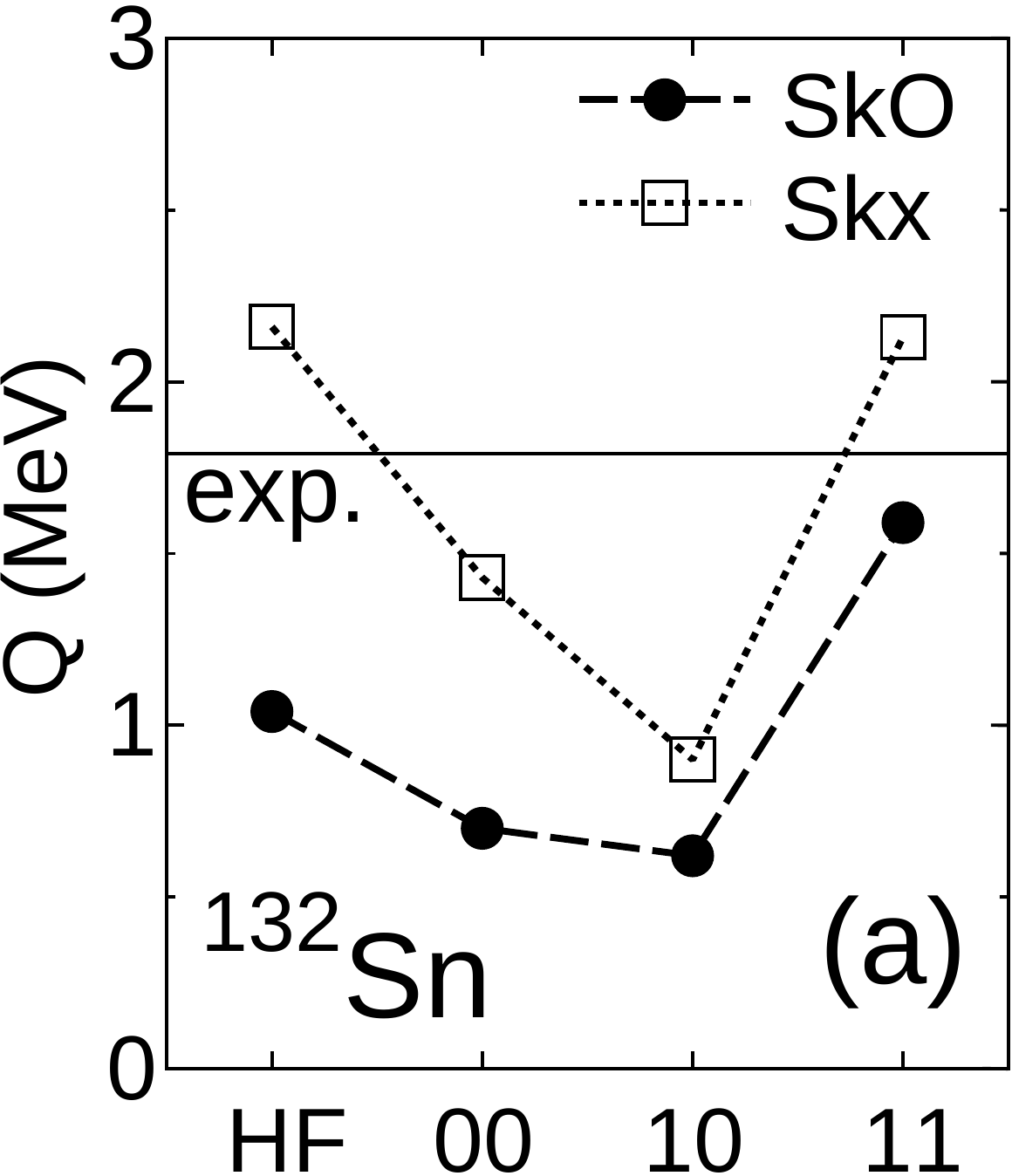}
\includegraphics[width=6cm,clip]{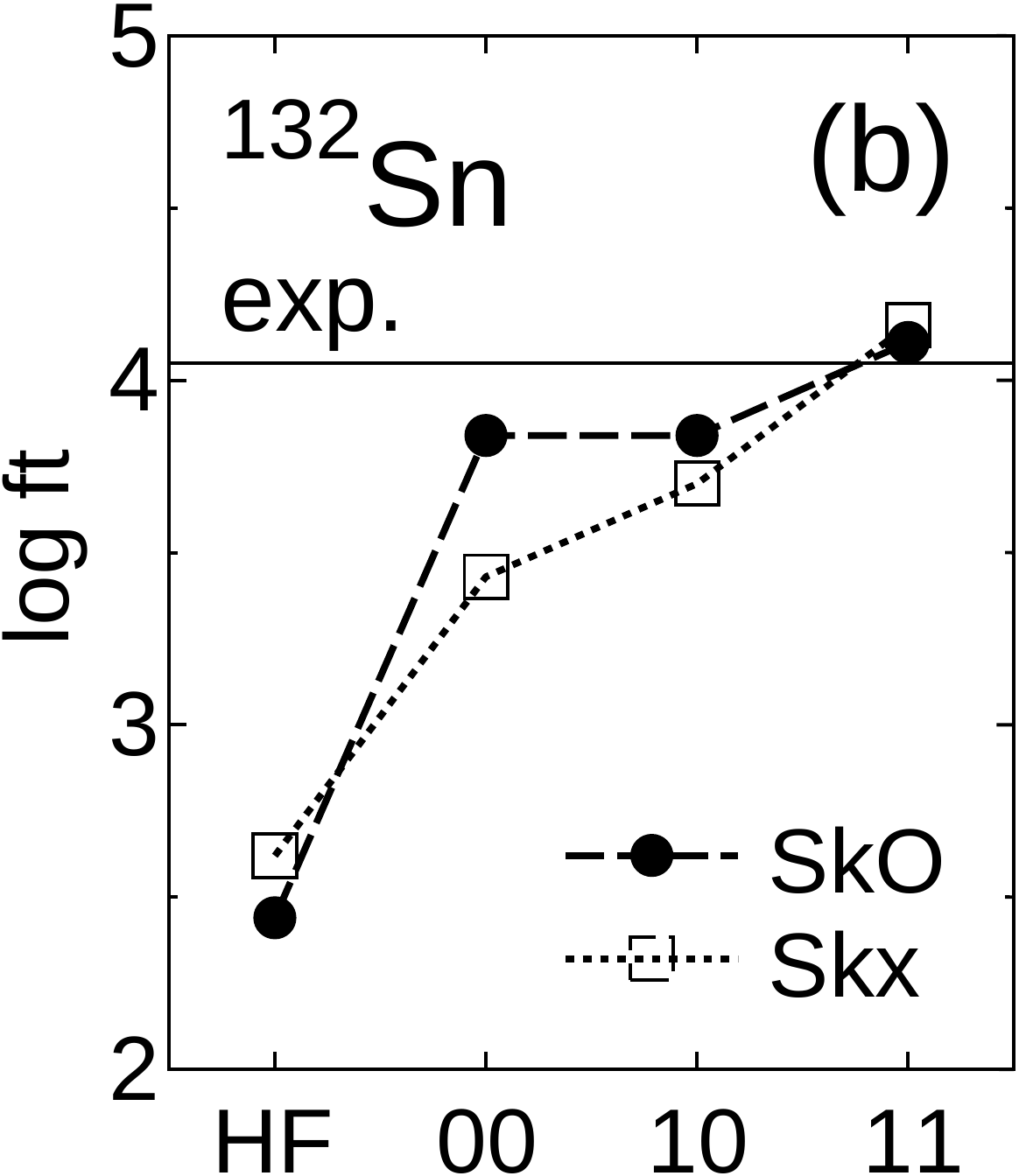}
\caption{\label{fig:Q-value}
Q-values and log($ft$) values of $^{132}$Sn obtained within
the same framework as in the previous figure. Figure taken
from \cite{Minato}.}
\end{figure}

Some of the results obtained in Ref. \cite{Minato} are 
displayed in Figs. \ref{fig:half_life} and \ref{fig:Q-value}. 
The strong effect of the tensor correlations on the half-lives
is visible and quite remarkable in case of both Skyrme sets
considered. In particular, in the case of SkX plus tensor, 
the result with tensor included can reproduce the experimental
values of the half-life in all the nuclei shown in Fig. 
\ref{fig:half_life}. 
These nuclei are not amenable to a
description including pairing, and their half-lives span
about three orders of magnitude. 
As is well known, both matrix elements and $Q$-values play
a role in such kind of calculations. 
Fig. \ref{fig:Q-value} demonstrates that both for the $Q$-value
and log($ft$) values it is crucial to take into account tensor
correlations both in the HF mean-field and residual interaction.

\section{Instabilities}
\label{instabilities}

Recently, there has been much interest in detecting possible 
instabilities associated with the widely used
energy density functionals. Instabilities could manifest themselves 
in several possible ways but in all cases the system
must be subject to an external perturbing field if one
wishes to detect those instabilities.

If the perturbation brings zero momentum (${\vec q}=0$), 
or infinite wavelength, we are in the so-called Landau limit.
This case has been studied within the Landau's theory of Fermi
liquids, that has been extended by Migdal and collaborators 
to the case of finite Fermi systems like atomic nuclei
\cite{Migdal}. In the Landau-Migdal's theory the key quantity
is the interaction $V$ acting among quasiparticles around the Fermi
surface, whose matrix elements are written in the momentum space
as
\begin{equation}
\langle {\vec k_1}{\vec k_2 }|V|{\vec k_1}{\vec k_2} \rangle =
N_0^{-1} ( F({\vec k_1}{\vec k_2 })
+F'({\vec k_1} {\vec k_2 })\tau_1\tau_2+G({\vec k_1} {\vec k_2 })\sigma_1\sigma_2
+G'({\vec k_1} {\vec k_2 })\tau_1\tau_2\sigma_1\sigma_2 ),
\label{eq:LM1}
\end{equation}
where $N_0=2k_Fm^*/\hbar^2\pi^2$ is the density of states per energy
at the Fermi surface and $m^*$ is the effective mass. The 
single-particle momenta $\vec{k_1}$ and $\vec{k_2 }$ are, in fact, 
taken exactly at the Fermi surface: then, in homogeneous matter, the 
so-called Landau parameters $F,F',G$ and $G'$ are only functions 
of the angle $\theta$ between $\vec{k_1}$ and $\vec{k_2 }$ and can be consequently 
expanded in Legendre polynomials:
\begin{equation}
F=\sum_lF_lP_l(cos\theta),
\label{eq:LM2}
\end{equation}
and likewise for $F',G$ and  $G'$. In order for a spherical Fermi surface
to be stable against any deformation, the parameters must satisfy
the criterion
\begin{equation}
F_l>-(2l+1),
\label{eq:LM3}
\end{equation}
and analogous criteria for all the other parameters. We remind that the
Landau parameters defined in this way have no dimension: the units are
chosen so that the change $F_l/(2l+1)$ in the potential energy
is accompanied by a change 1 in the kinetic energy. For any
given interaction, like the Skyrme or Gogny interaction, the
Landau parameters can be estimated by calculating matrix elements
on a plane wave basis and by comparing with Eq. (\ref{eq:LM1}). 
As we stress again, all states are considered at the Fermi surface
and the initial and final relative momenta $\vec k$ and $\vec k^\prime$
are equal in Eq. (\ref{eq:LM1}); therefore, the momentum transfer
${\vec q}\equiv {\vec k}-{\vec k^\prime}$ vanishes as we mentioned
at the start of this paragraph. The inequality (\ref{eq:LM3}) ensures
that the system is free from zero-momentum or long-wavelength instabilities.
This equality must be satisfied by the Landau parameters for every
value of $l$. For the Skyrme force, only $l=0,1$ Landau parameters
do not vanish and have to be considered. This is not the case
for the Gogny interaction (see below).

Tensor components can be added to $V$ and Eq. (\ref{eq:LM1}) has
to be modified accordingly. The effects of tensor forces within the
Landau-Migdal framework has been considered for the first time in 
Ref.~\cite{Dab76} (see also \cite{Back79,Olsson04}). Using the 
convention of Ref.~\cite{Dab76}, the tensor terms
\begin{equation}
\Bigl(\frac{{q}_k^2}{k_F^2}H(cos\theta)+
\frac{{q}_k^2}{k_F^2}H'(cos\theta)\vec\tau_1 
\vec\tau_2 \Bigr )S_{12}({{\hat q}_k})
\label{eq:LM4}
\end{equation}
are added to the interaction (\ref{eq:LM1}). Here
${\vec q_k}={\vec k_1}-{\vec k_2}$ with ${\vec k_1}$ and ${\vec k_2}$
again lying at the Fermi surface, and
\begin{equation}
S_{12}({\hat q}_k)=3\vec\sigma_1\cdot {\hat q}_k \vec\sigma_2
\cdot {\hat q}_k-\vec\sigma_1\cdot\vec\sigma_2,
\end{equation}
where ${\hat q}_k$ 
denotes the unit vector in the direction of
${\vec k_1}-{\vec k_2}$. $H$ and $H'$ can be expanded in principle on the
Legendre polynomials in the same way as in Eq. (\ref{eq:LM2}). 
The stability conditions of nuclear matter in the spin and spin-isospin
channels will be affected by the tensor interaction. One must
consider spin and isospin degrees of freedom, and impose
that the Fermi surface is stable under generalized
deformations. The resulting stability conditions
have been studied (see, e.g., Ref. \cite{Back79}). The associated
equations, that generalize Eq. (\ref{eq:LM3}), can be found 
in Refs. \cite{Back79,Gang10} and will not be reported here for
the sake of brevity. However, it is useful to remind that
since the tensor force couples $\vec l$ and $\vec s$ (the spin-orbit
force does not act in uniform matter), the deformations of
the Fermi surface have $J^\pi$ as quantum numbers. For $l=1$, 
one has three independent deformations associated with 
$0^-,1^-$ and $2^-$. For $1^+$ one has two coupled equations
associated with $l=0$ and $l=2$. All these equations that
guarantee stability must be checked separately for the
isoscalar and isovector case. 

In Ref. \cite{Gang10} a very careful study of the stability
of a large set of Skyrme forces plus tensor has been performed.
The main conclusions of this work are: (i) instabilities
occur for all the considered sets at some value of critical
density $\rho_C$; (ii) if applications for finite nuclei are
envisaged (excluding systems like neutron stars), pushing the
value of the critical density above $\approx$ 1.5 or 2 times
the saturation density may be satisfactory enough, and this
can be obtained for some limited values of the parameters
in ($\alpha,\beta$) plane; (iii) a full variational procedure
to determine the Skyrme parameters is preferable to a
perturbative adding of the tensor terms also from the point
of view of avoiding instabilities.

Later, the question has been raised above the finite-$\vec q$
instabilities. This sort of discussion has been initially driven by some sparse findings that are not, at first sight, easy to
be cast in a unifying picture. One of such findings has been that in HF calculations with
some Skyrme forces, even when performed for standard double magic nuclei, after a sufficiently long
number of iterations the system converges to an unphysical state in which proton densities and
neutron densities are separated apart \cite{Lesinski:2006}. Another case of instability
occurred in cranked-HFB calculations performed in $^{194}$Hf, where it has been found
that in certain cases the system was converging to a polarised state, namely a spin phase-transition
was taking place \cite{Hellemans}. One may be tempted to associate these situations with the instabilities
that manifest themselves through imaginary eigenvalues in the response of the system to
an external field, although no formal proof exist, to our best knowledge, of a relationship between
the aforementioned situations and the imaginary eigenvalues.

In order to explore systematically the finite-$\vec q$ instabilities, 
thanks to the Lyon group \cite{Davesne:2011}, a general response
function formalism has been developed for a Skyrme force including central, spin-orbit
and tensor terms. 
This formalism has been applied to symmetric unpolarised nuclear matter with the purpose of
detecting the appearance of imaginary eigenvalues. This work has 
generalised the previous
works of Refs. \cite{Garcia-Recio:1992,Margueron:2006}. The 
response function of uniform matter is labelled by the indices
corresponding to the total spin and isospin ($S$ and $I$) as well as by those corresponding
to their projection on the quantisation axis ($M$ and $Q$). The quantisation axis is chosen in the
direction of the transferred momentum $\vec q$. The label $\alpha$ denotes the set ($S$, $M$; $I$,
$Q$). To find the response function $\chi^{(\alpha)}(q,\omega)$ one starts by solving the
Bethe-Salpeter equation,
\begin{equation}
G^{(\alpha)}_{\rm RPA} (q,\omega,{\vec k_1}) =
G^{(\alpha)}_{\rm HF} (q,\omega,{\vec k_1}) +
G^{(\alpha)}_{\rm HF} (q,\omega,{\vec k_1})
\sum_{\alpha'} \int \frac{d^3k_2}{(2\pi)^3} V_{\rm ph}^{(\alpha,\alpha')}
(q,{\vec k_1},{\vec k_2})
G^{(\alpha')}_{\rm RPA} (q,\omega,{\vec k_2}).
\end{equation}
This equation is written in terms of the HF Green's function $G_{\rm HF}$ and of the
particle-hole interaction $V_{\rm ph}$ derived consistently from the starting Skyrme force. Once
this is solved and the RPA Green's function $G_{\rm RPA}$ is known, the response
function can be easily found as
\begin{equation}\label{chi}
\chi^{(\alpha)}_{\rm RPA}(q,\omega) = g \int\frac{d^3k_1}{(2\pi)^3}
G^{(\alpha)}_{\rm RPA} (q,\omega,{\vec k_1}),
\end{equation}
where $g$ is the degeneracy factor (4 in the case of symmetric unpolarised nuclear
matter). Finally, the main quantity of interest is the strength function defined
as
\begin{equation}
S^{(\alpha)}(q,\omega)  = -\frac{1}{\pi}\ Im\ \chi^{(\alpha)}(q,\omega).
\end{equation}

The main finding of Ref. \cite{Davesne:2011} has been that the tensor
force significantly affects the $S=1$ strength functions. At a critical
density $\rho_c$, the strength function is found to diverge 
for finite values
of $q$. This critical density may be lower than the critical density
at which the strength function diverges for $q=0$ (this latter divergence
coincides, at zero momentum, with the instability defined in terms
of the Landau parameters and discussed in the first part of this
Section). Intuitively, while the strength function divergence ({\it viz.} 
instability) at zero momentum can be thought to concern the whole
uniform medium, the same phenomenon at finite $q$ can be associated
with a finite size instability taking place in a domain whose scale is
$\Delta R\approx 2\pi/q$. On a principle basis, this could be tolerable if the momentum
scale (the real space scale) are much larger (much smaller) than that which matters
for low-energy nuclear physics. However, for practical purposes, it may be
of great advantage to dispose of functionals that are free from those pathologies.

\begin{figure}[htb]
\centering
\includegraphics[width=8cm]{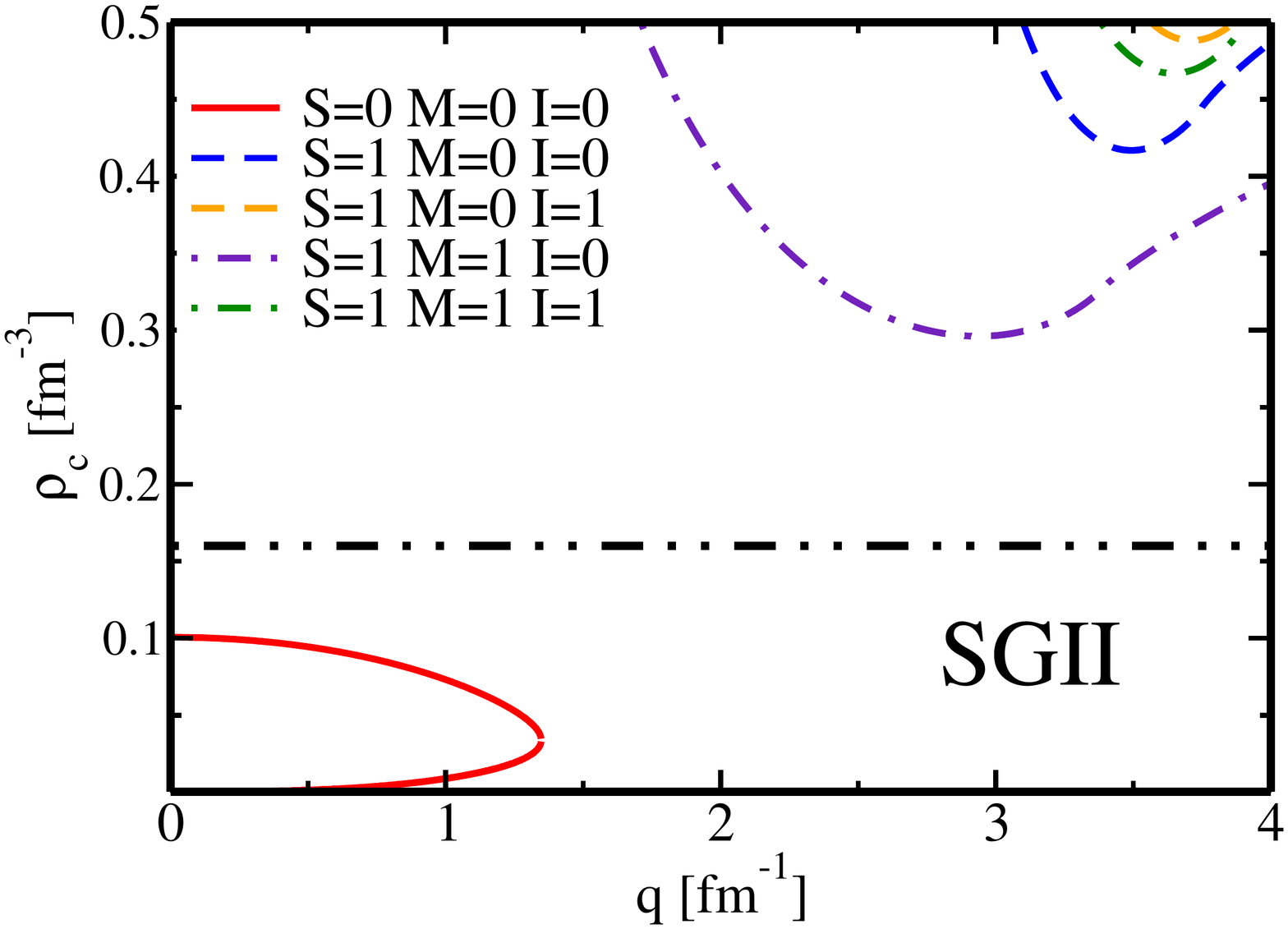}
\includegraphics[width=8cm]{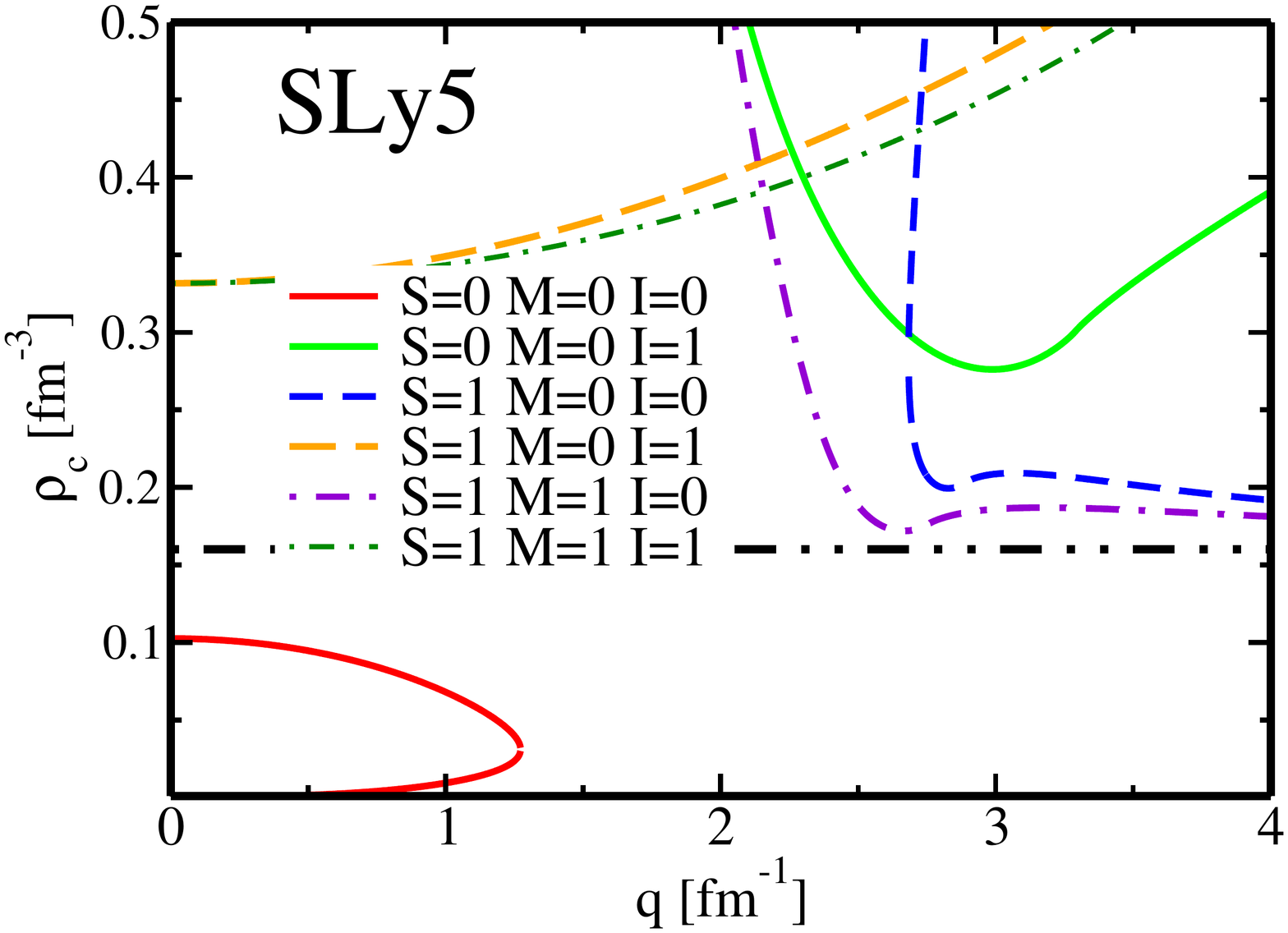}
\includegraphics[width=8cm]{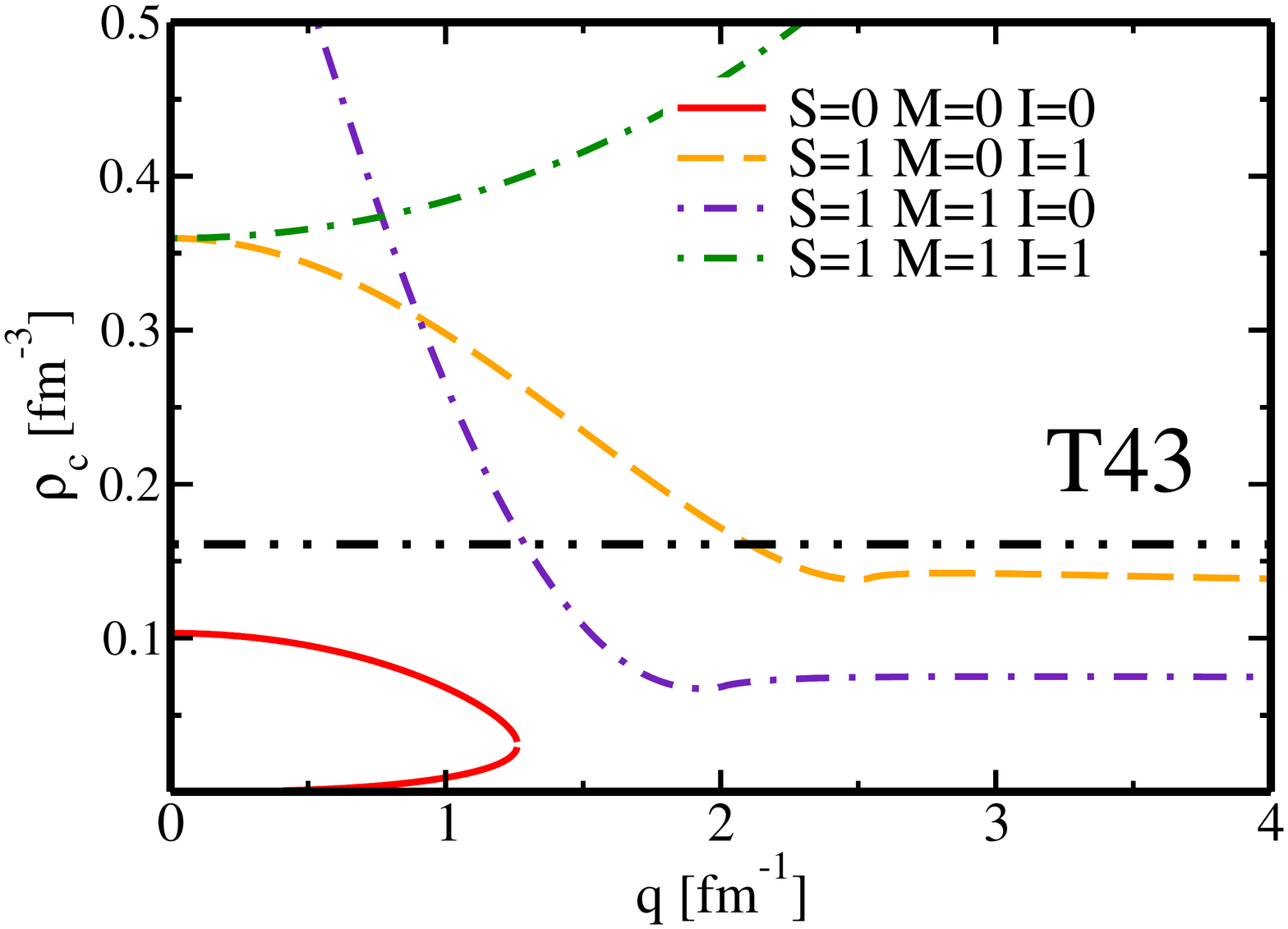}
\includegraphics[width=8cm]{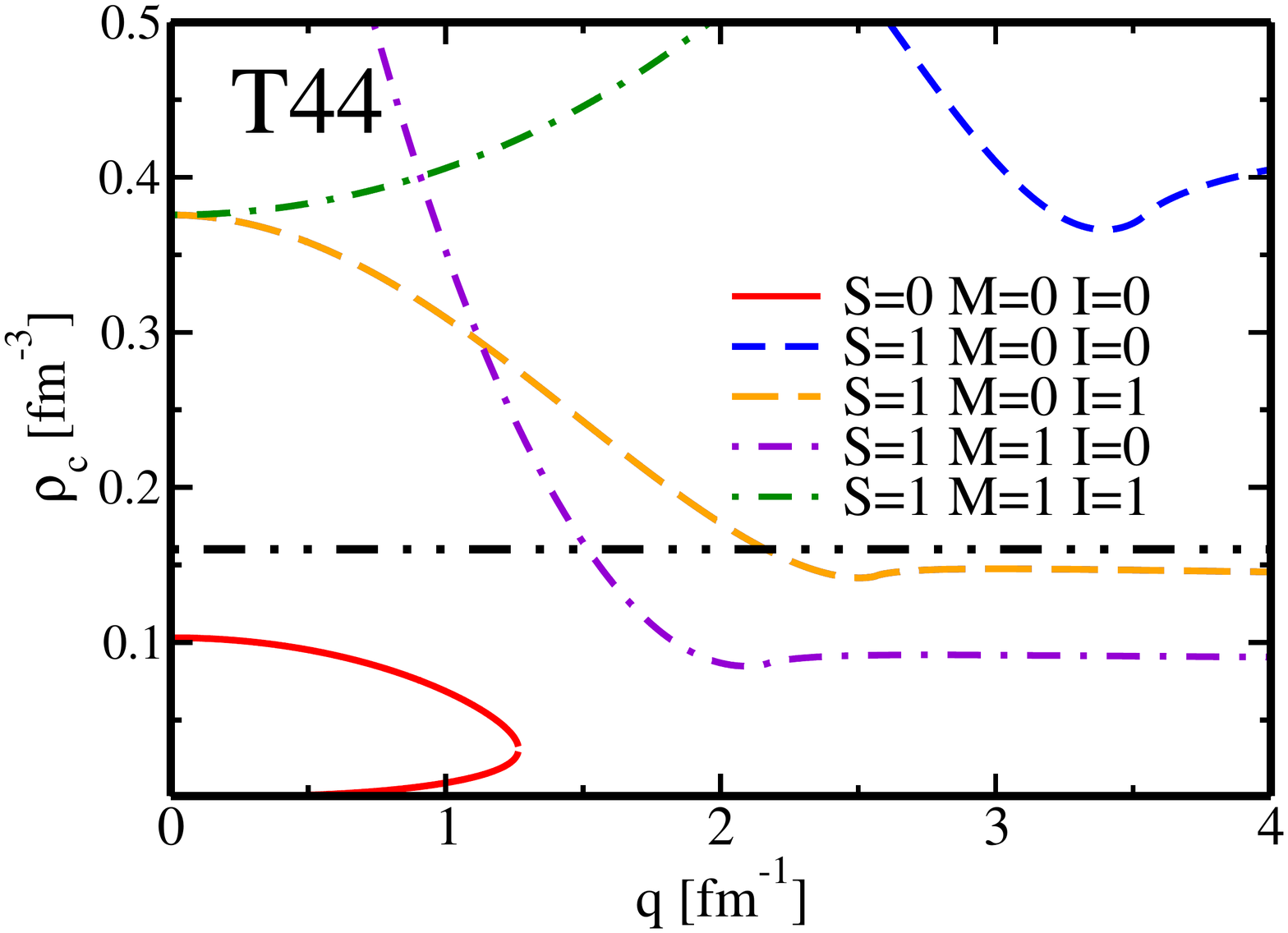}
\caption{\label{fig:instab1}
Critical densities $\rho_C$ as functions of the transferred 
momentum q, in symmetric nuclear matter. These critical densities
have been extracted from the poles of the inverse energy-weighted
sum rule. They are displayed for different channels, and compared
with the saturation density (horizontal line). This figure 
\cite{Pastore:2013} is analogous to Figs. 6 and 7 of Ref. 
\cite{Pastore:2012a}. See the text for a short discussion. 
}
\end{figure}

\begin{figure}[htb]
\centering
\includegraphics[width=8cm]{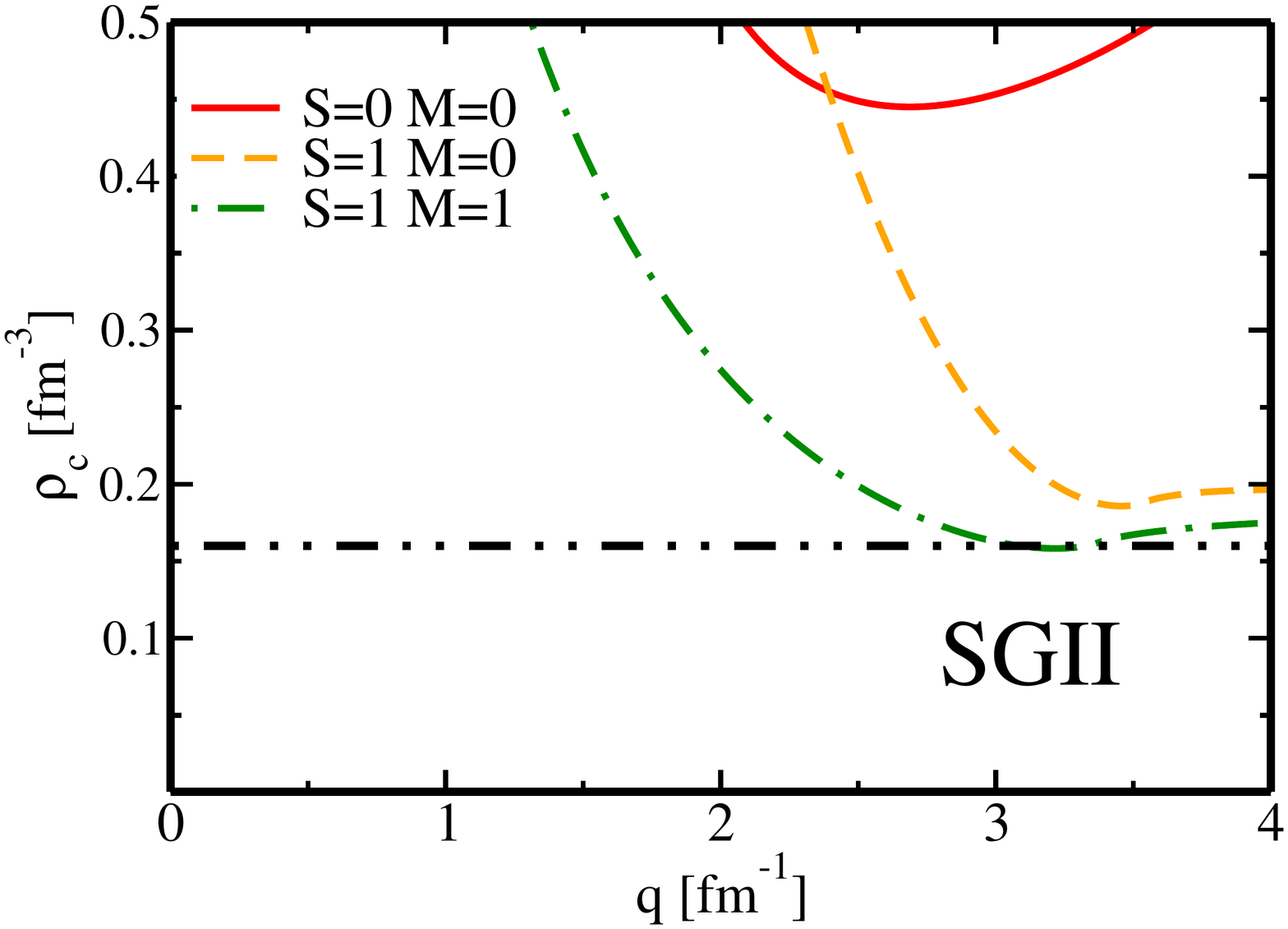}
\includegraphics[width=8cm]{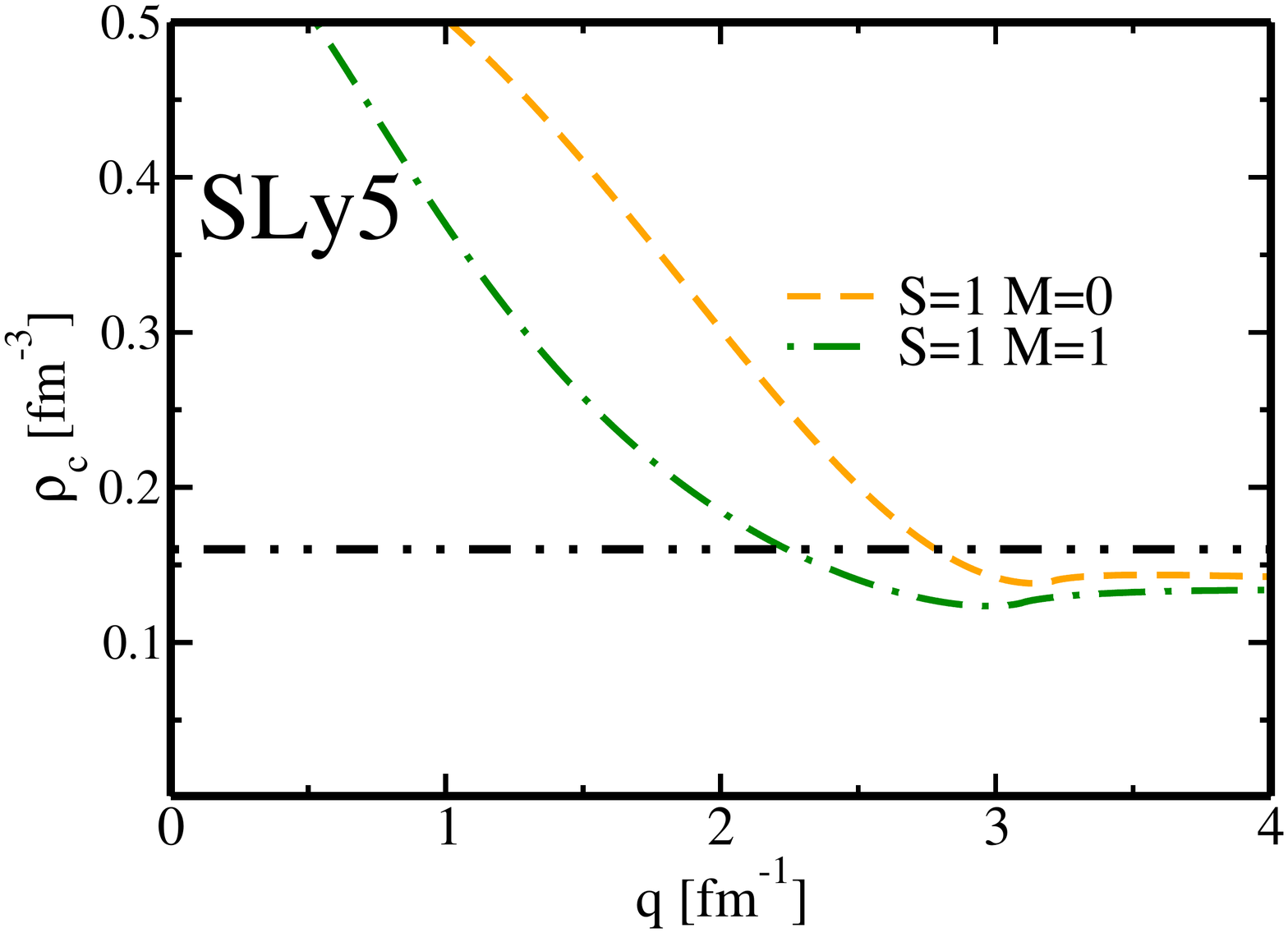}
\includegraphics[width=8cm]{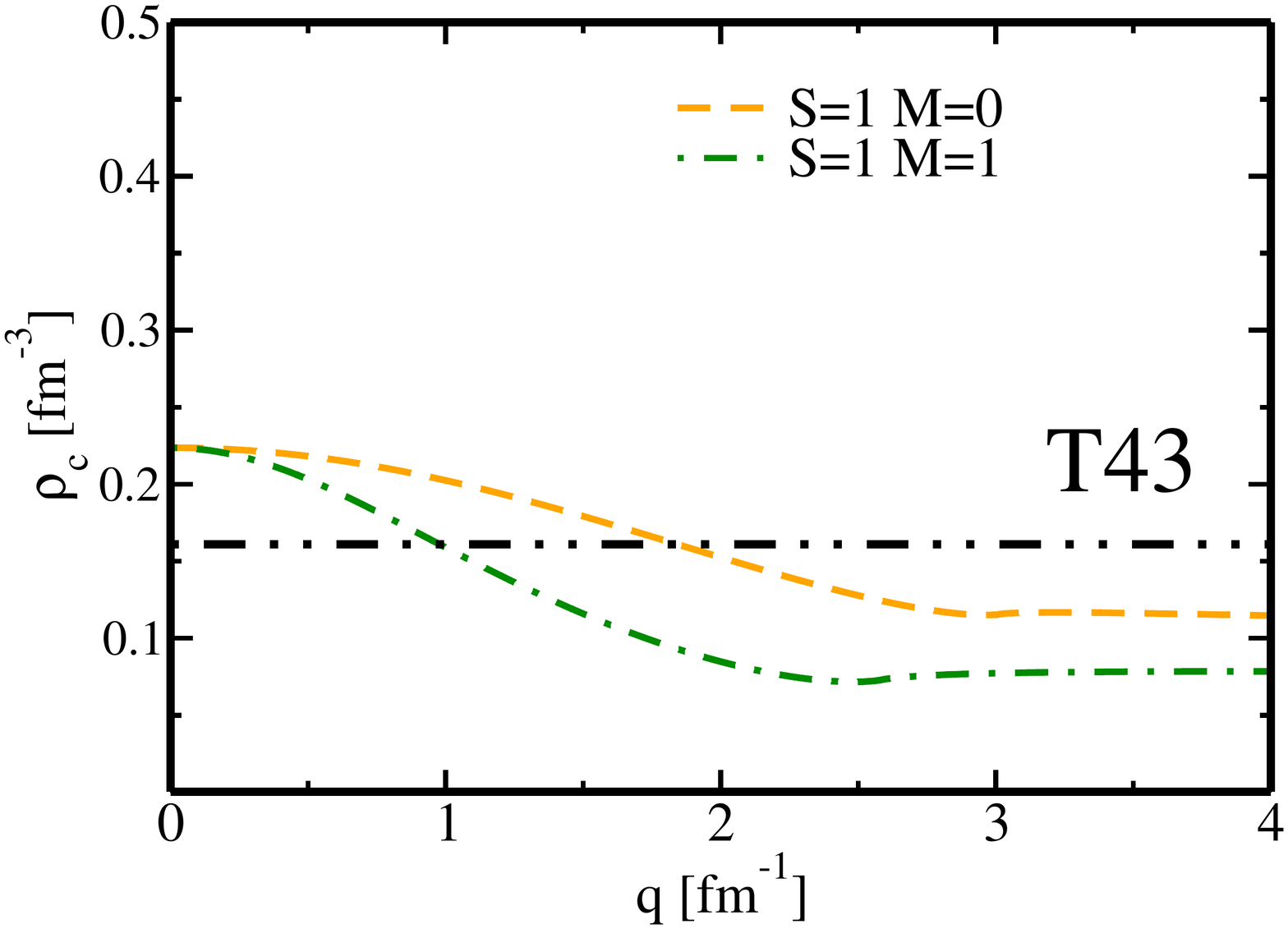}
\includegraphics[width=8cm]{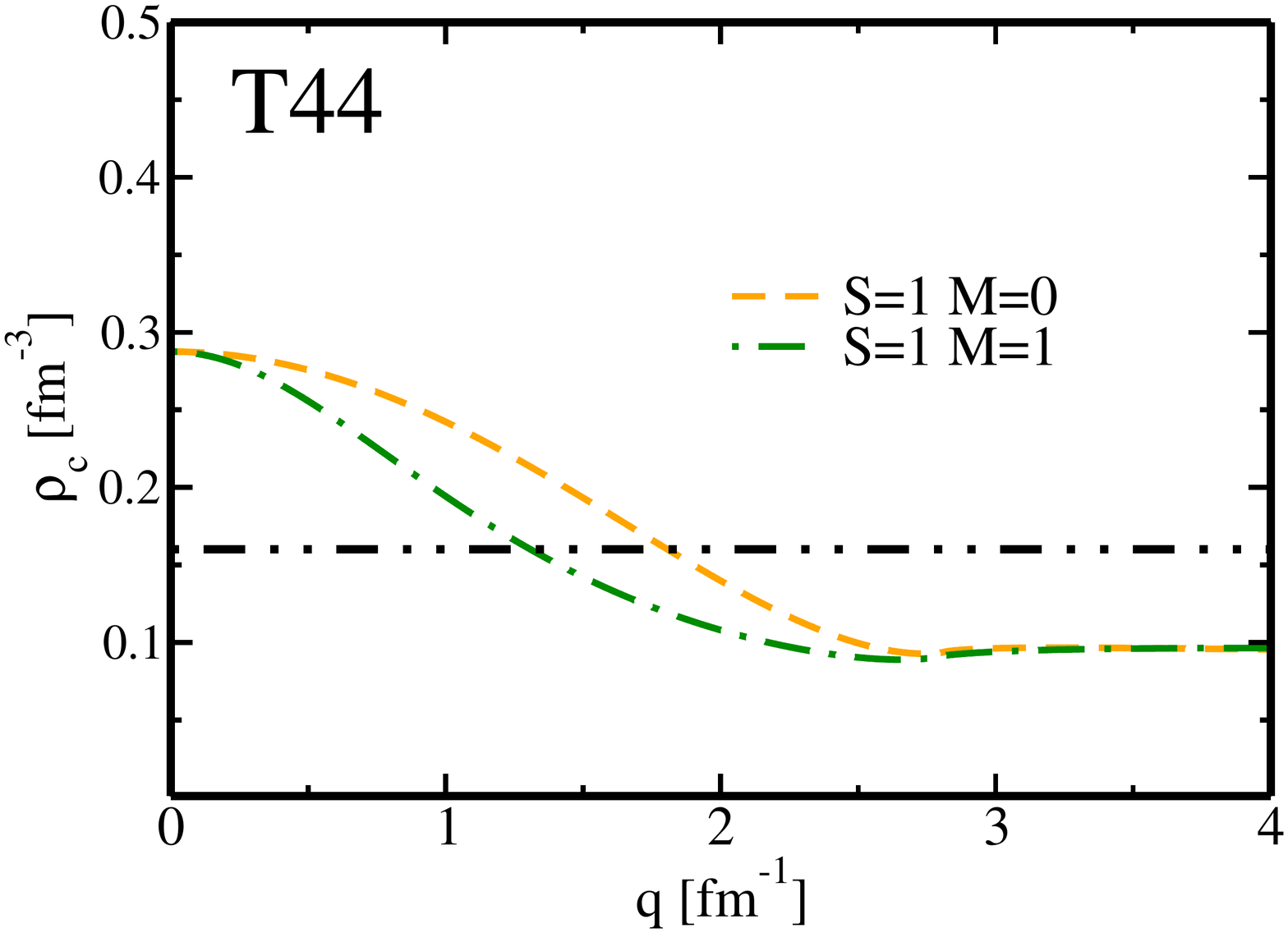}
\caption{\label{fig:instab2}
The same as Fig. \ref{fig:instab1} in the case of pure 
neutron matter \cite{Pastore:2013}.
}
\end{figure}

The work of Ref. \cite{Davesne:2011} has been extended to a functional (not
necessarily derived from an Hamiltonian) in Ref. \cite{Pastore:2012a}. One of the main
goals of this paper is also to find an efficient way to detect the instabilities. It has
been found that the poles in the response function manifest themselves in the
inverse energy-weighted sum rule $m_{-1}$. We remind that, given the
strength function $S(\omega)$, the sum rule $m_k$ is defined as
\begin{displaymath}
m_k=\int d\omega\ \omega^k S(\omega). 
\end{displaymath}
Thus, if instabilities manifest themselves as eigenvalues that
become zero or imaginary, the associated inverse-energy weighted
sum rule will have a pole. The advantage of seeking a pole by
directly analyzing the inverse energy-weighted sum rule is 
that this has an analytical expression which can be calculated 
quite fast. Using this method, in Ref. \cite{Pastore:2012a} a thoroughly
analysis of the poles revealing instabilities in the known Skyrme functionals, has been
undertaken. We display 
in Figs. \ref{fig:instab1} and \ref{fig:instab2} some results obtained
with the method of Ref. \cite{Pastore:2012a}, in the case of the 
Skyrme forces SGII \cite{SGII}, SLy5 \cite{Chabanat}, T43 and T44 
\cite{Les07}. In symmetric nuclear matter, one finds the previously
known spinoidal (i.e., mechanical) instability that corresponds to
the $S=0, M=0, I=0$ channel. The presence of tensor terms favours
the rise of instabilities, as is visible in Fig. \ref{fig:instab1}
by comparing T43 and T44 (that have been discussed in the previous
Sections) with SLy5 and SGII. This feature remains true if we 
look at the case of neutron matter. Thus, the functionals 
T$IJ$ are, generally speaking, very much plagued by instabilities. 
The Lyon group is presently working to include in the fitting protocol 
of a Skyrme functional the
requirement that no instability should manifest at least at a density smaller
than $\approx$ 1.2 times the saturation density. 
Probably the question about a maximum $q$ to be considered should be
raised.

\begin{figure}[htb]
\centering
\includegraphics[width=12cm]{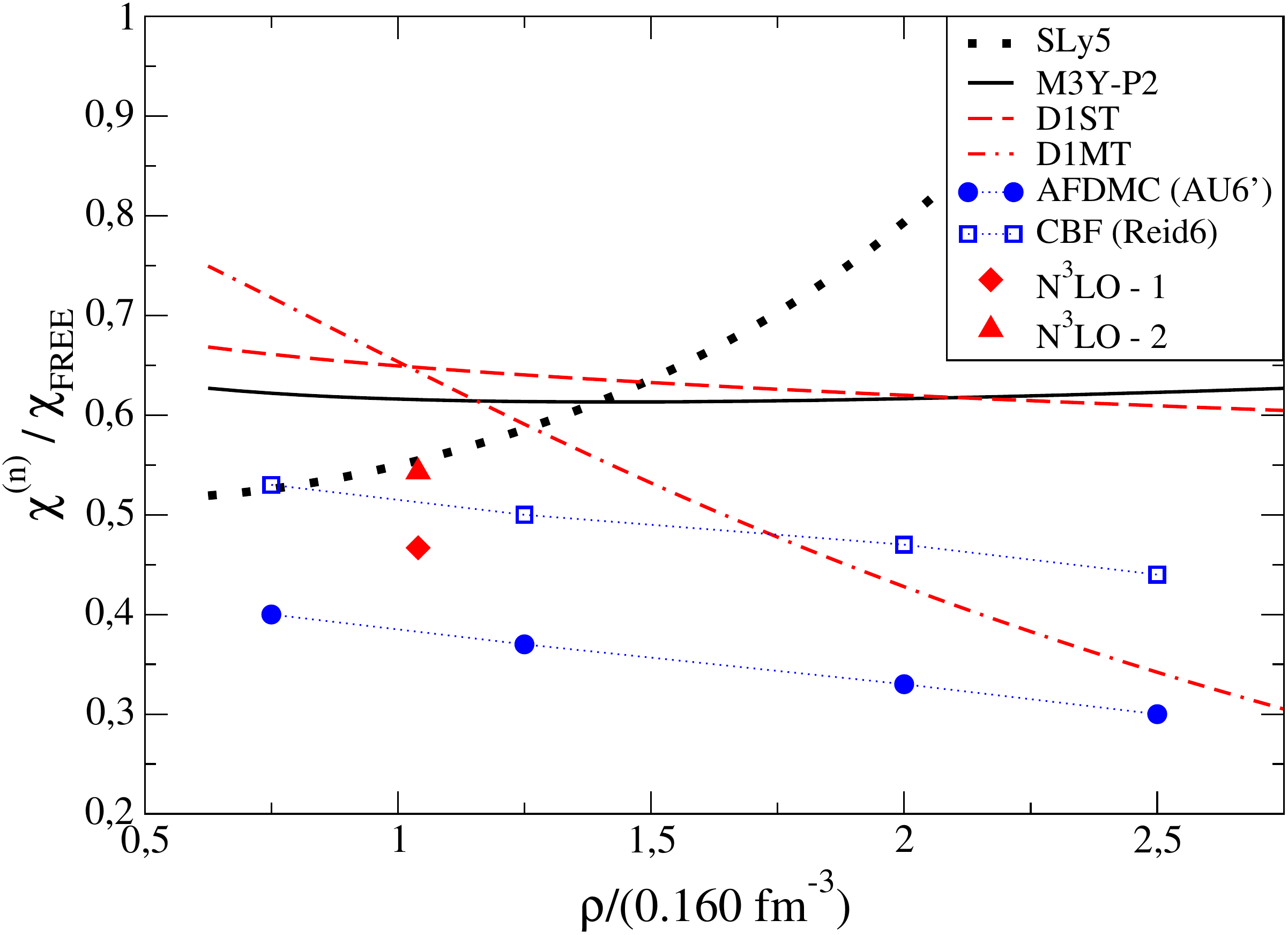}
\caption{\label{fig:chi}
Spin susceptibility of neutron matter (expressed as a ratio of the
free susceptibility) as a function of the density. Results for
both phenomenological and realistic interactions are shown. All of
them include tensor terms. Figure taken from Ref. \cite{Navarro}. 
}
\end{figure}

In the work of Ref. \cite{Navarro} an interesting comparison
between instabilities induced by zero-range and finite-range interactions
is carried out. It has been shown that in the case of zero-range
interactions the addition of tensor terms favours the appearance
of instablities but this is not the case for the finite-range
forces. For instance, the force M3Y-P2 \cite{Nakada1} is quite free from
instabilities although containing a genuine tensor part. 
Another interesting point of Ref. \cite{Navarro} is that the authors
have shown that instabilities manifest themselves in another
quantity that is easier to calculate than the full
response function, namely the spin susceptibility that
is obtained as a proper limit to zero energy and momentum
transfer of Eq. (\ref{chi}). The spin susceptibility 
can be written as $\chi_{\rm RPA}(0)$. Its values, as a 
function of the density, are displayed in Fig. \ref{fig:chi} 
in the case of neutron matter. Given that realistic interactions
can be employed to perform reliable {\em ab-initio} calculations
of neutron matter, and these calculations do not show any trace
of ferromagnetic instablities, it is interesting to compare
the associated values of spin susceptibility with those
exracted from effective interactions that can also include tensor
terms. The values of spin susceptibility from realistic
interactions have still some error bar, as it is evident
from Fig. \ref{fig:chi}; however, the trend of the spin
susceptibility is better followed by finite-range 
forces like M3Y-P2 and D1MT. 

\section{Conclusions and future perspectives}


There has been  a lively debate on the role of the tensor
force in nuclear structure. While in light nuclei the tensor
force manifests itself in the form of the bare force, which
is well known from textbooks in the deuteron case, the 
situation is less clear in medium-mass or heavy nuclei. 
The question about the role of the tensor force in the
effective theories for medium-heavy nuclei, like
self-consistent mean-field or DFT, has been raised
because of the suggestion that the tensor force plays
a crucial role for the shell evolution far from the valley
of stability, towards the proton or neutron drip lines.
At the same time, it is expected that the tensor force is
also important for spin and spin-isospin states.
In this review article, we discussed extensively the role of tensor interactions not only on  the ground state
properties (the binding energies, the single-particle states and the deformations), but also on the excited states (M1 states, GT states, 
SD states and the rotational bands).  
We compared also several different approaches to disentangle the effect 
of tensor correlations on nuclear structure problems, i.e., 
Skyrme Hamiltonians, Skyrme EDFs, RHF and finite-range Gogny tensor 
interactions.

Several groups have attacked the problem of the tensor
force. In some cases, the characteristic feature  of the bare proton-neutron tensor
force has been used as a guideline for
the tensor interaction in the nuclear medium: we refer to 
its attractive (repulsive) 
character in the case 
in which the spins are aligned with  (perpendicular to) 
the relative distance.
 In other cases,
the philosophy has been completely different: based
on the idea that in complex nuclei,  the major effect of the
tensor force is often taken care of by central terms,
the parameters of the tensor force have been treated
as completely free. We have tried, in the present review
paper, to discuss with some care the relationships
between the bare tensor force, and the effective
tensor implemented either in relativistic Hartree-Fock
or in the non-relativistic zero-range and finite-range
frameworks.

The main difficulty faced by the works of many authors
has been that of finding unambiguous signatures
of the tensor force. In fact, in self-consistent mean-field
or DFT, strictly speaking all parameters are coupled
and the specific effects of e.g. the tensor terms 
are not at all easy to disentangle.

Based on the work of Ref. \cite{Otsuka}, many papers
have tried to find a signature of the tensor force 
in the evolution of the single-particle states along
isotopic or isotonic chains. We have reviewed all these 
activities,  that has led to two main conclusions. In
most of the cases, evidence has been found of a
strong attraction between neutrons and
protons lying respectively in $j=l+\frac{1}{2}$ and
$j=l-\frac{1}{2}$ orbits; this interaction becomes
instead  
repulsive when the neutron and the proton lie both in the same 
$j=l+\frac{1}{2}$ orbit, or in the same $j=l-\frac{1}{2}$ orbit.
The interaction between equal particles is far less
clearly established, although it seems that it has the
opposite sign. If one tries to explain systematically
the single-particle spectra based on this picture, however, 
many contradictions show up. It is not completely clear
to which extent these contradictions depend on the
specific ansatz (e.g., the Skyrme ansatz), and to
which extent they point to more general problems.

Certainly, the single-particle states do not belong,
strictly speaking, to the DFT framework, 
i.e., beyond mean field effects such as the particle-vibration coupling could be very important.
 Therefore,
several groups have tried to pay attention to excited states
like spin or spin-isospin modes. The unnatural parity
modes, like excited 0$^-$ states, are quite sensitive to
the tensor force. Also the 1$^+$ (M1) resonance and the
charge-exchange Gamow-Teller resonance are significantly
affected by the tensor force, although these effects cannot be
easily separated from other effects. In other words, one could
pin down precise values of the tensor force parameters provided
the uncertainty on the other terms is much smaller.
A very specific signature of the tensor force is the splitting
between the different components (0$^-$, 1$^-$ and 2$^-$) of the spin-dipole
charge-exchange resonance, which has been recently
measured in $^{208}$Pb and which has to be mainly attributed to the
tensor force. Recently, also the $\beta$-decay in 
exotic magic nuclei has been highlighted. These charge-exchange
transitions point to values of the tensor force parameters that
turn out to be in reasonable agreement with what 
extracted from the single-particle
states (see the previous paragraph).

In the language which is familiar to Skyrme practitioners,
the overall conclusion seems to be that the parameter
$\beta$ [see Eq. (\ref{eq:dWT})] should be definitely positive. The
conclusion on $\alpha$ [see Eq. (\ref{eq:dWT})]  is less clear although
a negative value may be preferable. These signs can be also
extracted from Gogny or other finite-range forces as well as
from RHF Lagrangians.
So far, the isovector-type pion tensor and the $\rho$ tensor interactions are introduced in RHF Lagrangians.
It was found that the strength of these tensor interactions are strongly suppressed in the nuclear media to obtain good binding energy systematics and the shell structure of
heavy nuclei. It is a challenge to include 
the isoscalar-type $\omega$ tensor interaction to improve further the 
RHF model \cite{Long2013}.

Unfortunately, although there is a general consensus on the fact
that a tensor component should be added to all parameter sets,
attention has been drawn on the fact that some implementation
of the tensor force lead to instabilities of the energy functionals.
Spin and spin-isospin instabilities (i.e., transitions to
configurations that display spontaneous ferromagnetization) 
plague many paramatrizations of EDFs that include the tensor
force. This is especially true for the Skyrme EDFs. Care should
be taken to avoid this.

Ultimately, the introduction of the tensor force has opened
several new lines of investigation. The tensor terms in the EDFs
cannot be discarded but further analysis of the instabilities, 
of the properties of spin-isospin transitions and of the
correlations associated with single-particle states are called
for.

\section*{Acknowledgments}

This work stems from several previous papers and collaborations
and it is our pleasure to acknowledge these collaborations here,
in particular with C.L. Bai, P.F. Bortignon, L. G. Cao, 
S. Fracasso, A. Li, Z.Y. Ma, L. Sciacchitano, F.R. Xu, 
H.Q. Zhang, X.Z. Zhang, X.R. Zhou and W. Zou.
We also acknowledge fruitful discussions with D. M. Brink, G. Co', T. Duguet, M. 
Grasso, M. Ichimura, H.Z. Liang, W.H. Long, T. Otsuka, H. Sakai, T. Suzuki, N. Van Giai, K. Yako and T. Wakasa.  
We would also like to warmly thank all those who helped by
providing or preparing figures for this paper, namely 
T. Abe, M. Anguiano, C.L. Bai, G. Co', T. Duguet,
M. Grasso, T. Lesinski, W.H. Long, F. Minato,  
H. Nakada, J. Navarro, T. Otsuka, A. Pastore, A. Polls, and V. Hellemans.

\end{document}